\documentclass[lettersize,journal]{IEEEtran}

\usepackage{amsmath,amsfonts}

\usepackage[ruled,linesnumbered]{algorithm2e}

\usepackage[caption=false,font=footnotesize]{subfig}

\usepackage{array}
\usepackage{textcomp}
\usepackage{stfloats}
\usepackage{booktabs}
\usepackage{url}
\usepackage{siunitx}   
\usepackage{verbatim}
\usepackage{graphicx}
\usepackage{multirow}
\usepackage{cite}

\begin{document}

\title{Agon: A Semi-Supervised Framework for Robust Satellite Interference Detection}

\author{Boyu Yang,~\IEEEmembership{Student Member,~IEEE},
        Chunyu Yang,
        Zhe~Chen,~\IEEEmembership{Member,~IEEE},\\
        Kun~Qiu,~\IEEEmembership{Senior Member,~IEEE},
        and Yue~Gao,~\IEEEmembership{Fellow,~IEEE}
\thanks{A preliminary version of this work appeared in part at IEEE/CIC ICCC 2025 \cite{11148638}. This work was supported by the National Natural Science Foundation of China under Grant U25A20396. 
Boyu Yang is with the College of Computer Science and Artificial Intelligence, Fudan University, Shanghai 200438, China, and also with the Space Internet Research Institute, Fudan University, Shanghai 200438, China.(Corresponding author: Kun Qiu.) (email: 24110240144@m.fudan.edu.cn).}%

\thanks{Zhe Chen, Kun Qiu and Yue Gao are with the Space Internet Research Institute, Fudan University, Shanghai 200438, China. (email: zhechen@fudan.edu.cn; qkun@fudan.edu.cn; gao.yue@fudan.edu.cn).}%
\thanks{Chunyu Yang is with the College of Computer Science and Artificial Intelligence, Fudan University, Shanghai 200438, China (email: 22307140114@m.fudan.edu.cn).}%
}

\markboth{IEEE TRANSACTIONS ON MOBILE COMPUTING}%
{Shell \MakeLowercase{\textit{et al.}}: A Sample Article Using IEEEtran.cls for IEEE Journals}

\maketitle

\begin{abstract}
The rapid expansion of non-geostationary orbit (NGSO) satellites alongside existing geostationary orbit (GSO) systems has intensified spectrum congestion and inter-system interference, placing stringent demands on real-time interference management to sustain reliable coexistence in next-generation communication networks. While existing machine learning (ML)--based reconstruction models have made strides, they remain constrained to an area under the curve (AUC) of 0.83 due to fixed thresholds, causing unacceptable false alarm rates that undermine critical link reliability. Additionally, their decoupled training paradigm neglects cross-domain dependencies, limiting time and frequency-domain AUCs to 0.83 and 0.71, respectively. To address these limitations, this paper introduces a semi-supervised satellite interference detection framework named Agon, employing a novel two-stage hybrid learning paradigm. Agon integrates masked autoencoder (MAE) pre-training of a dual attention transformer (DAT) with multi-task fine-tuning to optimize a direct binary classifier, effectively eliminating unstable thresholds. Furthermore, it incorporates high-order statistics (HOS)-augmented attention and wavelet regularization to bolster noise robustness and structural fidelity. Extensive validation on public NGSO-GSO dataset and a high-fidelity NGSO-NGSO dataset demonstrates that Agon achieves state-of-the-art (SOTA) detection performance, with a 25.3\% improvement in AUC. Moreover, the multi-task learning (MTL) framework facilitates accurate modulation classification with accuracies exceeding 90\%, while simultaneously maintaining optimal detection performance across diverse scenarios characterized by varying off-axis angles and interference-to-noise ratios (INRs).
\end{abstract}


\section{Introduction}
With the exponential expansion of geostationary orbit (GSO) and non-geostationary orbit (NGSO) satellite constellations, global connectivity is undergoing a profound transformation, making low-latency and high-throughput broadband services possible \cite{yahia2024evolution}. However, the deployment of tens of thousands of satellites has not only intensified competition for limited spectrum resources but also introduced interference issues between satellites, further complicating spectrum management \cite{kodheli2020satellite}. This interference, particularly in shared frequency bands, is causing increasing potential conflicts between NGSO systems, traditional GSO systems, and ground networks. To ensure the effective coexistence of future non-terrestrial networks (NTNs) and terrestrial networks, as well as the reliability of critical communication services, dynamic spectrum sharing \cite{gu2021dynamic} and efficient interference management \cite{ruan2019energy} mechanisms are crucial. In this context, the development of advanced interference detection technologies that ensure service quality while complying with international regulations has become the foundation for the stable operation of next-generation satellite communication systems.

However, achieving such stability is increasingly difficult as the rapid expansion of NGSO constellations profoundly reshapes global communications and intensifies spectrum complexity \cite{al2022survey}. Despite coordination efforts by the International Telecommunication Union (ITU) \cite{ITU}, its static frameworks struggling with over 123,000 annual filings and a standard 7-year regulatory cycle lag behind the sub-second interference dynamics of dense constellations, leaving a critical gap in real-time management. With constellations projecting to exceed 100,000 satellites by 2030, interference probability has risen sharply. Traditional physical link budget calculations fail to meet the millisecond-level latency requirements of dynamic NGSO systems \cite{giggenbach2023link}. Additionally, the generally low signal-to-noise ratio (SNR) and high temporal variability of satellite channels further exacerbate the difficulty of robust detection. Therefore, there is an urgent need for new interference detection solutions that are capable of efficient computation, meeting real-time demands, and ensuring reliable performance.

In recent years, interference detection has evolved from traditional analytical techniques \cite{hao2020interference} to learning-based paradigms. Since satellite mobility introduces stochastic optimization challenges, reinforcement learning (RL) \cite{wang2024toward} and multi-agent RL (MARL) \cite{wang2025novel} have been widely investigated for adaptive control and optimization in dynamic mobile networks. While general signal detection remains a thoroughly investigated academic field, the rapid expansion of NGSO constellations introduces entirely new dynamic complexities. However, RL-based schemes excel primarily in control policy optimization, whereas detection tasks in such stochastic environments often rely on reconstruction-based models to identify signal deviations \cite{liu2025enhancing}. Advanced attention architectures like the transformer based interference detector \cite{saifaldawla2024genai} achieve state-of-the-art (SOTA) status yet remain fundamentally hindered by unstable reconstruction error thresholds \cite{baraniuk2017exponential}. They strictly depend on exactly 17281 fully annotated snapshots from a specific public benchmark dataset which severely limits practical application.

Beyond these data dependencies, the research problem requires handling sub-second spectral variations and extreme low SNR conditions inherent in dynamic satellite links. The critical research gap therefore lies in achieving reliable real-time interference detection without relying on highly unstable fixed thresholds under such stochastic channel conditions \cite{vazquez2021machine}. To evaluate current limitations, the receiver operating characteristic (ROC) curve and the area under the curve (AUC) serve as primary metrics \cite{fawcett2006introduction}. The ROC curve shows the trade-off between detection probability and false alarms, while the AUC measures the overall ability to distinguish interference from noise. Evaluated by these standard metrics, existing Transformer based decoupled training paradigms often process time and frequency domains in isolation. By ignoring essential cross-domain information, this isolation caps their AUC at 0.8318 and their F1 score at 0.8321. These isolated Transformer models also impose high computational burdens resulting in latencies exceeding 100 ms which remain incompatible with rapid NGSO link adaptation \cite{ITU_2024}.

To address these specific challenges, this paper proposes a semi-supervised framework named Agon designed to achieve high-precision and robust interference detection. This framework demonstrates superior performance on the public NGSO-GSO dataset yielding AUC and F1 scores of 0.9327 and 0.8351 while delivering robust results on the high-fidelity NGSO-NGSO dataset with scores of 0.9085 and 0.9055. The architecture employs a novel two-stage hybrid paradigm where the core dual attention transformer (DAT) encoder first undergoes pre-training via a masked autoencoder (MAE) task on unlabeled data to learn universal signal structures. Subsequently a multi-task learning (MTL) strategy fine-tunes the model on limited labeled data optimizing a direct binary classifier to eliminate threshold dependency. By integrating high-order statistics (HOS) augmented dual attention alongside a wavelet regularization loss Agon significantly boosts noise discrimination and preserves multi-scale structural fidelity.
The key contributions of this paper are summarized as follows:
\begin{itemize}
    \item This paper proposes Agon, a novel semi-supervised framework that transitions from self-supervised structural pre-training to multi-task fine-tuning leveraging unlabeled data to learn robust signal representations and fundamentally addressing decision threshold instability.
    \item Agon incorporates a HOS-augmented dual attention mechanism that reformulates the attention scoring function by incorporating second-order statistical priors, allowing the framework to identify complex statistical signatures typically ignored by linear attention projections.
    \item Agon implements a novel wavelet regularization loss within the objective function by applying multi-scale constraints in the discrete wavelet domain, which ensures the precise recovery of transient spectral details and preserves structural fidelity during signal reconstruction.
    \item A comprehensive empirical validation framework systematically substantiates the architectural generalizability across diverse signal environments with exceptional out-of-distribution robustness and cross-band adaptability of Agon against conventional baseline methodologies.
\end{itemize}

This paper is organized as follows. Section~\ref{sec2} reviews related works and technical limitations. Section~\ref{sec3} motivates the design of Agon by revealing the challenges faced by satellite interference detection. Section~\ref{sec4} presents the design overview of Agon. Section~\ref{sec5} introduces the training and detection algorithm design of Agon. Section~\ref{sec6} details the system implementation and experiment setup, followed by performance evaluation. Finally, conclusions and future research are presented in Section~\ref{sec7}.

\section{Related Work}
\label{sec2}
In this section we provide a concise taxonomical review of these technical milestones categorizing the existing literature into traditional interference detection methods alongside early deep learning methods and the frontiers in generative artificial intelligence (AI) to highlight the fundamental research gaps.

\subsection{Traditional Interference Detection Methods} 
\label{sec21}
Interference detection in satellite communications has traditionally relied on conventional signal processing methods. Among these, energy detectors (EDs) are widely used due to their low computational complexity and effectiveness in detecting random signals under Gaussian white noise \cite{steven1993fundamentals}. Lipski et al. \cite{lipski2020practical} developed and evaluated a real-time adaptive threshold energy detector based on software-defined radio, while Sobron et al. \cite{sobron2015energy} applied energy detection for both cooperative and non-cooperative spectrum sensing. However, in dense modern NGSO environments, EDs exhibit limited sensitivity to weak interference, and their performance depends heavily on precise time-window configurations and unstable power thresholds. As an alternative, cyclostationary feature detection exploits the periodic statistical properties of signals to mitigate the challenges posed by low SNR \cite{enserink1994cyclostationary}. Dimc et al. \cite{dimc2015experimental} demonstrated that this approach provides greater robustness than energy-based detection in low-SNR conditions. Nevertheless, traditional methods entail the computation of all periodic frequencies, resulting in high computational complexity and limited adaptability to the rapidly time-varying propagation characteristics of NGSO satellites.

\subsection{Early Deep Learning Methods} 
\label{sec22}
To address the limitations of traditional approaches, machine learning (ML) techniques have been increasingly adopted for interference detection, offering augmented accuracy and adaptability to complex signal environments. Zhang et al. \cite{usynin2021adversarial} employed deep autoencoders for end-to-end communication over two-user interference channels, while Liu et al. \cite{liu2022deep} utilized fully connected deep neural networks for interference detection in communication systems. Given the stringent real-time performance and data processing requirements of satellite communications, deep learning models face additional challenges \cite{vazquez2021machine}. Pellaco et al. \cite{pellaco2019spectrum} proposed an autoencoder-based long short-term memory model to detect intentional interference in NGSO communication signals. However, such early sequential models suffer from high inference latency due to their recursive structures and lack of parallelism, rendering them unsuitable for dynamic satellite systems that demand real-time responsiveness.

\subsection{Frontiers in Generative AI and Self-Supervised Learning} 
\label{sec23}
In recent years, generative AI models have achieved notable advances in interference detection. Zhao et al. \cite{zhao2021variational} employed variational autoencoders to develop a variational optimization framework for efficient signal detection. Zhang et al. \cite{zhang2023self} introduced a self-supervised variational graph autoencoder for system-level anomaly detection. In the context of satellite communications, Mascher et al. \cite{mascher2023hybrid} applied an autoencoder to quantify the severity or level of concern of detected interference. Saifaldawla et al. \cite{saifaldawla2024convolutional} leveraged a variational autoencoder to learn the latent representation of normal data distributions, thereby generating interference-free signals and using reconstruction error as a noise indicator. The TrID model \cite{saifaldawla2024genai} further integrates a Transformer-based self-attention mechanism to enhance long-term sequence dependency modeling, enabling the generation of expected GSO signal samples to address NGSO-to-GSO interference. Despite these advancements, existing models share a fundamental limitation: their reliance on reconstruction error thresholds for final decision-making. This dependence on empirically determined thresholds leads to unstable detection performance, heightened sensitivity to environmental variations, and challenges in accurately calibrating false positive and false negative rates in real-world applications.

\section{Motivation and Background}
\label{sec3}
In this section we establish the analytical foundation for satellite interference detection by characterizing the evolving landscape of NGSO interference alongside the mathematical modeling of dynamic signals and the resulting regulatory standards to highlight the fundamental deficiencies in current detection paradigms.

\subsection{The Evolving Landscape of NGSO Interference}
\label{sec31}
\begin{figure}[t]
    \includegraphics[width=\columnwidth]{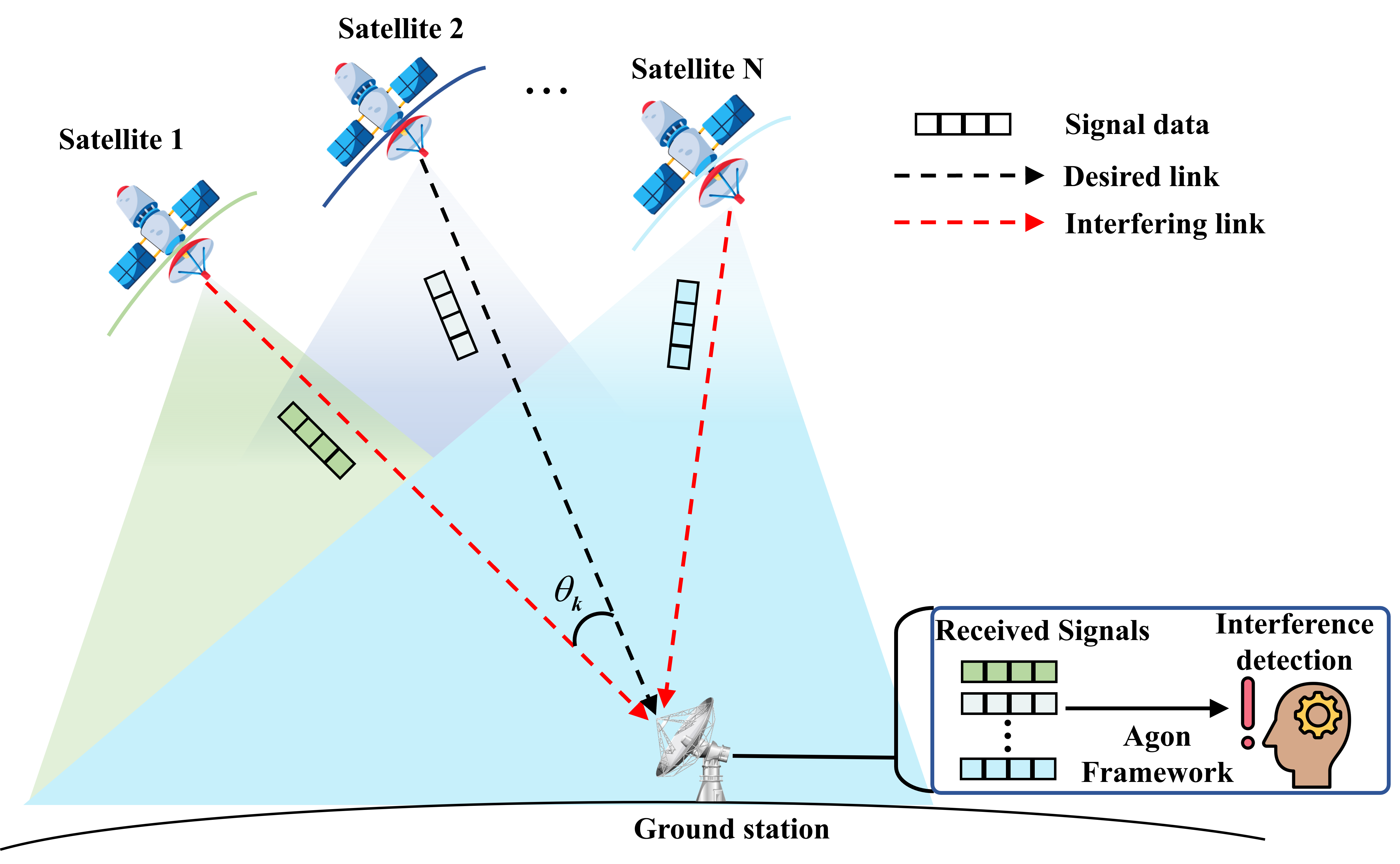}
    \caption{An example scenario for satellite interference detection.}
    \label{fig1}
\end{figure}

The telecommunications sector is currently witnessing a paradigm shift driven by the exponential proliferation of NGSO megaconstellations \cite{al2022survey}. The operational environment has transitioned from static, predictable GSO topologies to highly complex, multi-layer dynamic networks. In this evolving context, as illustrated in Fig.~\ref{fig1}, a ground station aiming to maintain a reliable link with a primary service satellite is frequently subjected to co-channel interference from a rapidly time-varying set of visible NGSO satellites \cite{kodheli2020satellite}. The high relative velocities of NGSO satellites induce interference patterns that fluctuate on a millisecond timescale. Consequently these transient signal conflicts render traditional static analysis ineffective for real time operations. Static analysis in this context refers to conventional offline link budget evaluations \cite{ITU_S1503} which focus on long term cumulative impact but fail to capture sub second signal dynamics such as rapid Doppler frequency shifts and millisecond level power fluctuations. Compounding this physical complexity is a substantial regulatory lag that reflects mounting coordination burdens and an increasingly inadequate institutional response. Projections suggest active satellites will exceed 100,000 by 2030 and generate rapid high-speed operational conflicts. The ITU framework \cite{ITU} cannot manage such dynamics due to 123,000 annual filings and a 7-year cycle. This widening gap between static regulation and dynamic operational reality necessitates the development of autonomous, real-time detection mechanisms at the receiver end to ensure system coexistence.

\subsection{Mathematical Modeling of Dynamic Signals}
\label{sec32}
To scientifically address the detection problem, it is essential to rigorously model the signal characteristics governed by the dynamic link geometry, the key notations and descriptions used throughout this paper are summarized in Table~\ref{tab_notation}. First, we quantify the geometric relationship using the off-axis angle $\theta_k(t)$. This angle represents the angular separation between the ground station's boresight direction aligned with the vector to the desired satellite $\vec{V}_{GS \to d}(t)$ and the vector to the $k$-th interfering satellite $\vec{V}_{GS \to k}(t)$:
\begin{equation}
\label{equ_theta}
\theta_k(t) = \arccos\left( \frac{\vec{V}_{GS \to d}(t) \cdot \vec{V}_{GS \to k}(t)}{||\vec{V}_{GS \to d}(t)|| \cdot ||\vec{V}_{GS \to k}(t)||} \right).
\end{equation}

\begin{figure}[t]
  \centering
  \subfloat[Ground station antenna gain pattern. The antenna's high directivity, with a sharp main lobe at $0^{\circ}$ boresight and rapidly decaying side lobes, which means even strong interference sources can be significantly attenuated if they fall into the antenna's side lobes or nulls.]{
    \includegraphics[width=0.46\columnwidth]{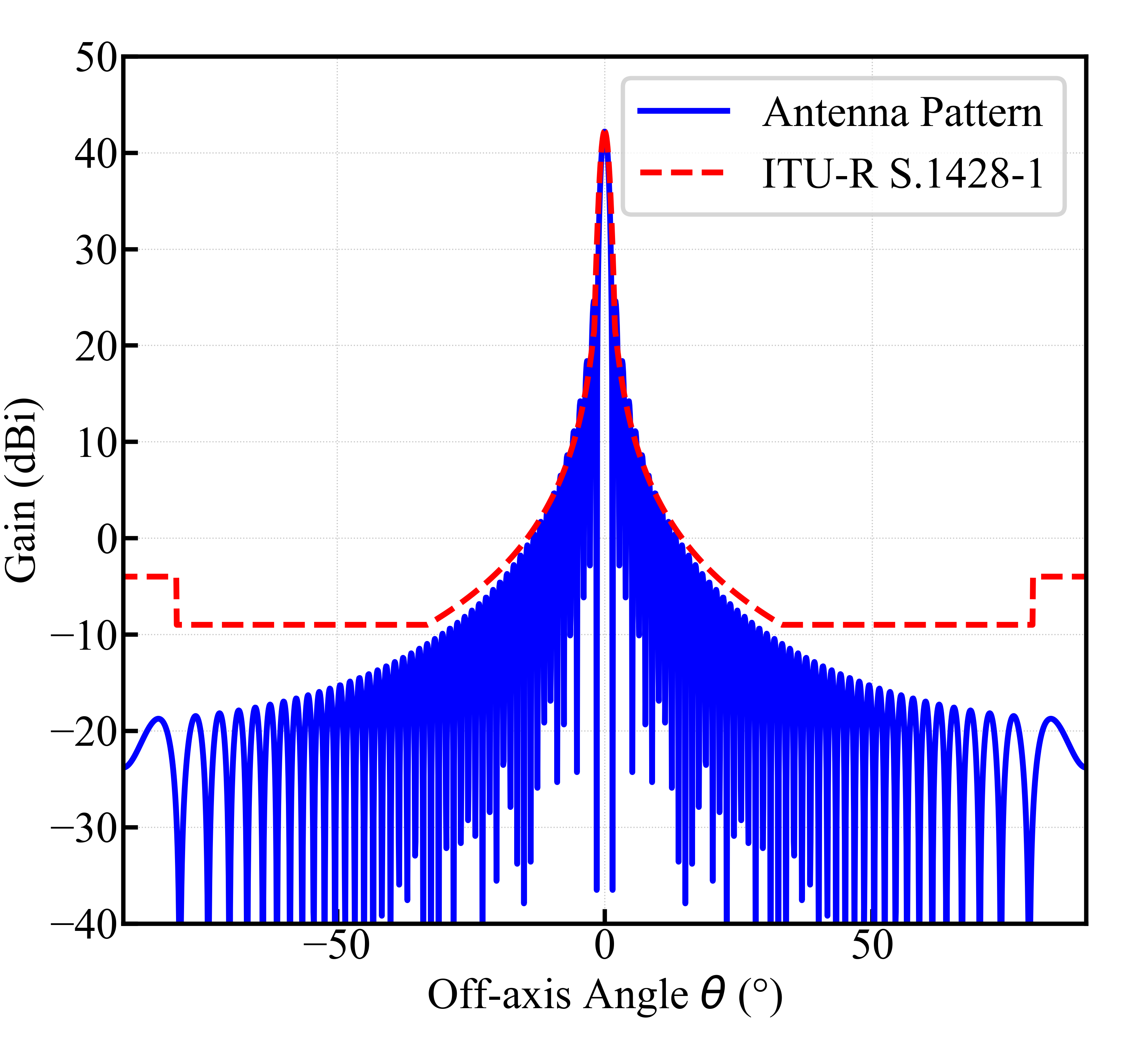}
    \label{fig2a}
  } \hspace{0.01\linewidth}
  \subfloat[Example heatmap of aggregate EPFD. The heatmap reveals spatial non-uniformity in interference, with localized hotspots where the aggregate power exceeds the regulatory threshold, defining our detection task as regions where spectral integration breaches the ITU limit.]{
    \includegraphics[width=0.46\columnwidth]{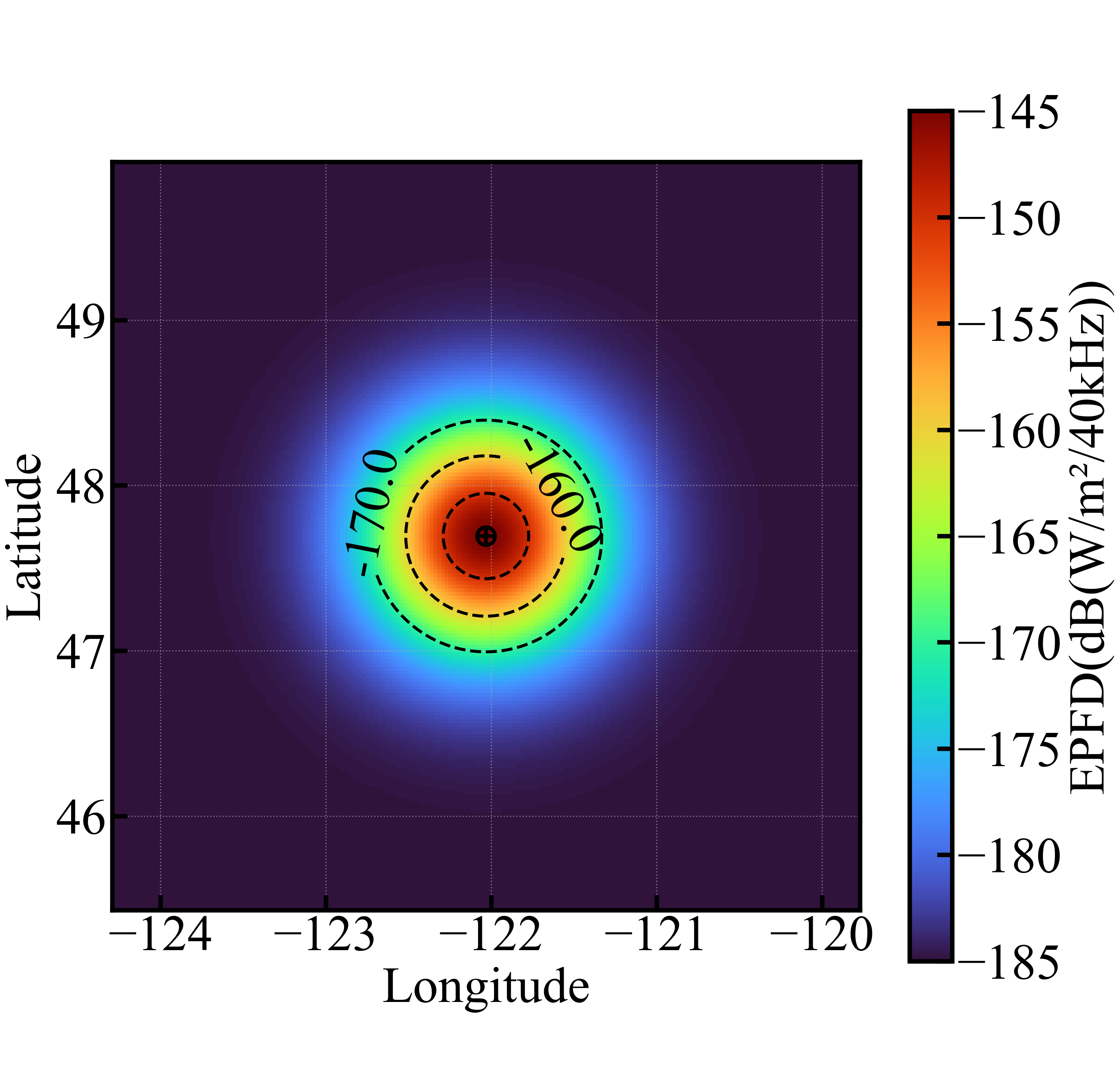}
    \label{fig2b}
  }
  \caption{Physical characteristics of the ground station antenna and the resulting interference spatial distribution.}
  \label{fig2}
\end{figure}

This geometric parameter is critical as it determines the received interference power $I_k$ via the antenna gain pattern $G_r(\theta_k)$. As shown in Fig.~\ref{fig2}\subref{fig2a}, we adopt the standard ITU-R S.1428-1 reference pattern \cite{ITU-R_S1428-1}, where the steep gain roll-off creates significant spatial selectivity. Based on this, the instantaneous Interference-to-Noise Ratio (INR) is derived as:
\begin{equation}
\label{equ_inr}
I_k = \frac{EIRP_k \cdot G_r(\theta_k)}{L_{FS,k} \cdot L_{add}}, \quad INR_{k,n} = \frac{I_{k}}{N_0 B},
\end{equation}
where $EIRP_k$ denotes the equivalent isotropically radiated power of the $k$-th interferer, $L_{FS,k}$ is the free-space path loss, and $L_{add}$ accounts for additional atmospheric losses. The noise floor is defined by the noise power spectral density $N_0$ and system bandwidth $B$. Similarly, the Carrier-to-Noise Ratio (CNR) for the desired link is formulated as:
\begin{equation}
\label{equ_cnr}
CNR_n = \frac{EIRP_d \cdot G_r(0)}{L_{FS,d} \cdot L_{add} \cdot N_0 B},
\end{equation}
where $EIRP_d$ represents the radiated power of the desired satellite, $L_{FS,d}$ is the corresponding path loss, and $G_r(0)$ denotes the maximum receive gain assuming perfect tracking alignment. Incorporating these parameters, the composite baseband signal $y_n(t)$ is modeled as a linear superposition of the desired signal, aggregate interference, and noise:
\begin{equation}
\begin{split}
&y_n(t)=\\
&x_{d,n}(t) \sqrt{CNR_n}+\sum_{k=1}^{K_n} x_{k,n}(t) e^{j2\pi \Delta f_{k,n} t} \sqrt{INR_{k,n}}+\zeta_n(t),
\end{split}
\label{equ7}
\end{equation}
where $K_n$ denotes the number of visible interfering satellites while $x_{d,n}$ and $x_{k,n}$ represent the normalized baseband waveforms. For the purpose of establishing a consistent power reference across diverse link conditions we assume the complex additive white Gaussian noise $\zeta_n(t)$ has zero mean and unit variance. By multiplying the desired signal and the $k$-th interferer by $\sqrt{CNR_n}$ and $\sqrt{INR_{k,n}}$ respectively we precisely calibrate their power levels relative to this unit noise floor. The term $\Delta f_{k,n}$ accounts for the Doppler frequency shift induced by the high relative velocity of the NGSO satellites. Given the non-stationary nature of these signals caused by dynamic Doppler shifts and power variations, traditional 1D time-domain analysis is often insufficient. Therefore, we adopt the Pseudo Wigner-Ville Distribution (PWVD) to generate a high-resolution Time-Frequency representation $\mathbf{Y} \in \mathbb{R}^{T \times F \times C}$:
\begin{equation}
\label{equ_pwvd}
y_n^{\mathrm{PWVD}}(t,f)=\int_{-\infty}^\infty y_n(t+\frac\tau2)y_n^*(t-\frac\tau2)h(\tau)e^{-j2\pi f\tau}d\tau,
\end{equation}
where $\tau$ represents the time lag, $h(\tau)$ is the smoothing window function, $T$ and $F$ denote the number of time frames and frequency bins, respectively. Unlike standard Fourier transforms which smear transient features, this 2D representation for in-phase and quadrature (I/Q) components effectively concentrates energy, capturing transient spectral overlaps essential for distinguishing interference from noise \cite{shin1993pseudo}.

\begin{table}[t]
\centering
\caption{Notation and Description}
\label{tab_notation}
\renewcommand{\arraystretch}{1.3}
\setlength{\tabcolsep}{2.5pt} 
\begin{tabular}{l|p{0.81\columnwidth}} 
\hline
Notation & Description \\
\hline
$\theta_k(t)$ & Off-axis angle for $k$-th interferer at time $t$ \\
$d_k$ & Slant range distance to $k$-th satellite \\
$G_{t,k}(\cdot)$ & Transmit gain for $k$-th interfering satellite \\
$G_r(\cdot)$ & Receive antenna gain pattern at ground station \\
$EIRP_k$ & Equivalent isotropically radiated power (EIRP) for satellite $k$ \\
$INR_{k,n}$ & Interference to noise ratio (INR) for interferer $k$ at snapshot $n$ \\
$CNR_n$ & Carrier to noise ratio (CNR) for desired link at snapshot $n$ \\
$y_n(t)$ & Composite baseband signal at snapshot $n$ \\
$\mathbf{Y}$ & Pseudo Wigner Ville Distribution (PWVD) representation \\
$\text{EPFD}_{total}$ & Aggregate equivalent power flux-density (EPFD) \\
$B_{ref}$ & Reference bandwidth for EPFD calculation \\
$\mathcal{H}_0, \mathcal{H}_1$ & Interference-free and interference hypotheses \\
$\mathbf{\Sigma}$ & Feature covariance matrix for HOS augmentation \\
$\mathcal{L}_{\mathrm{Total}}$ & Composite multi-task objective function \\
$N_b$ & Mini-batch size used during training \\
\hline
\end{tabular}
\end{table}

\begin{figure}[t]
  \centering
  \subfloat[Reconstruction error distribution shift. The error distribution for weak interference signals (green) heavily overlaps with that of normal signals (blue), making it impossible for a fixed threshold to reliably separate the two classes.]{
    \includegraphics[width=0.46\columnwidth]{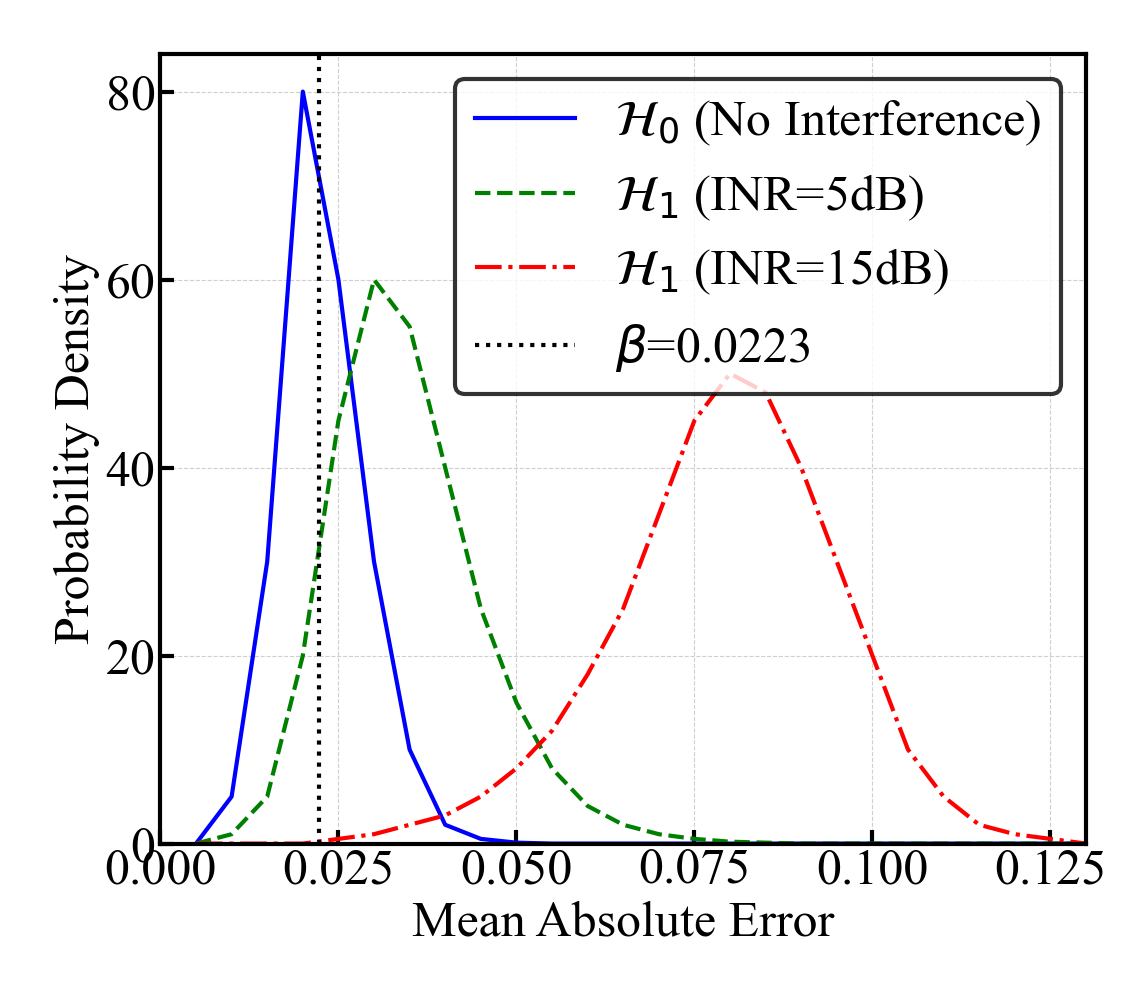}
    \label{fig3a}
  } \hspace{0.01\linewidth}
  \subfloat[ROC Comparison: Time vs. Frequency Domain Features. The performance gap between the time-domain model and the frequency-domain model show that analyzing domains in isolation misses key signal characteristics.]{
    \includegraphics[width=0.46\columnwidth]{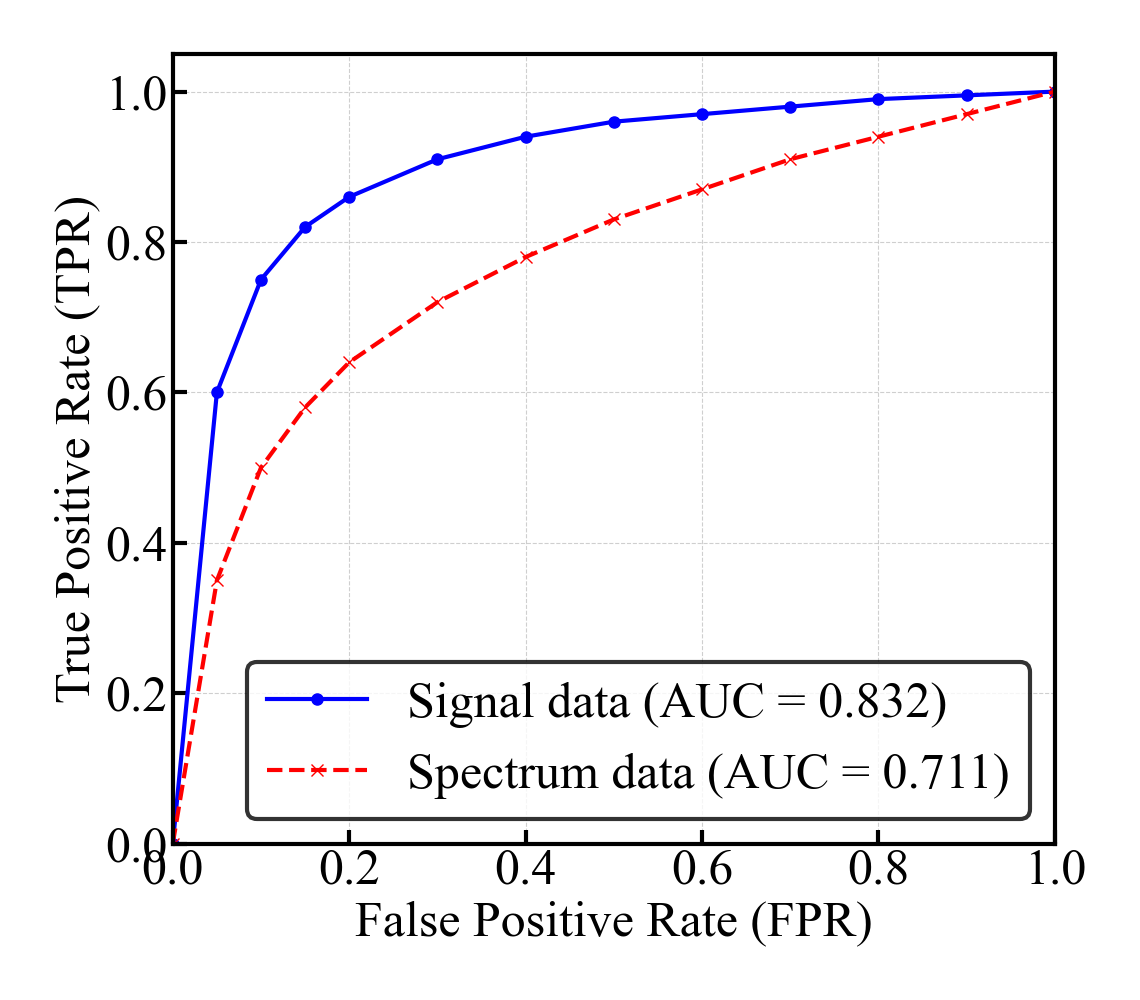}
    \label{fig3b}
  }
  \caption{Illustration of key challenges for existing interference detection methods.}
  \label{fig3}
\end{figure}

\subsection{Regulatory Standards and Ground Truth}
\label{sec33}
To establish a rigorous ground truth for the detection task, the equivalent power flux-density (EPFD) metric is adopted. This metric is officially mandated by the ITU as the primary regulatory standard for evaluating aggregate interference compliance between satellite systems. As visualized in the heatmap of Fig.~\ref{fig2}\subref{fig2b}, the aggregate EPFD exhibits strong spatial non-uniformity, forming localized hotspots of interference. The contribution of the k-th interfering satellite, denoted as $\text{EPFD}_k$, is calculated based on the transmit power $P_k$, transmit gain $G_{t,k}$, receive gain $G_{r,ngso}$, bandwidth $B_{ref}$ and distance $d_k$:
\begin{equation}
\label{equ_epfd_k}
\text{EPFD}_k = 10\log_{10}\left(\frac{P_k \cdot G_{t,k}(\phi_k) \cdot G_{r,ngso}(\psi_k)}{4\pi d_k^2 \cdot B_{ref}}\right),
\end{equation}
where $P_k$ is the transmit power and $d_k$ is the distance. The terms $G_{t,k}(\phi_k)$ and $G_{r,ngso}(\psi_k)$ represent the transmit gain at the off-axis angle $\phi_k$ and the receive gain at the off-axis angle $\psi_k$, respectively. Subsequently, the total EPFD is calculated by integrating the spectral power contributions from all $K$ visible interfering satellites over a reference bandwidth $B_{ref,Hz}$:
\begin{equation}
\label{equ_epfd_total}
\text{EPFD}_{total}=10\log_{10}\left( B_{ref} \cdot \sum_{k=1}^K 10^{(\text{EPFD}_k/10)}\right).
\end{equation}
The detection problem is formulated as a binary classification task rooted in regulatory compliance: Hypothesis $\mathcal{H}_0$ (Interference-Free) applies when $\text{EPFD}_{total} \leq \text{EPFD}_{limit}$, where the limit is set to -174.5 dB(W/m$^2$·40kHz) according to ITU regulations for Ku-band. Conversely, Hypothesis $\mathcal{H}_1$ (Interference) applies when this threshold is breached \cite{ITU_EPFD_support}. 

\begin{figure}[t]
    \centering
    \includegraphics[width=\columnwidth]{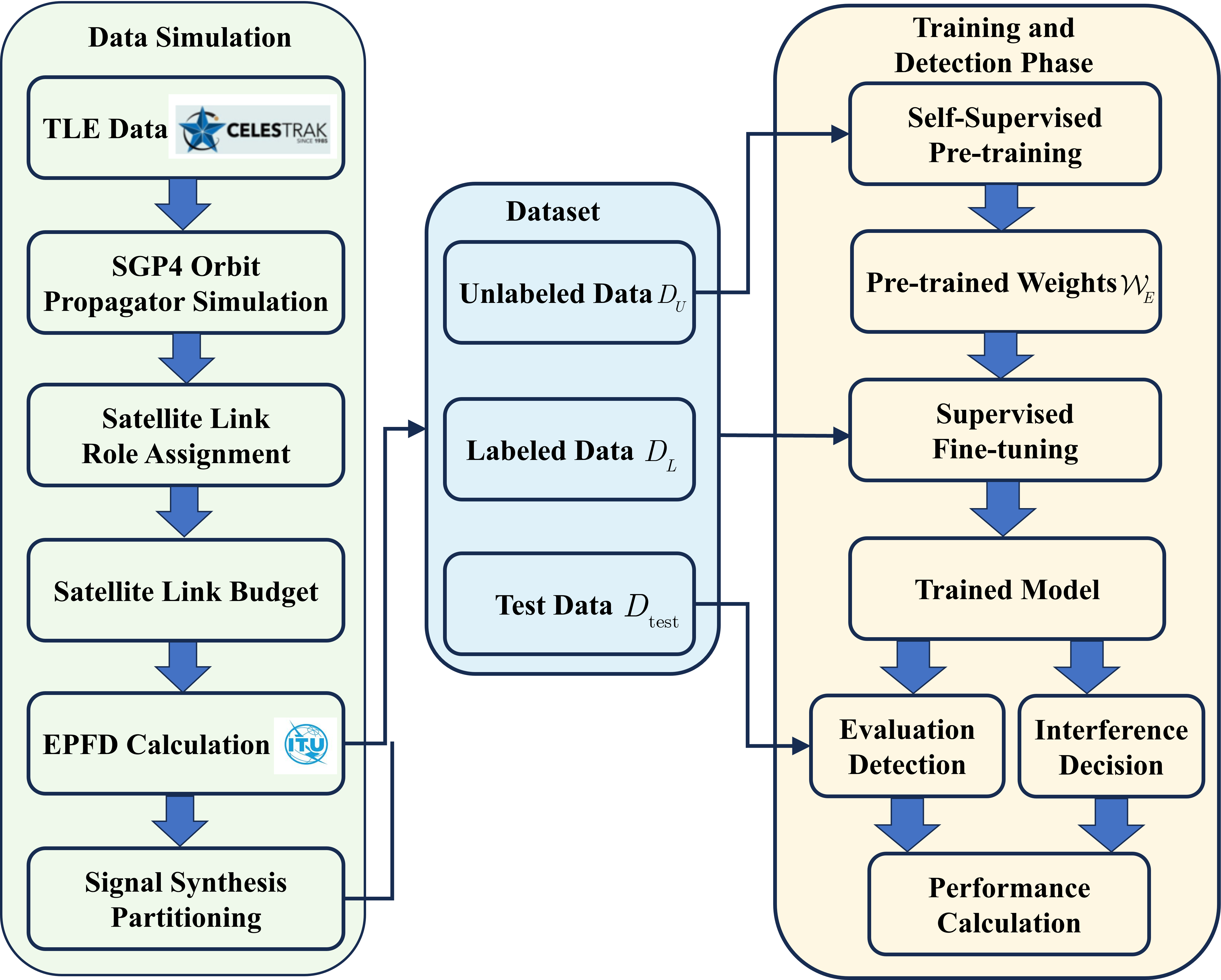}
    \caption{Schematic overview of the proposed Agon framework, illustrating the end-to-end workflow from data simulation to two-stage model training and final interference detection.}
    \label{fig4}
\end{figure}

\subsection{Deficiencies in Current Detection Paradigms}
\label{sec34}
Despite these theoretical foundations, practical detection remains hindered by the fundamental limitations of existing methodologies. Traditional energy detection relies on power thresholds but lacks sensitivity in the low-SNR regimes typical of satellite downlinks \cite{digham2003energy}. Although recent ML-based reconstruction models such as TrID \cite{saifaldawla2024genai} mark a notable advancement they still exhibit critical limitations. First the reconstruction error distributions highlighted in Fig.~\ref{fig3}\subref{fig3a} represent the statistical frequency of mathematical differences between actual input signals and outputs generated by a network trained exclusively on interference-free data. The graphical representation clearly demonstrates that the resulting distributions for normal $\mathcal{H}_0$ and weak interfering $\mathcal{H}_1$ signals overlap significantly. Such extensive overlap makes anomaly detection based on fixed thresholds \cite{baraniuk2017exponential} highly unstable causing a high false positive rate (FPR) of 17.63\% for baselines like TrID and ultimately leading to unnecessary resource allocation. Second, current architectures often process time and frequency domains in isolation \cite{cordonnier2020multi}, failing to leverage the synergistic cross-domain information illustrated in Fig.~\ref{fig3}\subref{fig3b}. Such isolation limits detection performance, evidenced by a time-domain AUC of 0.832 versus a frequency-domain AUC of 0.711. Finally, the strict reliance on large-scale labeled datasets restricts the scalability of current approaches. Moreover, standard architectures typically lack specialized mechanisms to explicitly model higher-order statistical dependencies or enforce multi-scale structural fidelity. This deficiency compromises their robustness against the complex noise profiles and transient dynamics inherent in satellite channels \cite{saifaldawla2024convolutional}.

\begin{figure}[t]
  \centering
  \subfloat[Orbital configuration of the Starlink constellation.]{
    \includegraphics[width=0.49\linewidth]{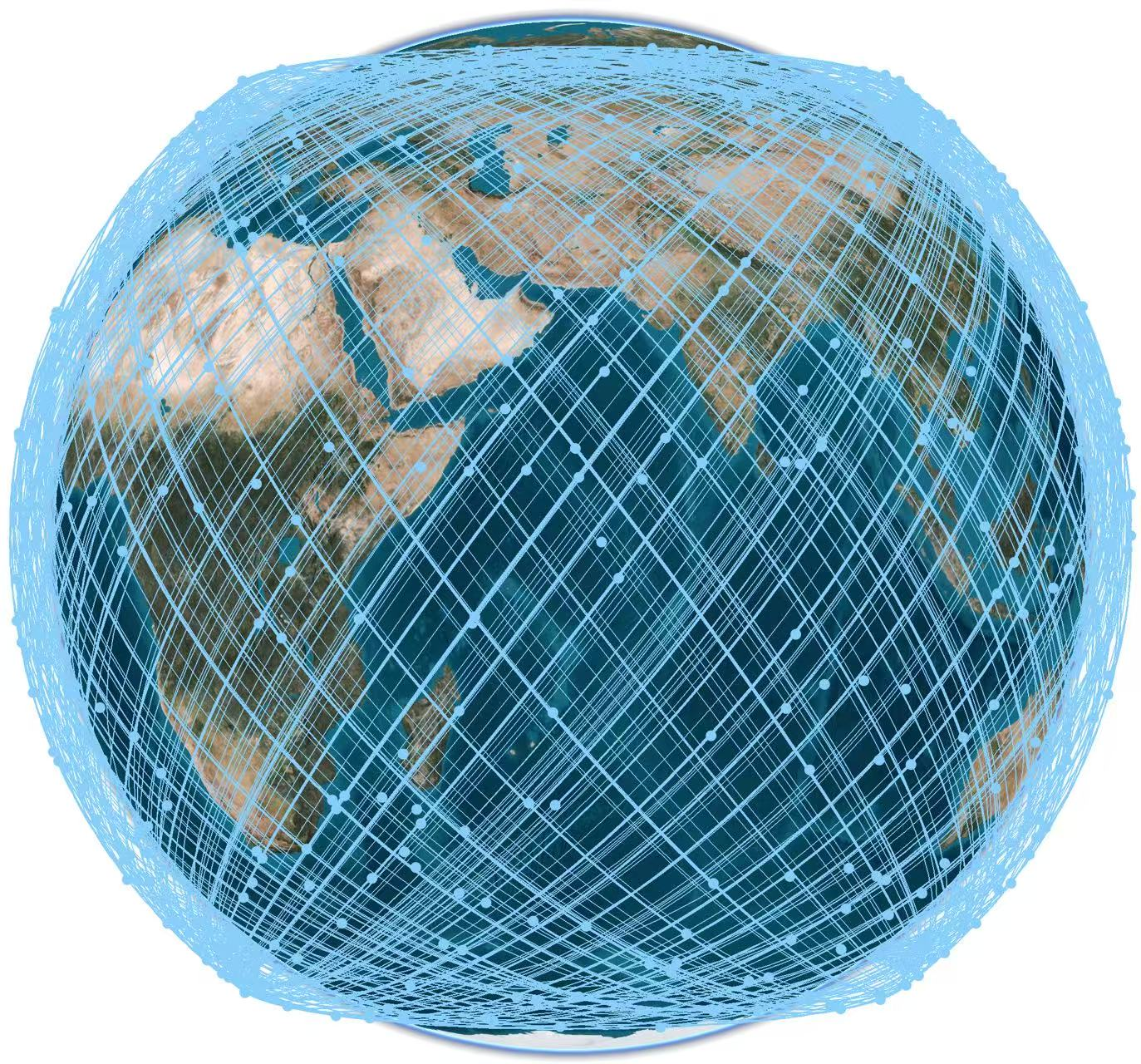}
    \label{fig5a}
  }
  \subfloat[Orbital configuration of the OneWeb constellation.]{
    \includegraphics[width=0.49\linewidth]{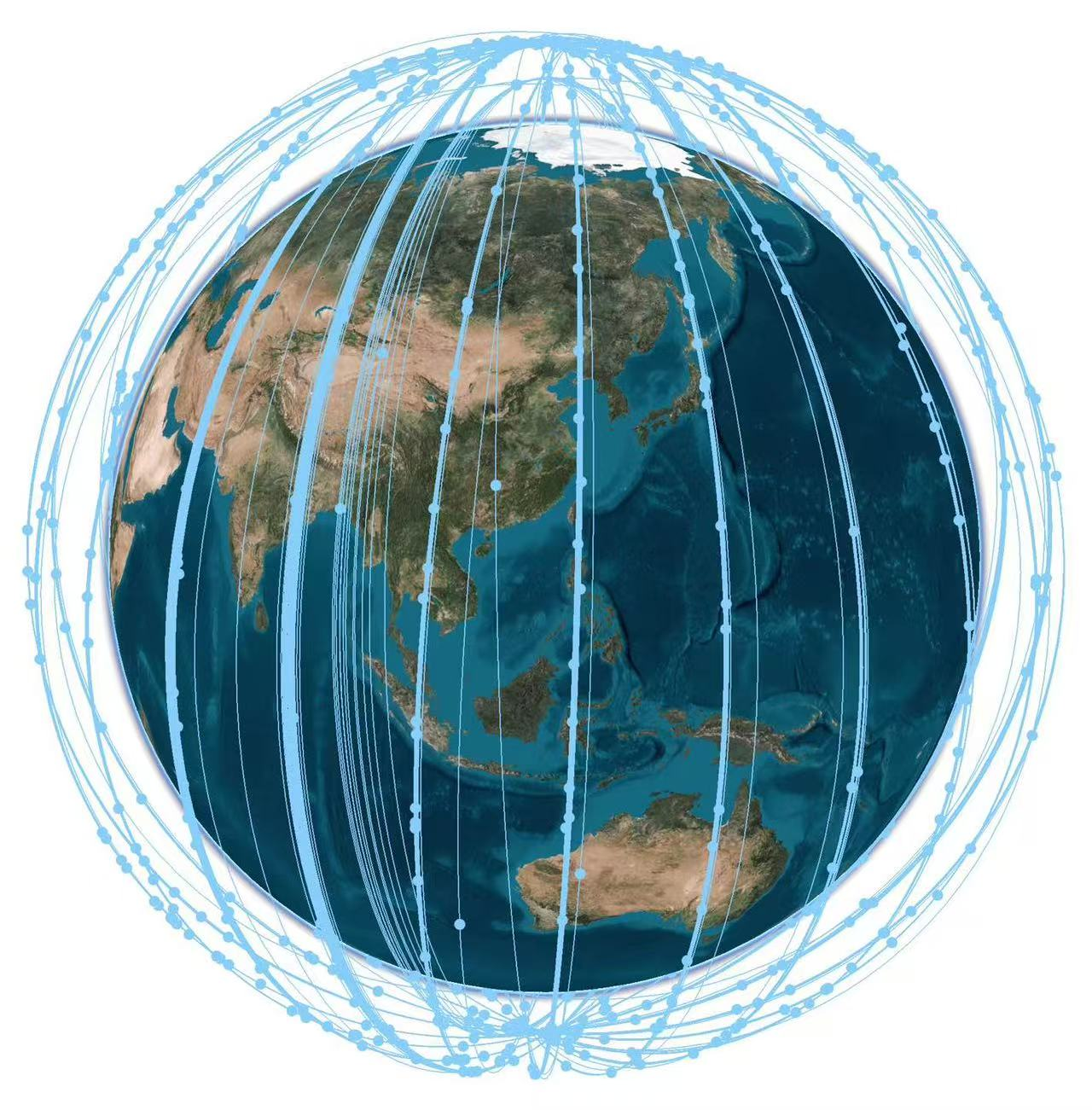}
    \label{fig5b}
  }
  \caption{Visualization of the NGSO constellations used in this paper.}
  \label{fig5}
  \vspace{-3ex}
\end{figure}

\section{Design Overview}
\label{sec4}
To effectively address the complexities associated with detecting interference in dynamic NGSO satellite environments, we have developed a comprehensive framework that bridges rigorous physical simulations with a semi-supervised learning approach. The overall workflow of the proposed Agon framework is conceptually illustrated in Fig.~\ref{fig4}. This high-level design is systematically organized into three interconnected phases: data simulation, dataset partitioning, and the core training and detection process.

\subsection{Data Simulation}
\label{sec41}
The workflow commences with the generation of a high-fidelity dataset firmly rooted in orbital physics. We utilize publicly available two-line element (TLE) data \cite{CelesTrak2025} to model dense satellite constellations, specifically Starlink and OneWeb, whose orbital configurations are visualized in Fig.~\ref{fig5}. By employing a simplified general perturbations 4 (SGP4) propagator, we simulate precise satellite trajectories to capture realistic orbital dynamics. A critical step in this phase is the dynamic assignment of link roles, which determines the desired and interfering links for each time snapshot based on elevation angles. This geometric configuration drives the calculation of link budgets and aggregate EPFD values, which are subsequently used to synthesize realistic signal samples through the PWVD transformation.

\subsection{Framework Overview}
\label{sec42}
To effectively address the complexities of dynamic NGSO environments, Agon establishes a comprehensive semi-supervised framework. The architecture integrates high-fidelity physical simulations with a sequential two-stage training paradigm. This process begins by partitioning the simulated data into a large-scale unlabeled dataset $D_U$ and a limited labeled dataset $D_L$. The training pipeline first executes a self-supervised pre-training phase using $D_U$ to initialize the encoder via a MAE task, allowing the network to capture universal signal structures and physical dependencies. Subsequently, the model transitions to a supervised fine-tuning phase using $D_L$, where these pre-learned features are optimized to adapt to specific interference environments.

\subsection{Task-Specific Classification and Execution}
\label{sec43}
The core objective of the framework is to enable robust task-specific classification and real-time decision-making. Following the training phase, the fine-tuned model functions as a direct execution engine capable of handling multiple downstream tasks simultaneously. This includes binary interference detection to identify the presence of jamming signals and modulation classification to categorize specific signal types. By integrating HOS augmentation and wavelet regularization, the system directly outputs classification probabilities during the inference stage. 

\begin{figure*}[t]
    \centering
    \includegraphics[width=\textwidth]{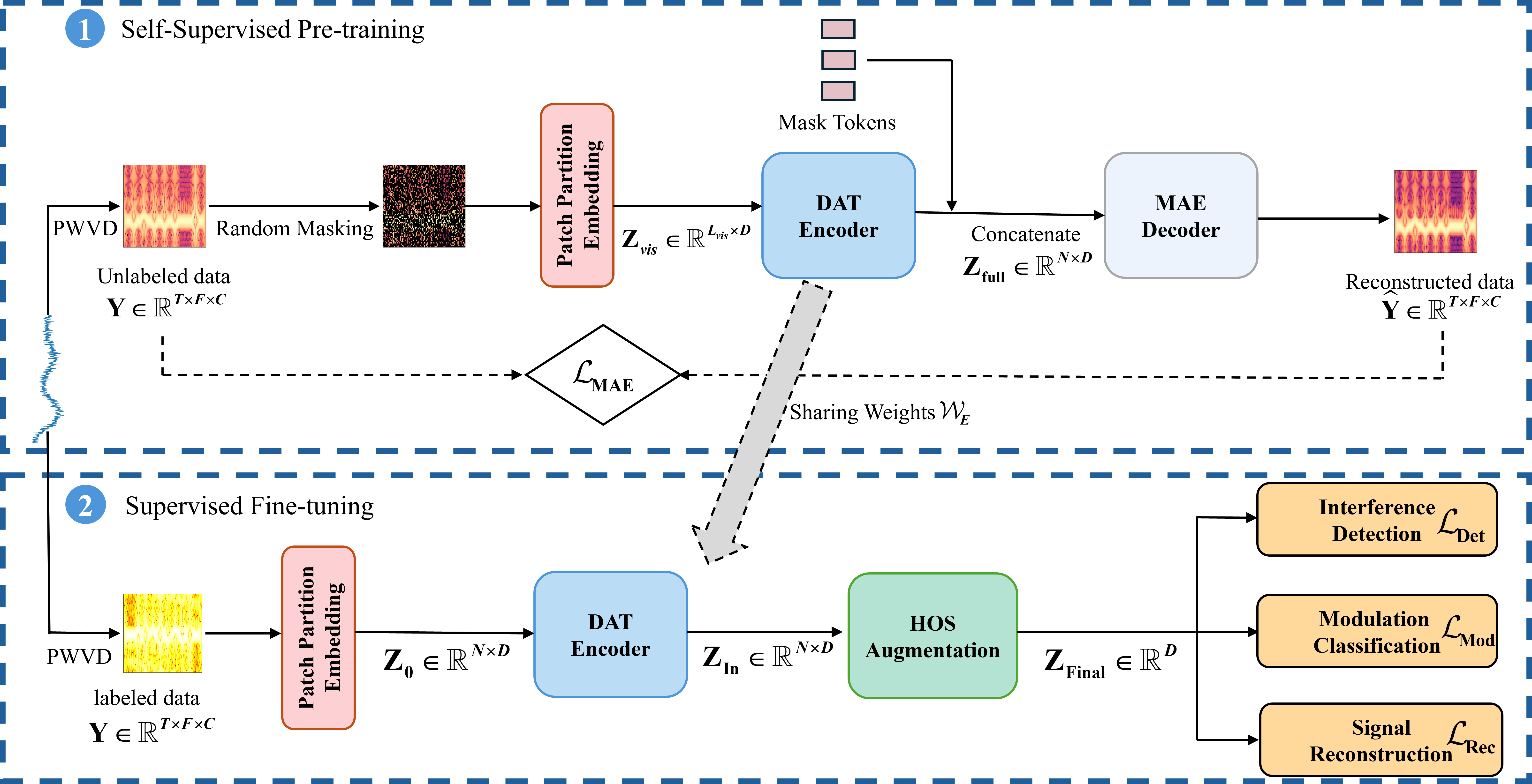}
    \caption{Schematic illustration of the Agon training and detection process, detailing the transition from self-supervised pre-training to supervised fine-tuning.}
    \label{fig6}
\end{figure*}

\section{Training and Detection Algorithm Design}
\label{sec5}
Building upon the design overview, this section details the training and detection algorithm design. This paper proposes a structural reformulation of interference detection, transitioning from passive reconstruction-error mapping to a two-stage discriminative paradigm. As illustrated in Fig.~\ref{fig6}, Agon leverages a DAT encoder tailored to extract deep spectral features through a hierarchical optimization sequence. Self-supervised structural pre-training initiates this process by mapping unlabeled spectral data into a robust latent manifold via a MAE objective, effectively capturing the intrinsic physical constraints of satellite links. Subsequently, the framework undergoes multi-task fine-tuning to supervise the convergence of a direct binary decision boundary on limited labeled data. Incorporating a HOS augmentation mechanism empowers the encoder to distinguish complex signal correlations from background noise, while multi-scale wavelet regularization preserves structural fidelity, collectively ensuring high classification accuracy and algorithmic robustness.

\subsection{Self-Supervised Pre-training}
\label{sec51}
To effectively mitigate the dependency on large-scale annotated datasets, we first introduce a robust self-supervised pre-training strategy rooted in the MAE paradigm. The training process commences by initializing the network parameters, specifically for the DAT Encoder $g_E(\cdot)$ and the MAE Decoder $g_D(\cdot)$, using a strategy optimized for layers followed by ReLU activations \cite{he2015delving} to ensure stable gradient propagation. Unlike conventional symmetric autoencoders, we explicitly design an asymmetric architecture to decouple feature extraction from reconstruction. To initiate this generative process, the continuous time-frequency data must first be adapted into a discrete format suitable for the transformer-based architecture.

\textit{Patch Partition Embedding:} The input spectrogram $\mathbf{Y} \in \mathbb{R}^{T \times F \times C}$ is reshaped into a sequence of flattened 2D patches $\mathbf{Y}_p \in \mathbb{R}^{N \times (P^2 \cdot C)}$, where $T, F, C$ denote time frames, frequency bins, and channels, respectively. Here, $P$ is the patch size and $N=TF/P^2$ is the sequence length. These patches are mapped to latent embeddings via a linear projection $\mathbf{E} \in \mathbb{R}^{(P^2 \cdot C) \times D}$ and augmented with learnable positional embeddings $\mathbf{E}_{pos} \in \mathbb{R}^{N \times D}$:
\begin{equation}
\label{equ_patch_embed}
\mathbf{Z}_0 = \mathbf{Y}_p \mathbf{E} + \mathbf{E}_{pos},
\end{equation}
where $D$ represents the latent embedding dimension, and $\mathbf{Z}_0 \in \mathbb{R}^{N \times D}$ serves as the input to the encoder. To initiate the MAE task, we sample a subset of indices $\mathcal{M}$ and construct the visible sequence $\mathbf{Z}_{vis} = \{ \mathbf{Z}_{0,i} \,|\, i \notin \mathcal{M} \}$ with length $L_{vis}$, effectively removing trivial redundancy to compel the model to learn global contextual dependencies.

\begin{figure}[t]
    \centering
    \includegraphics[width=0.8\columnwidth]{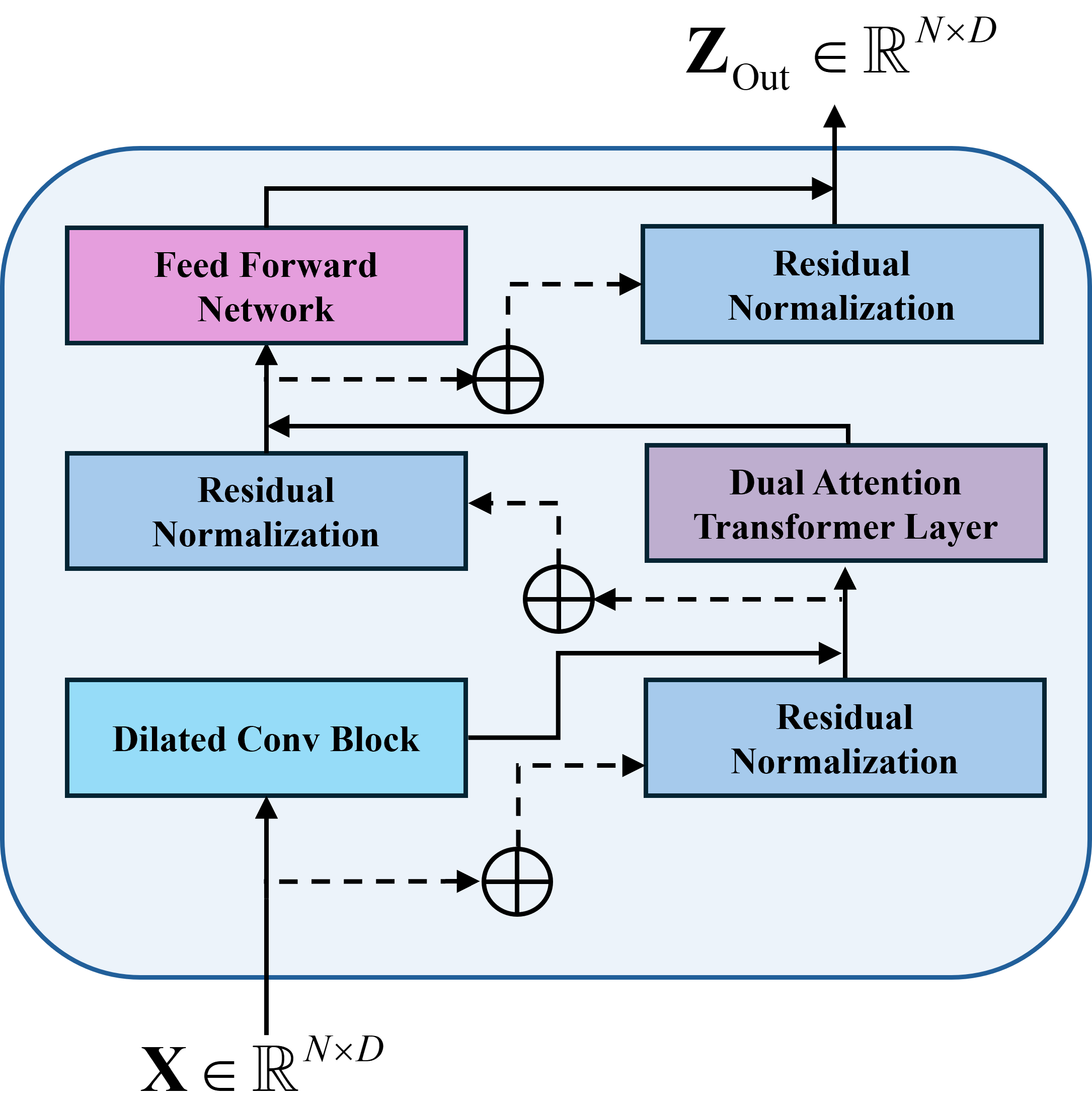}
    \caption{Detailed architecture of the DAT Encoder Block. The input representation $\mathbf{X}$ is directly processed by a Dilated Conv Block to extract local spectral features and a Dual Attention Layer to capture global dependencies. A Feed Forward Network then refines the features, yielding the block output $\mathbf{Z}_{\text{Out}}$ via residual connections.}
    \label{fig7}
\end{figure}

\textit{DAT Encoder Core Architecture:} The DAT Encoder $g_E(\cdot)$ functions as the unified feature extraction engine. As detailed in Fig.~\ref{fig7}, the data flow first passes through the Dilated Conv Block. Given the latent input $\mathbf{X}$, this module exponentially expands the receptive field to compute the local feature representation $\mathbf{Z}_{\text{loc}}$:
\begin{equation}
\label{equ_dconv}
\mathbf{Z}_{\text{loc}} = \sigma\left(\text{DConv}_d\left(\text{LN}(\mathbf{X})\right)\right),
\end{equation}
where $d$ denotes the dilation rate, $\text{LN}(\cdot)$ represents Layer Normalization, and $\sigma(\cdot)$ is the GeLU activation function \cite{hendrycks2016gaussian}. Complementing this, the model applies self-attention across time-frequency axes. We generate projections $\mathbf{Q}, \mathbf{K}, \mathbf{V}$ from $\mathbf{Z}_{\text{loc}}$ to compute the global dependency map $\mathbf{Z}_{\text{glo}}$:
\begin{equation}
\label{equ_dat_attn}
\mathbf{Z}_{\text{glo}} = \text{Softmax}\left(\frac{\mathbf{Q}\mathbf{K}^T}{\sqrt{D}}\right)\mathbf{V}.
\end{equation}
Here, $\sqrt{D}$ serves as the scaling factor based on the embedding dimension $D$. To reconcile local details with global coherence, the local and global features are fused via a residual connection before entering the position-wise Feed Forward Network. The final block output $\mathbf{Z}_{\text{Out}}$ is derived as:
\begin{equation}
\label{equ_ffn}
\mathbf{Z}_{\text{Out}} = \text{LN}\left(\mathbf{Z}_{\text{mix}} + \left(\sigma(\mathbf{Z}_{\text{mix}}\mathbf{W}_1 + \mathbf{b}_1)\mathbf{W}_2 + \mathbf{b}_2\right)\right),
\end{equation}
where $\mathbf{Z}_{\text{mix}} = \mathbf{Z}_{\text{loc}} + \mathbf{Z}_{\text{glo}}$ denotes the fused feature stream, and parameters $\{\mathbf{W}_1, \mathbf{b}_1, \mathbf{W}_2, \mathbf{b}_2\}$ represent the learnable weights and biases of the Feed Forward Network.

\textit{MAE Mechanism:} The reconstruction task is handled by a lightweight MAE Decoder, $g_D(\cdot)$, which completes the asymmetric architecture. Unlike the heavy encoder, the decoder is designed to be shallow to ensure that the semantic abstraction capability resides primarily in the encoder. The decoder receives a full sequence $\mathbf{Z}_{\mathrm{full}} \in \mathbb{R}^{N \times D}$. Following the standard MAE paradigm \cite{he2022masked}, this sequence is constructed by integrating the encoder's latent output $\mathbf{Z}_{vis}$ with a globally shared, learnable mask token $\mathcal{T}_{\mathrm{mask}} \in \mathbb{R}^{D}$. To preserve spatial awareness, the token at sequence index $i$ is mathematically formulated by injecting the corresponding positional embedding $\mathbf{E}_{pos, i}$:
\begin{equation}
\label{equ_mask_token}
\mathbf{Z}_{\mathrm{full}, i} = 
\begin{cases} 
\mathbf{Z}_{vis, i}, & i \notin \mathcal{M} \\ 
\mathcal{T}_{\mathrm{mask}} + \mathbf{E}_{pos, i}. & i \in \mathcal{M} 
\end{cases}
\end{equation}
Rather than being heuristically extracted from the input data, $\mathcal{T}_{\mathrm{mask}}$ functions as an intrinsic network parameter initialized from a continuous uniform distribution. It acts as a structural placeholder representing the unobserved time-frequency patches, enabling the heavy encoder to process only the visible patches for computational efficiency. During the self-supervised pre-training phase, $\mathcal{T}_{\mathrm{mask}}$ is optimized jointly with the network weights. At the $k$-th training iteration, its update rule is rigorously governed by gradient descent:
\begin{equation}
\label{equ_mask_update}
\mathcal{T}_{\mathrm{mask}}^{(k+1)} = \mathcal{T}_{\mathrm{mask}}^{(k)} - \eta \nabla_{\mathcal{T}_{\mathrm{mask}}} \mathcal{L}_{\mathrm{MAE}},
\end{equation}
where $\eta$ denotes the learning rate. Since $\mathcal{T}_{\mathrm{mask}}$ is broadcasted to all masked positions $i \in \mathcal{M}$ via linear addition, its exact gradient is accumulated across the masked subset through the chain rule:
\begin{equation}
\label{equ_mask_grad}
\nabla_{\mathcal{T}_{\mathrm{mask}}} \mathcal{L}_{\mathrm{MAE}} = \sum_{i \in \mathcal{M}} \frac{\partial \mathcal{L}_{\mathrm{MAE}}}{\partial \mathbf{Z}_{\mathrm{full}, i}}.
\end{equation}

The objective is to reconstruct the pixel values of the original spectrogram $\hat{\mathbf{Y}}$ from this combined representation. As illustrated in the pre-training phase of Fig.~\ref{fig6}, predicting these unobserved patches serves as the core self-supervised pretext task. From a probabilistic perspective, reconstructing the missing regions forces the network parameters $\theta$ to model the conditional distribution $p_{\theta}(\mathbf{Y}_{\mathcal{M}} | \mathbf{Z}_{vis})$. Therefore, minimizing the reconstruction error effectively serves as a surrogate for maximizing the conditional log-likelihood:
\begin{equation}
\label{equ_mae_motivation}
\max_{\theta} \mathbb{E} \left[ \log p_{\theta}(\mathbf{Y}_{\mathcal{M}} | \mathbf{Z}_{vis}) \right].
\end{equation}
To enforce this deep structural learning, the optimization minimizes the Mean Squared Error (MSE) calculated exclusively on the masked patches, as defined by the self-supervised loss function:
\begin{equation}
\label{equ_mae}
\mathcal{L}_{\mathrm{MAE}}=\sum_{(t,f) \in \text{Masked}} \|\mathbf{Y}_{t,f}-\hat{\mathbf{Y}}_{t,f}\|_2^2,
\end{equation}
where $\|\cdot\|_2^2$ denotes the squared L2 norm \cite{solomon1991psd}, and $(t,f) \in \text{Masked}$ represents the specific pixel coordinates belonging to the masked patches derived from $\mathcal{M}$. By focusing the loss solely on these unseen regions, the DAT encoder is mathematically compelled to learn the intrinsic physical dependencies, temporal continuity, and cross-domain spectral correlations of the satellite signals, ultimately driving the model to recover complex, unobserved physical structures from minimal clues, yielding robust pre-trained weights $\mathcal{W}_E$ that are subsequently transferred to initialize the encoder for the fine-tuning phase.

\begin{figure}[t]
    \centering
    \includegraphics[width=0.76\columnwidth]{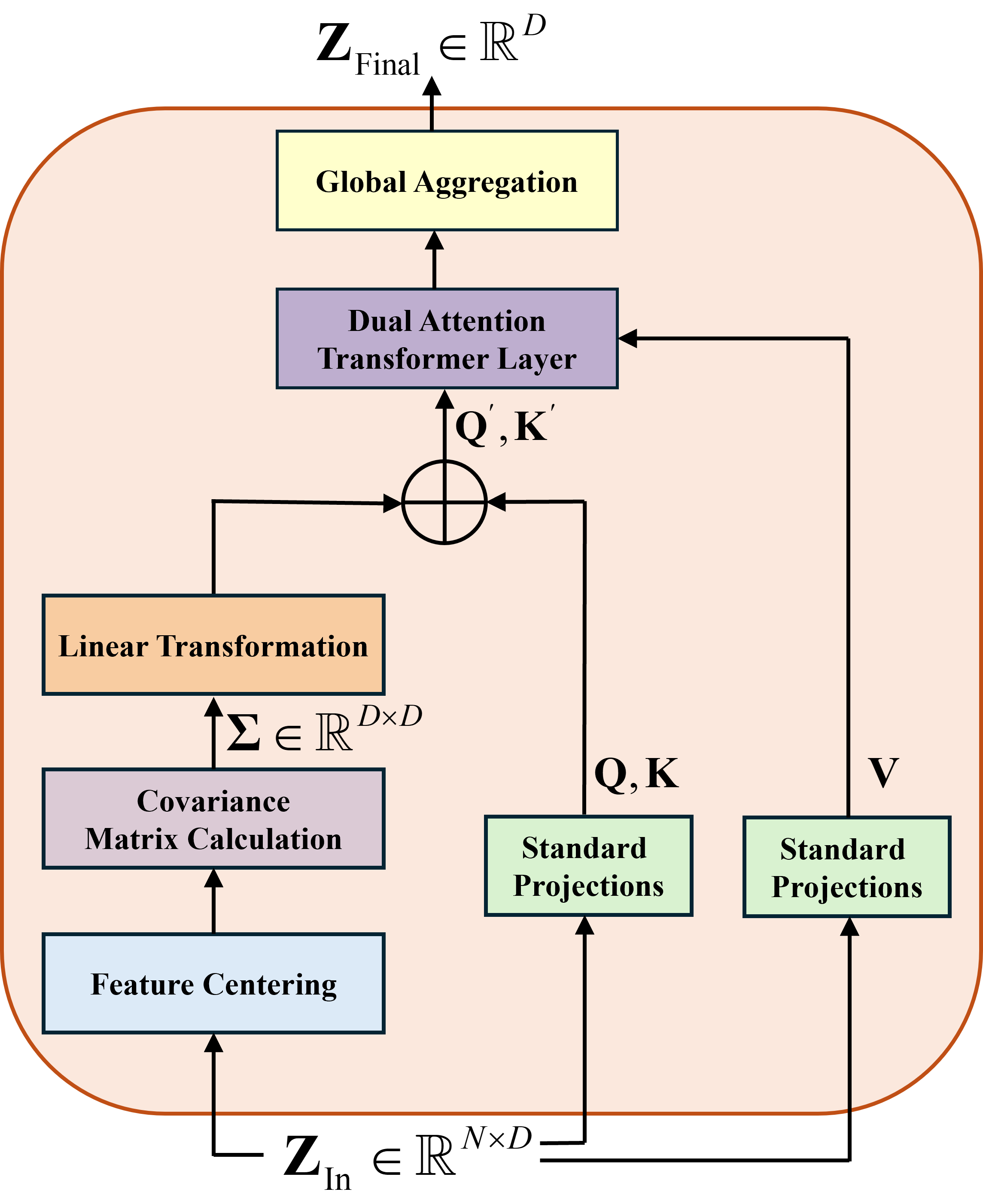}
    \caption{Architecture of the HOS-augmented Dual Attention Mechanism. The covariance matrix $\mathbf{\Sigma}$ derived from $\mathbf{Z}_{\mathrm{In}}$ modulates the query and key projections to incorporate second-order statistics, yielding the final output $\mathbf{Z}_{\mathrm{Final}}$.}
    \label{fig8}
\end{figure}

\subsection{Supervised Fine-tuning}
\label{sec52}
The supervised fine-tuning stage commences by initializing the DAT Encoder $g_E(\cdot)$ with the pre-trained weights $\mathcal{W}_E$. This step bridges the universal structural knowledge acquired from large-scale unlabeled data with the specific discriminative requirements of interference detection. Unlike standard fine-tuning that merely retrains the classifier, this phase introduces HOS augmentation and Wavelet regularization to rigorously address noise robustness and structural fidelity.

\textit{HOS-Augmented Dual Attention Mechanism:} A fundamental limitation of standard self-attention mechanisms is their reliance on first-order linear projections. In dynamic satellite channels, first-order statistics based on absolute amplitude and spectral energy are fundamentally insufficient for separating interference from noise. Mathematically, under the noise hypothesis $\mathcal{H}_0$ and the interference hypothesis $\mathcal{H}_1$, the received signal vectors are $\mathbf{y} = \mathbf{n}$ and $\mathbf{y} = \mathbf{s} + \mathbf{n}$ respectively. The expected signal energies are evaluated as:
\begin{equation}
\mathbb{E}[\|\mathbf{y}\|_2^2 | \mathcal{H}_0] = D \sigma_n^2, \quad \mathbb{E}[\|\mathbf{y}\|_2^2 | \mathcal{H}_1] = D (\sigma_s^2 + \sigma_n^2).
\end{equation}
When the interference is completely submerged beneath the thermal noise floor with $\sigma_s^2 \ll \sigma_n^2$, these first-order expectations converge, making amplitude-based separation intractable. To overcome this limitation, higher-order statistics become essential. Gaussian white noise is entirely uncorrelated and mathematically exhibits a diagonal covariance structure:
\begin{equation}
\mathbf{\Sigma}_{\mathcal{H}_0} = \mathbb{E}[\mathbf{n}\mathbf{n}^H] = \sigma_n^2 \mathbf{I}.
\end{equation}
In contrast, artificial communication signals possess inherent structural correlations that generate strong off-diagonal dependencies leading to:
\begin{equation}
\mathbf{\Sigma}_{\mathcal{H}_1} = \mathbb{E}[(\mathbf{s}+\mathbf{n})(\mathbf{s}+\mathbf{n})^H] = \mathbf{R}_s + \sigma_n^2 \mathbf{I},
\end{equation}
where $\mathbf{R}_s$ represents the signal correlation matrix \cite{mendel1991tutorial}. To practically exploit these statistical dependencies within the network architecture, as conceptually illustrated in Fig.~\ref{fig8}, the process commences by explicitly computing the feature covariance matrix $\mathbf{\Sigma} \in \mathbb{R}^{D \times D}$. Given the input feature sequence $\mathbf{Z}_{\text{In}}$ with channel wise mean $\boldsymbol{\mu}$, $\mathbf{\Sigma}$ is formally derived as:
\begin{equation}
\label{equ_cov}
\mathbf{\Sigma} = \frac{1}{L-1} (\mathbf{Z}_{\text{In}} - \boldsymbol{\mu})^T (\mathbf{Z}_{\text{In}} - \boldsymbol{\mu}).
\end{equation}
While strictly representing second order statistics, this covariance matrix theoretically serves as a computationally tractable proxy for full higher order capabilities. True higher order cumulants rigorously vanish for Gaussian noise but impose an intractable quartic processing overhead scaling with $\mathcal{O}(L D^4)$. By requiring only quadratic complexity $\mathcal{O}(L D^2)$, the covariance formulation successfully isolates the dominant off diagonal structural correlations inherent in artificial signals. This matrix encapsulates the global distribution of the signal in the feature space. To leverage such statistical priors, Agon reformulates the attention kernel beyond standard linear projections by injecting high order dependencies directly into the similarity computation. Specifically, weight matrices $\mathbf{W}_{\Sigma}^Q$ and $\mathbf{W}_{\Sigma}^K$ linearly transform $\mathbf{\Sigma}$ from the statistical manifold into the semantic attention space, facilitating the additive fusion with the standard Query $\mathbf{Q}$ and Key $\mathbf{K}$ projections:
\begin{equation}
\label{equ15}
\mathbf{Q}' = \mathbf{Q} + \mathbf{Z}_{\text{In}} \mathbf{W}_{\Sigma}^Q \mathbf{\Sigma},
\end{equation}
\begin{equation}
\mathbf{K}' = \mathbf{K} + \mathbf{Z}_{\text{In}} \mathbf{W}_{\Sigma}^K \mathbf{\Sigma},
\label{equ16}
\end{equation}
where the term $\mathbf{Z}_{\text{In}} \mathbf{W}_{\Sigma}^{(\cdot)} \mathbf{\Sigma}$ acts as a context-aware modulation, dynamically scaling the query and key vectors based on the global statistical significance of each feature channel. Expanding the resulting attention kernel enables the decomposition of the alignment score for each token pair $i$ and $j$ into local semantics and global statistical priors:
\begin{equation}
S_{i,j} = \mathbf{q}_i \mathbf{k}_j^T + \phi(\mathbf{z}_i, \mathbf{\Sigma}) + \psi(\mathbf{z}_j, \mathbf{\Sigma}) + \Omega(\mathbf{\Sigma}),
\end{equation}
where $\phi(\cdot)$ and $\psi(\cdot)$ represent the cross-modal interactions between local features and the covariance structure while $\Omega(\mathbf{\Sigma})$ acts as a statistical bias. Analyzing the expected statistical divergence $\Delta S$ between the interference hypothesis $\mathcal{H}_1$ and the noise hypothesis $\mathcal{H}_0$ reveals the fundamental theoretical advantage. Assuming uncorrelated background noise, the expected margin of the augmented attention score scales directly with the squared Frobenius norm of the signal correlation matrix $\mathbf{R}_s$ yielding:
\begin{equation}
\label{equ_delta_s}
\Delta S = \mathbb{E}[S_{i,j} | \mathcal{H}_1] - \mathbb{E}[S_{i,j} | \mathcal{H}_0] \propto \frac{\|\mathbf{R}_s\|_F^2}{\sigma_n^4}.
\end{equation}
In dense NGSO environments extreme path loss often buries interference signals entirely beneath the receiver thermal noise floor. By maximizing the statistical divergence against uncorrelated background noise this mechanism explicitly isolates structured modulation signatures embedded within the covariance matrix. Consequently the feature representation remains structurally identifiable even when severe satellite channel fading completely masks first-order spectral energy.

These augmented vectors $\mathbf{Q}'$ and $\mathbf{K}'$ synthesize both the local feature semantics from standard projections and the global statistical context from HOS injection. The attention scores are subsequently computed using these enhanced representations to generate the attention map:
\begin{equation}
\label{equ_hos_score}
\text{Att}_{\text{HOS}}(\mathbf{Q}', \mathbf{K}', \mathbf{V}) = \text{Softmax}\left(\frac{\mathbf{Q}'(\mathbf{K}')^T}{\sqrt{D}}\right)\mathbf{V},
\end{equation}
where the scaling factor $\sqrt{D}$ is applied to ensure gradient stability during backpropagation, the Softmax function is utilized here to transform the raw alignment scores into a normalized probability distribution. By incorporating second-order statistics into the dot-product mechanism, this design forces the model to prioritize signal structures that exhibit consistent correlation patterns over transient noise fluctuations. Finally, the resulting context-aware tokens are globally aggregated into a compact vector $\mathbf{Z}_{\text{Final}} \in \mathbb{R}^{D}$ to drive the subsequent multi-task classification heads.

\textit{Multi-Task Learning for Structural Regularization:} The fine-tuning objective is governed by MTL, strategically combining the primary detection goal with two auxiliary tasks to foster feature stability and contextual awareness. The framework utilizes three parallel heads: the Binary Detection Head $h_{\text{Det}}(\cdot)$, which outputs the interference probability and contributes $\mathcal{L}_\mathrm{Det}$ to the total loss; the Modulation Classification Head $h_{\text{Mod}}(\cdot)$, which predicts the modulation type and contributes $\mathcal{L}_{\mathrm{Mod}}$ as an auxiliary task; and the Signal Reconstruction Head $h_{\text{Rec}}(\cdot)$, which reconstructs the input signal and contributes $\mathcal{L}_\mathrm{Rec}$ as a structural regularization task. Each head takes the final feature vector $\mathbf{Z}_{\text{Final}}$ as input. The primary task optimizes the $\mathcal{H}_0/\mathcal{H}_1$ decision based directly on the EPFD compliance criterion, while the auxiliary tasks foster feature stability and contextual awareness, specifically preventing catastrophic forgetting of the low-level spectral details learned during pre-training.

\begin{figure}[t] 
    \centering
    \includegraphics[width=\columnwidth]{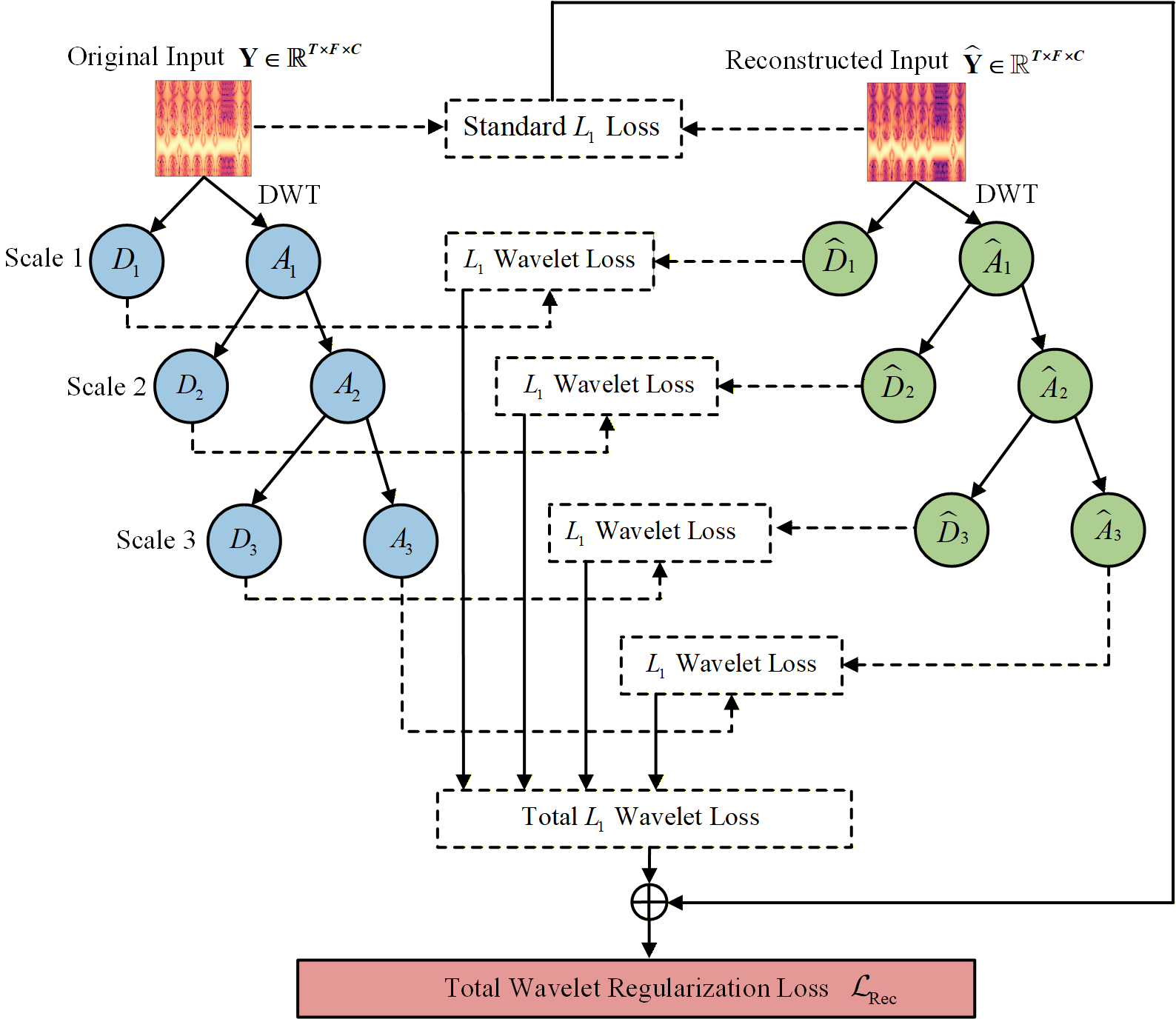} 
    \caption{Schematic of the Reconstruction Loss $\mathcal{L}_{\mathrm{Rec}}$ computation. It combines a standard $L_1$ loss for basic waveform similarity and a Total Wavelet Regularization Loss for multi-scale fidelity. The latter enforces structural consistency by summing $L_1$ errors between corresponding DWT coefficients derived from a 3-level decomposition using the Daubechies 4 basis.}
    \label{fig9}
\end{figure}

The final optimization objective is the total loss function $\mathcal{L}_{\mathrm{Total}}$ formulated as a linear scalarization of three task-specific components:
\begin{equation}
\label{equ18}
\mathcal{L}_{\mathrm{Total}}=\lambda_{\mathrm{Det}}\mathcal{L}_{\mathrm{Det}}+\lambda_{\mathrm{Mod}}\mathcal{L}_{\mathrm{Mod}}+\lambda_{\mathrm{Rec}}\mathcal{L}_{\mathrm{Rec}}.
\end{equation}
Theoretically, the multi-objective optimization framework \cite{sener2018multi} establishes that such a linear weighted-sum formulation is mathematically sound and appropriate for finding Pareto optimal solutions, provided the tasks exhibit structural compatibility. For shared network parameters $\theta$, joint optimization succeeds without catastrophic interference if the expected inner products of the gradients between the primary detection task and the auxiliary tasks remain non-negative:
\begin{equation}
\label{equ_gradient}
\mathbb{E} \left[ (\nabla_{\theta}\mathcal{L}_{\mathrm{Det}})^T \nabla_{\theta}\mathcal{L}_{\mathrm{Aux}} \right] \geq 0.
\end{equation}
In our framework, the tasks can be reliably expected to interact compatibly because they share a physically correlated foundation. Specifically, the auxiliary modulation classification and signal reconstruction tasks are designed to extract deterministic properties such as phase evolution and multi-scale wavelet transients from the exact same baseband I/Q signal manifold as the detection task. Consequently, their optimization trajectories are intrinsically bounded to align with the primary interference detection objective. By satisfying the gradient alignment condition in (\ref{equ_gradient}), these auxiliary tasks act as synergistic regularizers that enrich the feature space, rather than conflicting objectives. The precise weighting coefficients balancing these functional contributions are then empirically established by mapping the exact Pareto optimal front. Specifically, the component $\mathcal{L}_{\mathrm{Det}}$ acts as the primary loss for the binary interference detection task implemented as the standard binary cross-entropy loss. For a given batch of $N_b$ samples it is computed as:
\begin{equation}
\label{equ19} 
\mathcal{L}_{\mathrm{Det}} = -\frac{1}{N_b} \sum_{i=1}^{N_b} \left[ y_{\text{det},i} \log(\hat{y}_{\text{det},i}) + (1 - y_{\text{det},i}) \log(1 - \hat{y}_{\text{det},i}) \right],
\end{equation}
where $N_b$ denotes the mini-batch size, $y_{\text{det},i} \in \{0, 1\}$ is the ground truth label for the $i$-th sample and $\hat{y}_{\text{det},i}$ is the predicted probability from the detection head's Sigmoid output \cite{yin2003flexible}. $\mathcal{L}_\mathrm{Mod}$ serves as the auxiliary modulation classification loss, implemented as the categorical cross-entropy loss. For $M$ modulation classes, the loss is given by:
\begin{equation}
\label{equ20}
\mathcal{L}_{\mathrm{Mod}} = -\frac{1}{N_b} \sum_{i=1}^{N_b} \sum_{m=1}^{M} y_{\text{mod},i,m} \log(\hat{y}_{\text{mod},i,m}),
\end{equation}
where $y_{\text{mod},i,m}$ represents the categorical one-hot encoded ground truth vector. This specific vector maps the discrete modulation class into a binary format consisting of a single high bit alongside zeroed elements. Such a formulation aligns perfectly with established conventions for deep learning based modulation recognition \cite{o2016convolutional}. The variable $\hat{y}_{\text{mod},i,m}$ denotes the predicted probability for the $m$-th class derived directly from the auxiliary Softmax output \cite{vaswani2017attention}.

$\mathcal{L}_\mathrm{Rec}$ incorporates a Wavelet Regularization Loss to enforce multi-scale structural fidelity, as visually detailed in Fig.~\ref{fig9}. Agon imposes constraints in the discrete wavelet transform (DWT) domain. By decomposing both the input and the reconstruction into hierarchical frequency sub-bands, the model is compelled to match not just the global waveform envelope but also the fine-grained transient details at every resolution scale. The total reconstruction objective combines the standard pixel-wise error with this multi-scale wavelet penalty:
\begin{equation}
\label{equ21}
\mathcal{L}_{\mathrm{Rec}}=\mathcal{L}_{\mathrm{L}1}(\mathbf{Y},\hat{\mathbf{Y}})+\alpha\cdot\sum_{s=1}^S\left(\|\mathbf{W}_s(\mathbf{Y})-\mathbf{W}_s(\hat{\mathbf{Y}})\|_1\right),
\end{equation}
where $\mathcal{L}_{\mathrm{L1}}$ represents the standard $L_1$ loss \cite{shalev2009stochastic} measuring pixel-level fidelity between the input $\mathbf{Y}$ and reconstruction $\hat{\mathbf{Y}}$. The term $\mathbf{W}_s(\cdot)$ denotes the DWT operation at scale $s$ using the Daubechies 4 basis \cite{daubechies1992ten} to capture asymmetric transient features. Rapid NGSO satellite mobility induces severe Doppler shifts generating highly non-stationary transient interference bursts. Fundamental approximation theory within Besov spaces \cite{mallat1999wavelet} provides the analytical justification for capturing these rapid spectral fluctuations. For non-stationary interference possessing transient singularities the theoretical reconstruction error bound $\mathcal{E}_{bound}$ is strictly dictated by the decay rate of its multi-scale wavelet coefficients:
\begin{equation}
\label{equ_besov}
\mathcal{E}_{bound} \leq \lambda \sum_{s=1}^{S} 2^{-s \beta} \|\mathbf{W}_s(\mathbf{Y})\|_1,
\end{equation}
where $\lambda$ represents an absolute structural constant while $\beta$ denotes the Lipschitz regularity parameter defining inherent signal smoothness. This analytical bound proves that minimizing the $L_1$ norm of these wavelet coefficients across $S$ decomposition scales mathematically guarantees optimal preservation of sharp signal transitions. Regulated by the weighting factor $\alpha$ this multi-scale constraint theoretically ensures the precise recovery of sudden spectral overlaps typical of dynamic satellite links. Consequently the composite objective $\mathcal{L}_{\mathrm{Total}}$ directs the encoder to learn discriminative features $\mathbf{Z}_\mathrm{Final}$ for robust detection $\mathcal{L}_\mathrm{Det}$ while preventing the blurring of critical interference edges via $\mathcal{L}_\mathrm{Rec}$ regularized by $\mathcal{L}_{\mathrm{Mod}}$. The complete training procedure, encompassing both the self-supervised pre-training and supervised fine-tuning phases, is summarized in Algorithm~\ref{alg1}. 

To evaluate the practical feasibility of Agon, we analyze its computational complexity by separating the process into offline and online stages. The offline complexity is determined by the cumulative computational load during the weight convergence process. Let $|D_U|$ and $|D_L|$ be the sizes of the unlabeled and labeled datasets, $E$ be the training epochs, and $L$ be the number of DAT layers. The total offline complexity is expressed as $O(E \cdot (|D_U| + |D_L|) \cdot L \cdot (N^2 D + N D^2))$, which constitutes the one-time computational investment for self-supervised structural pre-training and multi-task optimization. In contrast, the online stage focuses on real-time interference detection using only a single forward pass through the fine-tuned DAT encoder and the binary detection head. For an input representation partitioned into $N$ tokens with an embedding dimension $D$, the online inference complexity is $O(L \cdot (N^2 D + N D^2))$. As the patch embedding process effectively reduces the sequence length $N$ by a factor of $P^2$, the online complexity is minimized, allowing Agon to support the low-latency operational requirements of next-generation satellite communications.

\begin{algorithm}[t]
\caption{Two-Stage Training Algorithm for the Agon Framework}
\label{alg1}
\KwIn{Labeled dataset $D_L = \{(\mathbf{Y}_n, y_{\text{det},n}, y_{\text{mod},n})\}$, Unlabeled dataset $D_U = \{\mathbf{Y}_n\}$, Hyperparameters $lr_U, lr_F, N_b$, Loss weights $\lambda_{\text{Det}}, \lambda_{\text{Mod}}, \lambda_{\text{Rec}}$.}
\KwOut{Final parameters of DAT Encoder $g_E(\cdot)$ and Detection Head $h_{\text{Det}}(\cdot)$.}
 
 Initialize parameters of $g_E(\cdot)$ and MAE Decoder $g_D(\cdot)$\;
 \Repeat{validation loss stabilizes}{ 
  \For{batch $\{\mathbf{Y}_i\}_{i=1}^{N_b}$ in $D_U$}{
   Define $\mathbf{Y}_{vis}$ after randomly masking $\mathbf{Y}_i$\;
   $\mathbf{Z}_{vis} = g_E(\mathbf{Y}_{vis})$\; 
   $\hat{\mathbf{Y}}_i = g_D(\mathbf{Z}_{vis}, \mathcal{T}_{\text{mask}})$\; 
   Calculate $\mathcal{L}_{\mathrm{MAE}}$ using~(\ref{equ_mae})\; 
   Update $g_E(\cdot)$ and $g_D(\cdot)$ with $lr_U$\;
  }
 }
 Save the pre-trained encoder weights $\mathcal{W}_E$\;
 
 Load pre-trained weights $\mathcal{W}_E$ into $g_E(\cdot)$\; 
 Initialize parameters of heads $h_{\text{Det}}(\cdot)$, $h_{\text{Mod}}(\cdot)$, $h_{\text{Rec}}(\cdot)$\;
 Set $epoch \leftarrow 0$\;
 \Repeat{$epoch = 30$}{
  $epoch \leftarrow epoch + 1$\;
  \For{batch $\{(\mathbf{Y}_i, y_{\text{det},i}, y_{\text{mod},i})\}_{i=1}^{N_b}$ in $D_L$}{
   $\mathbf{Z}_{\text{Final}} = g_E(\mathbf{Y}_i)$\; 
   $\hat{y}_{\text{det},i} = h_{\text{Det}}(\mathbf{Z}_{\text{Final}})$\; 
   $\hat{y}_{\text{mod},i} = h_{\text{Mod}}(\mathbf{Z}_{\text{Final}})$\; 
   $\hat{\mathbf{Y}}_i = h_{\text{Rec}}(\mathbf{Z}_{\text{Final}})$\; 
   Calculate $\mathcal{L}_{\text{Det}}$, $\mathcal{L}_{\text{Mod}}$, $\mathcal{L}_{\text{Rec}}$ using (\ref{equ19})--(\ref{equ21})\;
   Calculate $\mathcal{L}_{\text{Total}}$ using~(\ref{equ18})\;
   Update $g_E, h_{\text{Det}}, h_{\text{Mod}}, h_{\text{Rec}}$ with $lr_F$\;
  }
 }
 \KwRet{Final parameters of $g_E(\cdot)$ and $h_{\text{Det}}(\cdot)$}\;
\end{algorithm}

\section{Experiment Setup}
\label{sec6}
\subsection{Dataset}
\label{sec61}
The validation of the proposed Agon framework is conducted across two distinct datasets to ensure comprehensive evaluation: a publicly available dataset utilized for establishing comparison with prior art, and a self-generated NGSO-NGSO dataset used to assess the full capability of the proposed methodology.

\textit{Public NGSO-GSO Dataset:} For benchmarking against prior art, we employ a publicly available NGSO-GSO interference dataset simulating NGSO satellite interference on a GSO downlink \cite{Lagunas_FNR_SmartSpace}. This dataset comprises 17,281 temporal snapshots, provided as single-channel Time-Domain Magnitude $\mathbf{y}_{n}^{A}$ and Frequency-Domain PSD $\mathbf{y}_{n}^{F}$ representations, inherently lacking the phase information leveraged by Agon's core innovations. Consequently, when evaluating Agon on this dataset, we adapt our framework to use single-channel input and deactivate the phase-sensitive modulation classification loss $\mathcal{L}_{\mathrm{Mod}}$. Following standard anomaly detection protocols, the training set of 10,397 samples and the validation set of 1,175 samples contain exclusively interference-free data corresponding to hypothesis $\mathcal{H}_{0}$. The remaining 2,802 samples form a balanced testing set, including both $\mathcal{H}_{0}$ and interfering $\mathcal{H}_{1}$ instances, reserved for unbiased performance assessment.

\textit{NGSO-NGSO Dataset: }This dataset models a highly complex interference environment based on precise orbital dynamics derived from real-world orbital data. We provide the generated dataset alongside the comprehensive physical configurations, including specific ITU-R antenna patterns, continuous orbital epochs, and diverse modulation schemes, and make it publicly available at GitHub repository \cite{agon_dataset_2026}. We utilized the publicly available TLE data \cite{kelecy2007satellite} sourced from official online repositories for 6,800 Starlink and 650 OneWeb satellites as shown in Fig.~\ref{fig5}. These TLEs, which encode the orbital parameters for each satellite, were fed into the SGP4 orbit propagator \cite{vallado2008sgp4}, the standard model for predicting satellite positions from TLEs. A 48-hour simulation was then executed with a 10-second resolution. At each 10-second time step, the SGP4 propagator computed the exact geocentric position of all 7,450 satellites. This enabled us to derive the time-varying geometric parameters relative to the ground station. These geometric values are the critical inputs for the link budget and EPFD calculations used to determine the ground truth interference label ($\mathcal{H}_0$ or $\mathcal{H}_1$) for each snapshot.

\begin{figure}[t]
  \centering
  \subfloat[The fluctuating number of visible interfering satellite links over the 48-hour simulation period.]{
    \includegraphics[width=\columnwidth]{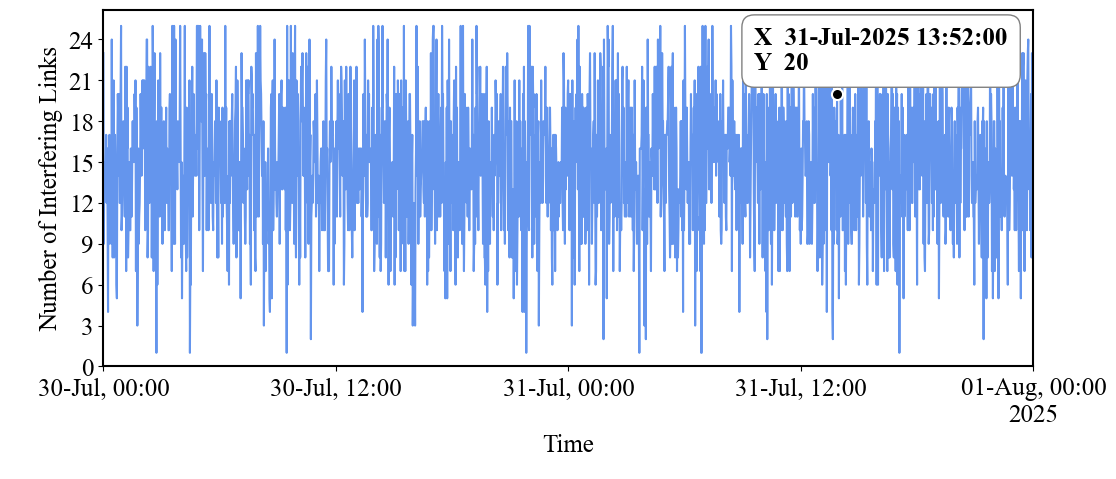}
    \label{fig10a}
  } \hspace{0.01\linewidth}
  \subfloat[The corresponding total EPFD received at the ground station compared against the ITU regulatory threshold.]{
    \includegraphics[width=\columnwidth]{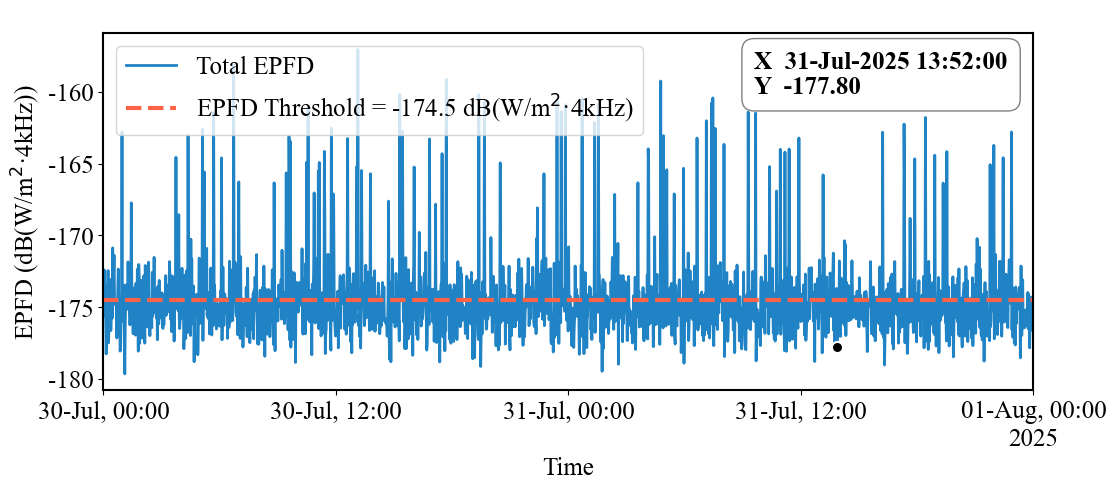}
    \label{fig10b}
  }
  \caption{Temporal dynamics of the simulated NGSO-NGSO interference environment.}
  \label{fig10}
\end{figure}

Fig.~\ref{fig10}\subref{fig10a} illustrates the temporal variation of visible interfering links over the 48-hour simulation while Fig.~\ref{fig10}\subref{fig10b} contrasts the corresponding aggregate EPFD against the ITU regulatory threshold. The framework ingests a high-resolution multi-channel time-frequency representation $\mathbf{Y}\in\mathbb{R}^{T\times F\times C}$ derived from the PWVD of complex I/Q components preserving critical phase information for the auxiliary modulation classification task $\mathcal{L}_{\mathrm{Mod}}$. Ground truth labels $\mathcal{H}_0$ and $\mathcal{H}_1$ are rigorously assigned by evaluating the SGP4-derived instantaneous $\text{EPFD}_{total}$ against the globally unified -174.5 dB(W/(m²·40 kHz)) threshold mandated by the ITU~\cite{ITU_EPFD_support}. To ensure a highly fair architectural evaluation across different interference topologies, we designated 11572 interference-free samples as the unlabeled dataset $D_U$ for Phase 1 pre-training, deliberately partitioning them into exactly 10397 training and 1175 validation instances to perfectly match the established public NGSO-GSO benchmark scale. Furthermore, we divided an additional 2802 perfectly balanced snapshots, allocating a balanced subset of 802 samples equipped with both binary interference labels $y_{\text{det}}$ and modulation labels $y_{\text{mod}}$ to form the labeled dataset $D_L$ for Phase 2 fine-tuning, while reserving the remaining 2000 balanced instances as the held-out test set $D_{\text{test}}$ for unbiased model evaluation.

\subsection{Implementation Details}
\label{sec62}
Experiments were implemented using PyTorch on an NVIDIA GeForce RTX 3060 GPU (6 GB VRAM). The received signal $y_n(t)$ for each snapshot was synthesized according to the normalized signal model in (\ref{equ7}), incorporating the CNR and INR values derived from the link budget and the unit-variance additive white Gaussian noise (AWGN) term $\zeta_n(t)$. Prior to network input, all signal data underwent zero-mean, unit-variance normalization based on statistics from the interference-free training data.
The Time-Frequency Representation $\mathbf{Y} \in \mathbb{R}^{128 \times 128 \times 2}$ was generated using the PWVD. To achieve this, the baseband signal, sampled at $f_s=100$ MHz to satisfy the Nyquist criterion for the 50 MHz bandwidth, was processed using a 256-point Hamming window with a hop length of 128 samples. This configuration yielded the $T=128$ time frames and $F=128$ frequency bins required for the network input. The underlying baseband signals utilized Root Raised Cosine (RRC) pulse shaping with a roll-off factor of 0.25. The DAT Encoder was configured with the model dimension $D$ set to 256. For MAE pre-training, a masking ratio of 60\% was applied, resulting in a reduced visible sequence length of $L_{vis}=16$. Supervised fine-tuning employed the Adam optimizer over 30 epochs with a batch size of 64. The learning rate for pre-training $lr_U$ is set to $1 \times 10^{-4}$, while the learning rate for fine-tuning $lr_F$ was set to $1 \times 10^{-5}$. The key simulation parameters and model configurations employed in this study are summarized in Table~\ref{tab2}.

\begin{table}[t]
\centering
\caption{Simulation Parameters}
\label{tab2}
\small
\begin{tabular}{@{}ll@{}}
\toprule
Parameter                      & Value \\
\midrule
Ground station location        & \ang{47.69}N, \ang{-122.03}W \\
Constellations simulated       & 6800 Starlink, 650 OneWeb \\
Simulation step $\Delta\tau$   & 10 s \\
Minimum elevation angle $\theta_{min}$ & \ang{10} \\
Center frequency $f_c$         & 11.75 GHz \\
Signal bandwidth $B$           & 50 MHz \\
Interfering satellite EIRP $EIRP_k$        & 44.7 dBW \\
Transmitted EIRP $EIRP_d$        & 48.7 dBW \\
Transmit gain $G_{t,k}$        & Based on ITU-R S.1587 \cite{ITU-R_S1587}\\
Antenna diameter $D_A$         & 1.2 m \\
Antenna efficiency $e_A$       & 0.6 \\
System noise temp $T_{sys}$    & 250 K \\
Boltzmann constant $k_B$      & -228.6 dBW/(K·Hz) \\
Additional loss $L_{add}$      & 2 dB \\
EPFD limit $EPFD_{limit}$      & -174.5 dB(W/(m²·40 kHz)) \\
Bandwidth $B_{ref}$            & 40 kHz \\
Receive gain $G_{r,ngso}$      & Based on ITU-R S.1428-1 \cite{ITU-R_S1428-1}\\
$\mathcal{L}_{Mod}$ modulations & BPSK, QPSK, 8PSK \\
                               & 8QAM, 16APSK, 32APSK \\
Spectrogram dimensions $T\times F$  & $128\times128$ \\
MAE Patch size & $16\times16$ \\
Attention heads $h$ & 4 \\
Wavelet loss $\alpha$          & 1.0 \\
Wavelet scales $S$     & 3 \\
Loss weight $\lambda_{Det}$, $\lambda_{Mod}$, $\lambda_{Rec}$ & 1.0, 0.5, 0.5 \\
Learning rate $lr_U$, $lr_F$ & $1 \times 10^{-4}$, $1 \times 10^{-5}$ \\
\bottomrule
\end{tabular}
\end{table}

\begin{figure}[t]
    \centering   
    \includegraphics[width=0.8\columnwidth]{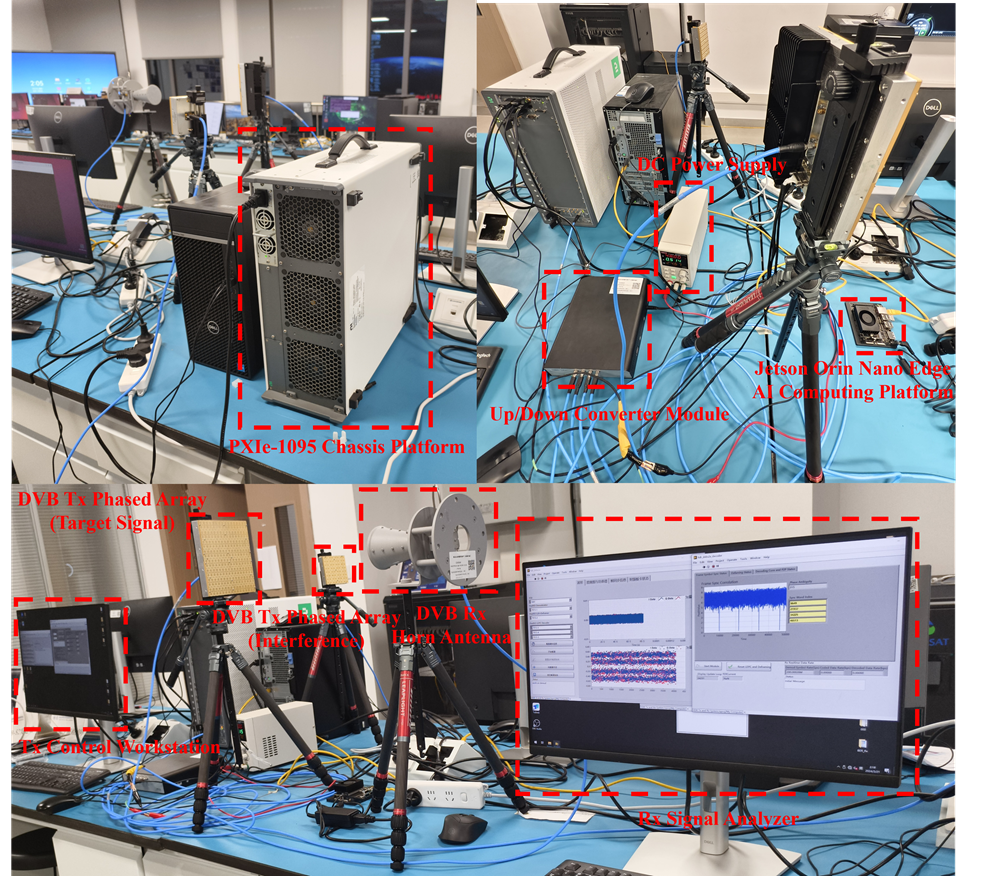} 
    \caption{The integrated indoor physical validation testbed architecture. The controlled indoor radio frequency transmission and reception environment (bottom) explicitly replicates authentic hardware impairments while maintaining absolute ground truth control. The physical reception and edge AI computing pipeline includes a DC power supply, an up/down converter module, and the Jetson Orin Nano edge platform (top right), with the PXIe-1095 data acquisition chassis (top left) providing the fundamental acquisition base.}
    \label{fig11}
\end{figure}

To evaluate practical applicability, we developed the integrated physical radio frequency testbed visualized in Fig.~\ref{fig11}. The infrastructure utilizes an indoor controlled physical transmission system to emit high-fidelity synthesized baseband waveforms from the NGSO-NGSO dataset. Specifically, the testbed incorporates two LEO-specific digital video broadcasting (DVB) transmit phased arrays to generate signals. One array generates the target signal while the other emits the aggregate interference waveform. These transmitter (Tx) arrays are managed by the Tx control workstation. The emitted signals are captured via a receiving horn antenna connected to the PXIe-1095 chassis platform and the up/down converter module, natively introducing authentic hardware phase noise and non-linear amplifier distortion while maintaining absolute analytical control. To comprehensively evaluate real-world deployment feasibility, we directly integrated a Jetson Orin Nano edge AI computing platform \cite{nvidia_jetson_orin_nano} into this receiver (Rx) pipeline to process the digitized baseband streams, representing a standard resource-constrained ground station environment. The processed signals and ground truths can be visually evaluated using the Rx signal analyzer.

\subsection{Baseline models}
\label{sec63}
To rigorously assess the performance gains of the proposed Agon framework against the SOTA, we perform a comprehensive comparison against models spanning traditional methods and contemporary deep learning architectures. We first establish the performance floor using the classical ED, which relies on simple signal power thresholding for anomaly detection. This is followed by a set of deep learning benchmarks adapted for reconstruction-based anomaly detection, including the convolutional variant convolutional neural network-autoencoder (CNN-AE) and the sequence-modeling long short-term memory autoencoder (LSTMAE). All these models are trained exclusively on interference-free data $\mathcal{H}_0$ and rely on the reconstruction error threshold for final $\mathcal{H}_1$ prediction. Finally, we include the contemporary GenAI models which represent the current SOTA in generative anomaly detection in this domain: the variational autoencoder (VAE) and the TrID \cite{saifaldawla2024genai}.The TrID is a critical benchmark as it achieved superior performance in recent comparative studies. These benchmarks are evaluated across both the public single-channel dataset and our self-generated multi-channel PWVD dataset to validate the generalizability of the Agon feature set.

\begin{table*}[t]
\centering
\caption{Performance Comparison of Different Models on Two Datasets}
\label{tab3}
\renewcommand{\arraystretch}{1.3} 
\setlength{\tabcolsep}{5pt}       

\begin{tabular}{l p{9cm} c c c c c} 
\toprule
Model & Dataset Details (Scale \& Type) & Accuracy & F1 Score & AUC Score & TPR & FPR \\
\toprule

\multirow{2}{*}{Agon (Proposed)} 
 & Public NGSO-GSO: 10 Starlink, 10 OneWeb, Single-Channel Magnitude & \textbf{0.9251} & \textbf{0.8351} & \textbf{0.9327} & \textbf{0.9355} & \textbf{0.0553} \\
 & NGSO-NGSO: 6,800 Starlink, 650 OneWeb, Dual-Channel PWVD & \textbf{0.9027} & \textbf{0.9055} & \textbf{0.9085} & \textbf{0.9122} & \textbf{0.1304} \\ 
\hline

\multirow{2}{*}{TrID} 
 & Public NGSO-GSO: 10 Starlink, 10 OneWeb, Single-Channel Magnitude & 0.8318 & 0.8321 & 0.8318 & 0.8399 & 0.1763 \\ 
 & NGSO-NGSO: 6,800 Starlink, 650 OneWeb, Dual-Channel PWVD & 0.7203 & 0.7201 & 0.7253 & 0.7255 & 0.2204 \\ 
\hline

\multirow{2}{*}{VAE} 
 & Public NGSO-GSO: 10 Starlink, 10 OneWeb, Single-Channel Magnitude & 0.5766 & 0.5742 & 0.5766 & 0.5692 & 0.3161 \\ 
 & NGSO-NGSO: 6,800 Starlink, 650 OneWeb, Dual-Channel PWVD & 0.5673 & 0.5655 & 0.5693 & 0.5705 & 0.3255 \\ 
\hline

\multirow{2}{*}{CNN-AE} 
 & Public NGSO-GSO: 10 Starlink, 10 OneWeb, Single-Channel Magnitude & 0.8020 & 0.8054 & 0.8825 & 0.7967 & 0.1754 \\ 
 & NGSO-NGSO: 6,800 Starlink, 650 OneWeb, Dual-Channel PWVD & 0.6903 & 0.6955 & 0.7003 & 0.7055 & 0.2403 \\ 
\hline

\multirow{2}{*}{LSTMAE} 
 & Public NGSO-GSO: 10 Starlink, 10 OneWeb, Single-Channel Magnitude & 0.5764 & 0.5744 & 0.5764 & 0.5715 & 0.3187 \\ 
 & NGSO-NGSO: 6,800 Starlink, 650 OneWeb, Dual-Channel PWVD & 0.5553 & 0.5605 & 0.5653 & 0.5705 & 0.3303 \\ 
\hline

ED & Public NGSO-GSO: 10 Starlink, 10 OneWeb, Single-Channel Magnitude & 0.5610 & 0.5320 & 0.5710 & 0.3950 & 0.2640 \\

\bottomrule
\end{tabular}
\end{table*}

\subsection{Overall Performance}
\label{sec64}
\begin{figure}[t]
  \centering
  \subfloat[ROC Curve on Public NGSO-GSO Dataset.]{
    \includegraphics[width=0.49\linewidth]{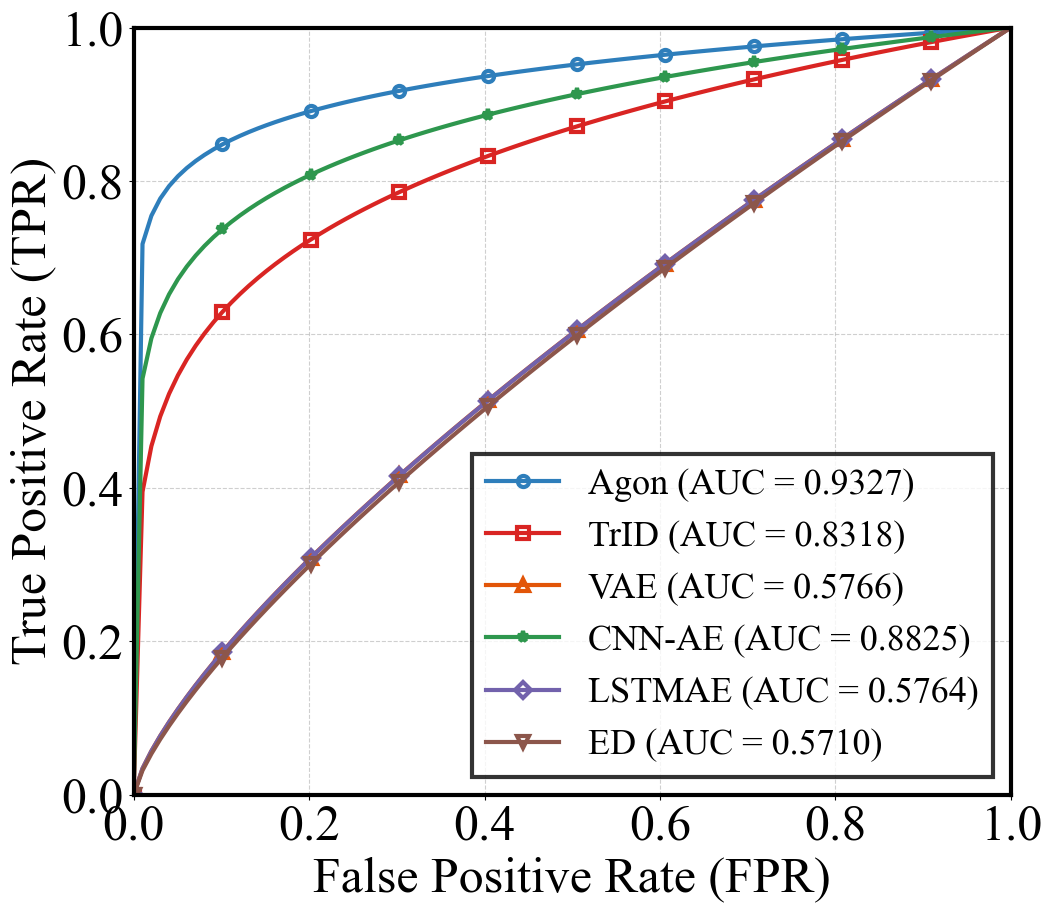}
    \label{fig12a}
  }
  \subfloat[ROC Curve on NGSO-NGSO Dataset.]{
    \includegraphics[width=0.49\linewidth]{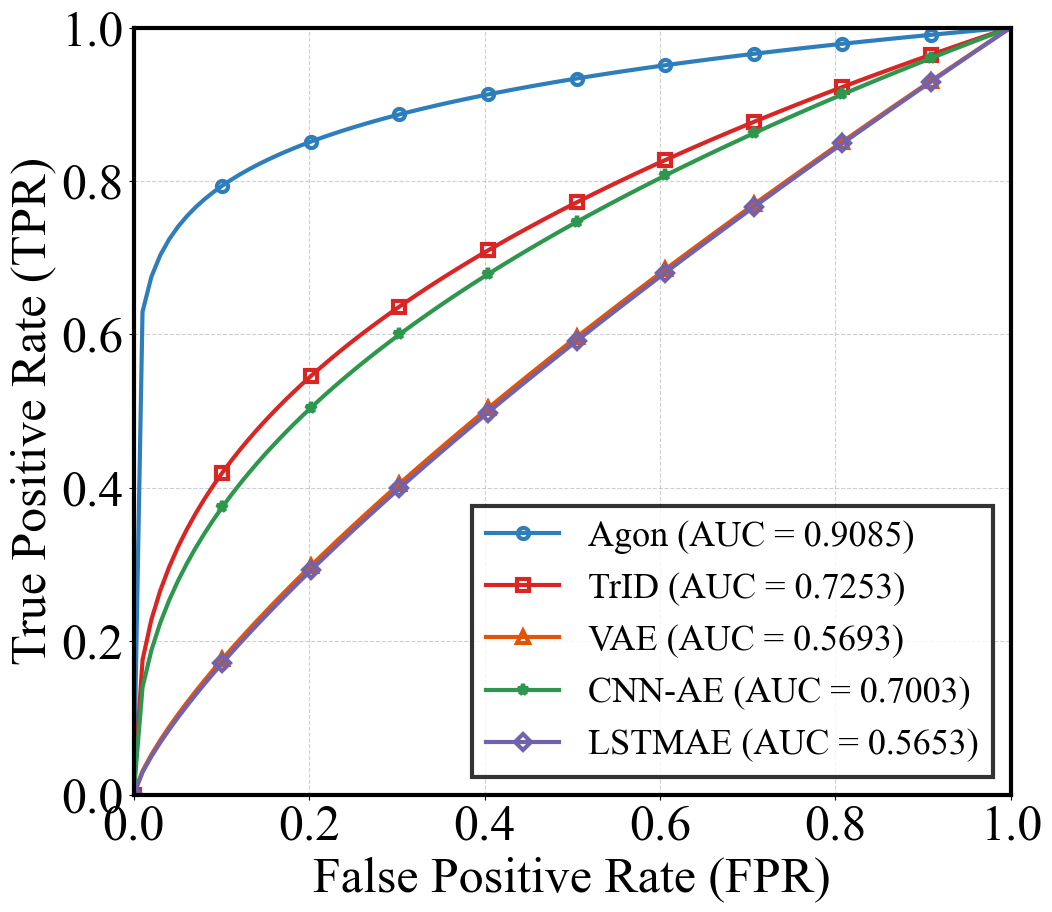}
    \label{fig12b}
  }
  \caption{ROC Curve Comparison across Datasets. Agon achieves an AUC of 0.9327 on public dataset and maintains a high AUC of 0.9085 on NGSO-NGSO dataset where baselines significantly degrade.}
  \label{fig12}
  \vspace{-3ex}
\end{figure}

The discriminative performance of the proposed framework is rigorously assessed using ROC curve analysis \cite{fan2006understanding}, with quantitative results detailed in Table~\ref{tab3}. On the Public NGSO-GSO benchmark, Agon demonstrates superior generalization, achieving an Accuracy of 0.9251, an AUC of 0.9327, and an F1 score of 0.8351. As illustrated in Fig.~\ref{fig12}\subref{fig12a}, Agon maintains a high True Positive Rate (TPR) of 0.9355 while suppressing the FPR to a negligible 0.0553. This represents a substantial improvement over the 0.1763 FPR observed in the TrID baseline and the 0.1754 FPR in CNN-AE. This performance confirms the efficacy of the direct binary classification strategy. Traditional reconstruction-based models (VAE with AUC 0.5766) struggle to distinguish anomalies from noise near the decision boundary. By eliminating reliance on heuristic reconstruction error thresholds, Agon establishes a precise, learnable decision boundary even with single-channel magnitude data, effectively circumventing the sensitivity-specificity trade-off that limits traditional models in standard scenarios.

The robustness of the framework is further substantiated on the high-fidelity NGSO-NGSO dataset, which introduces a complex and dense interference topology. As shown in Fig.~\ref{fig12}\subref{fig12b}, the baseline models exhibit substantial performance degradation, with TrID’s AUC falling from 0.8318 to 0.7253 and CNN-AE’s AUC declining to 0.7003. In contrast, Agon displays remarkable robustness, maintaining an AUC of 0.9085 along with an F1 score of 0.9055. Theoretically, this sustained margin of victory validates the proposed architectural innovations. In this complex scenario, the HOS augmentation explicitly captures statistical dependencies to differentiate structured interference from the increased background noise, while the Wavelet Regularization loss enforces multi-scale structural fidelity against transient signal variations. These mechanisms synergistically enable Agon to extract discriminative features from dynamic, low-SNR interference patterns that overwhelm standard baselines, maintaining a low FPR of 0.1304 compared to TrIDs 0.2204.

\subsection{Robustness analysis across critical physical parameters}
\label{sec65}

\begin{figure}[t]
  \centering
  \subfloat[$P_D$ vs. NGSO Off-axis Angle $\theta$.]{
    \includegraphics[width=0.49\linewidth]{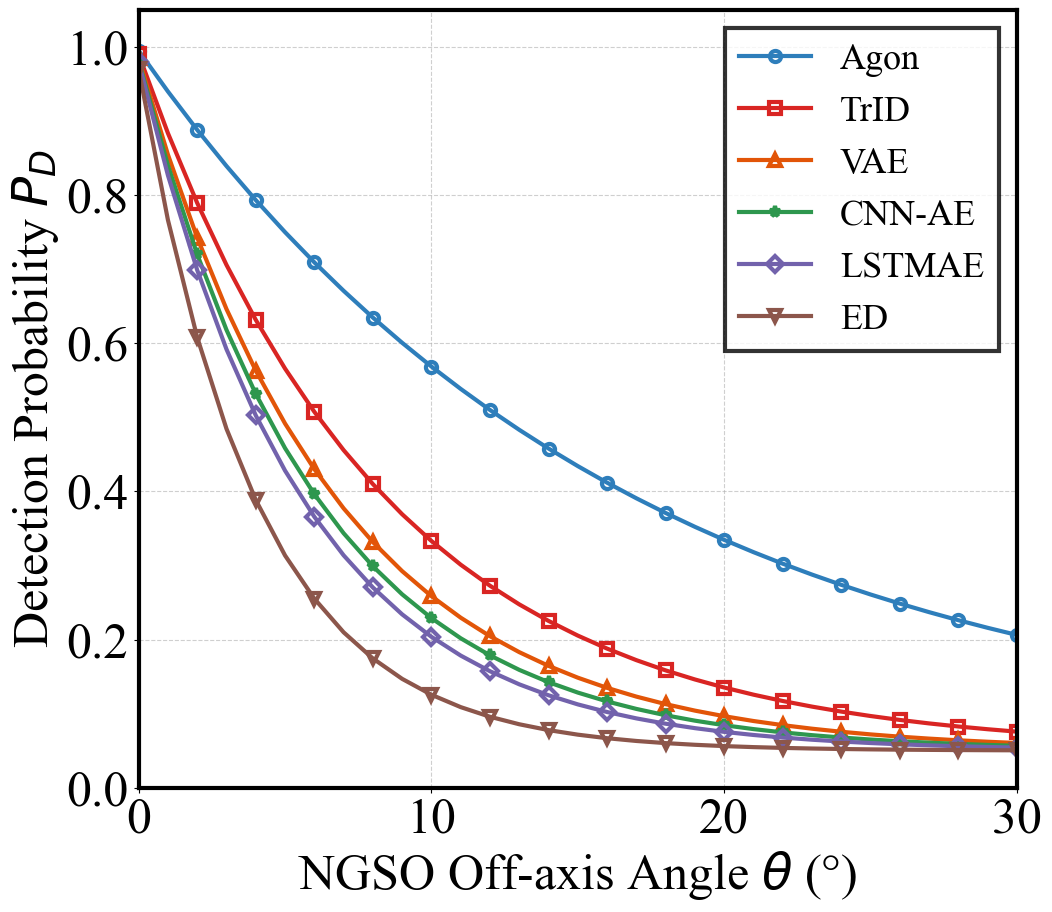}
    \label{fig13a}
  }
  \subfloat[$P_D$ vs. NGSO INR.]{
    \includegraphics[width=0.49\linewidth]{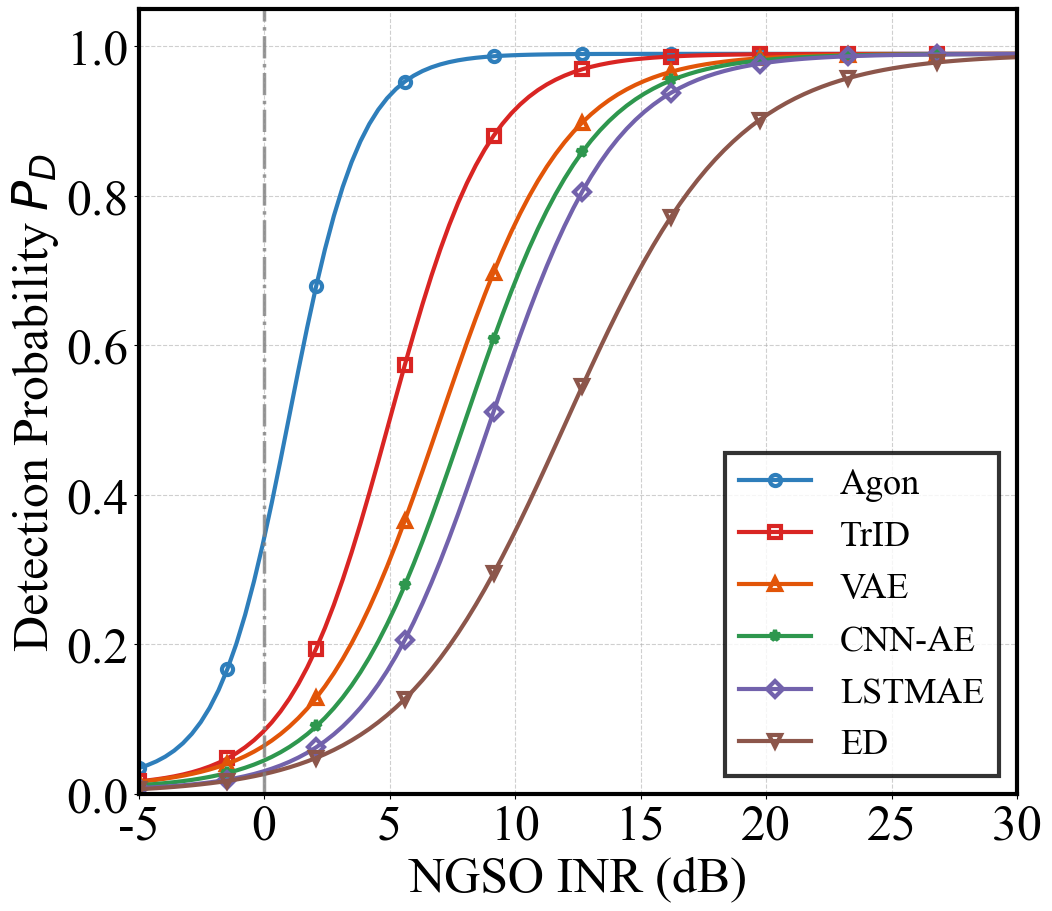}
    \label{fig13b}
  }
  \caption{Agon consistently outperforms baselines by maintaining higher detection probabilities under significant off-axis geometric distortions and achieving reliable detection at lower INR levels.}
  \label{fig13}
  \vspace{-3ex}
\end{figure}

We evaluate robustness by examining the detection probability $P_D$ under spatial and interference variations. As illustrated in Fig.~\ref{fig13}\subref{fig13a}, Agon maintains the highest detection performance across all off-axis angles and shows only a mild decline with increasing $\theta$. Around $10^\circ$, it still achieves a $P_D$ near 0.6, whereas TrID drops to roughly 0.4 and VAE decreases even more noticeably, indicating their vulnerability to side-lobe attenuation. Fig.~\ref{fig13}\subref{fig13b} further shows Agon's advantage across varying interference levels. Even in extreme low-INR regimes from 0 dB down to -5 dB, where detection inevitably degrades across all frameworks due to physical link limits, Agon maintains a substantial performance margin. As INR increases, Agon's curve rises quickly and approaches unity near 8 dB, compared to nearly 15 dB for TrID. These trends demonstrate that Agon preserves essential statistical structures despite severe power attenuation. Specifically, the proposed HOS augmentation enhances higher-order statistical cues that remain robust even when first-order spectral energy is completely masked by background noise. This allows Agon to separate structured interference from noise more effectively than traditional reconstruction-based methods, providing a richer representation space where coherent interference signatures remain identifiable even when first-order spectral features are severely weakened. This strengthened reliance on second-order statistics effectively increases Agon’s invariance to distortions inherent in NGSO links, enabling more consistent detection across a wide range of physical conditions.

\begin{figure}[t]
  \centering
  \subfloat[AUC Score on Public NGSO-GSO dataset.]{
    \includegraphics[width=0.49\linewidth]{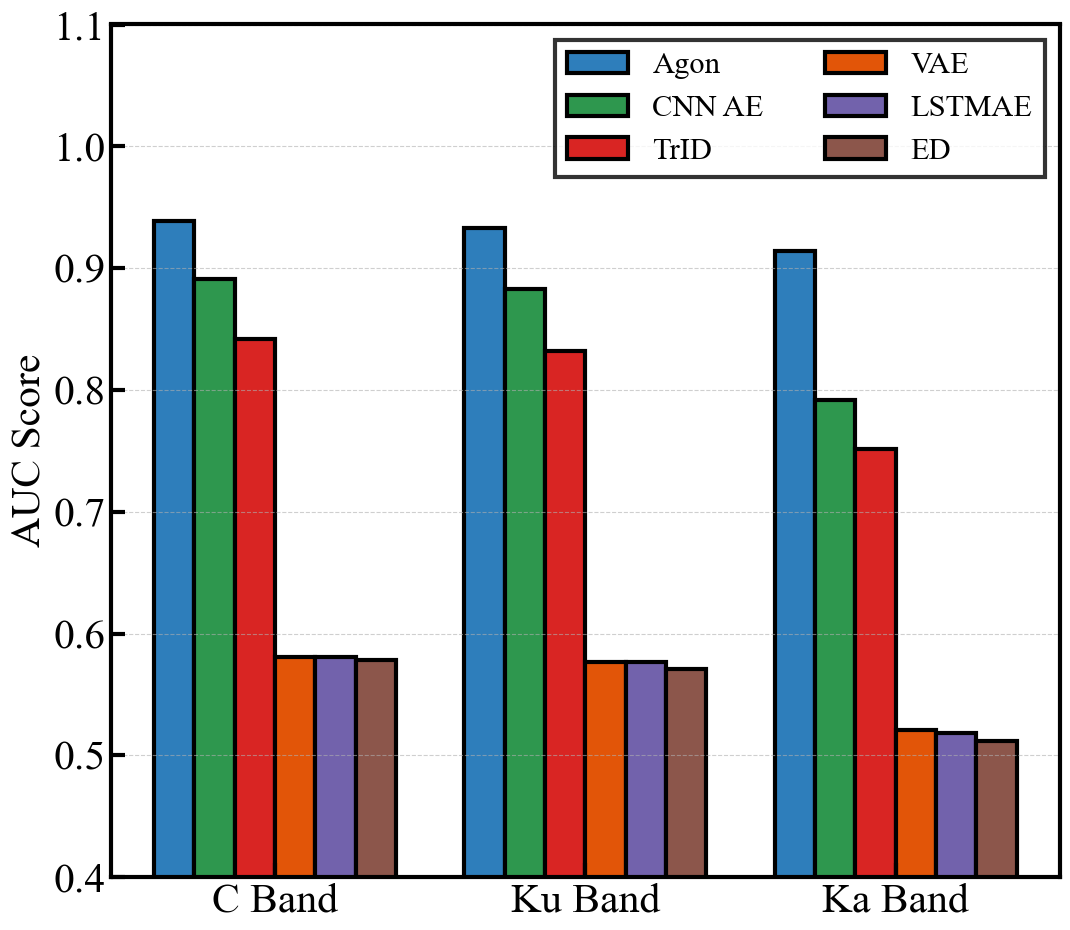}
    \label{fig14a}
  }
  \subfloat[AUC Score on NGSO-NGSO dataset.]{
    \includegraphics[width=0.49\linewidth]{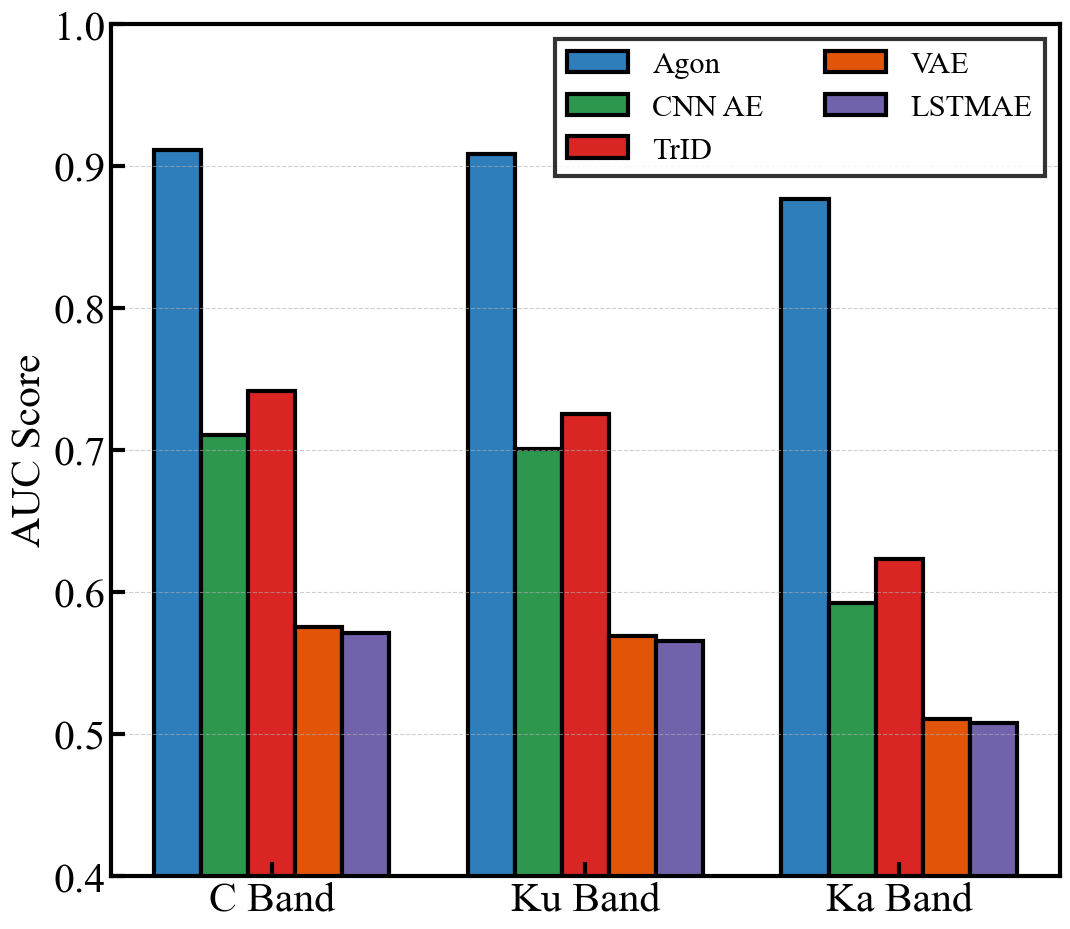}
    \label{fig14b}
  }
  \caption{Performance generalization across different frequency bands.}
  \label{fig14}
  \vspace{-3ex}
\end{figure}

Furthermore, to verify the generalization capability across mainstream spectrum allocations we rigorously evaluated the framework under C-band and Ka-band physical constraints against established baselines for each dataset. Mathematically the physical propagation disparities across these distinct carrier frequencies exclusively manifest as received power variations driven by frequency-dependent free space path loss and atmospheric attenuation such as severe Ka-band rain fade. Because the architecture processes normalized complex baseband representations post-downconversion the underlying statistical distributions governing uncorrelated thermal noise and structured interference remain fundamentally identical regardless of the initial radio frequency. As quantitatively demonstrated in Fig.~\ref{fig14}\subref{fig14a} for the public NGSO-GSO dataset Agon achieves exceptional generalization maintaining an AUC score above 0.91 even under severe Ka-band attenuation. Fig.~\ref{fig14}\subref{fig14b} illustrates consistent performance on the high-fidelity NGSO-NGSO dataset where Agon sustains superior detection capabilities across all evaluated spectrums. While the entire suite of generative and reconstructive baselines exhibits severe performance collapse under extreme Ka-band energy loss Agon successfully isolates the invariant off-diagonal correlation structures. By leveraging the high-order statistics module to capture universal baseband priors rather than carrier-specific artifacts the detection mechanism maintains robust generalization without requiring any architectural modifications.

\subsection{Multi-Task Efficacy and Feature Validation}
\label{sec66}
\begin{figure}[htbp]
\centering
\includegraphics[width=\columnwidth]{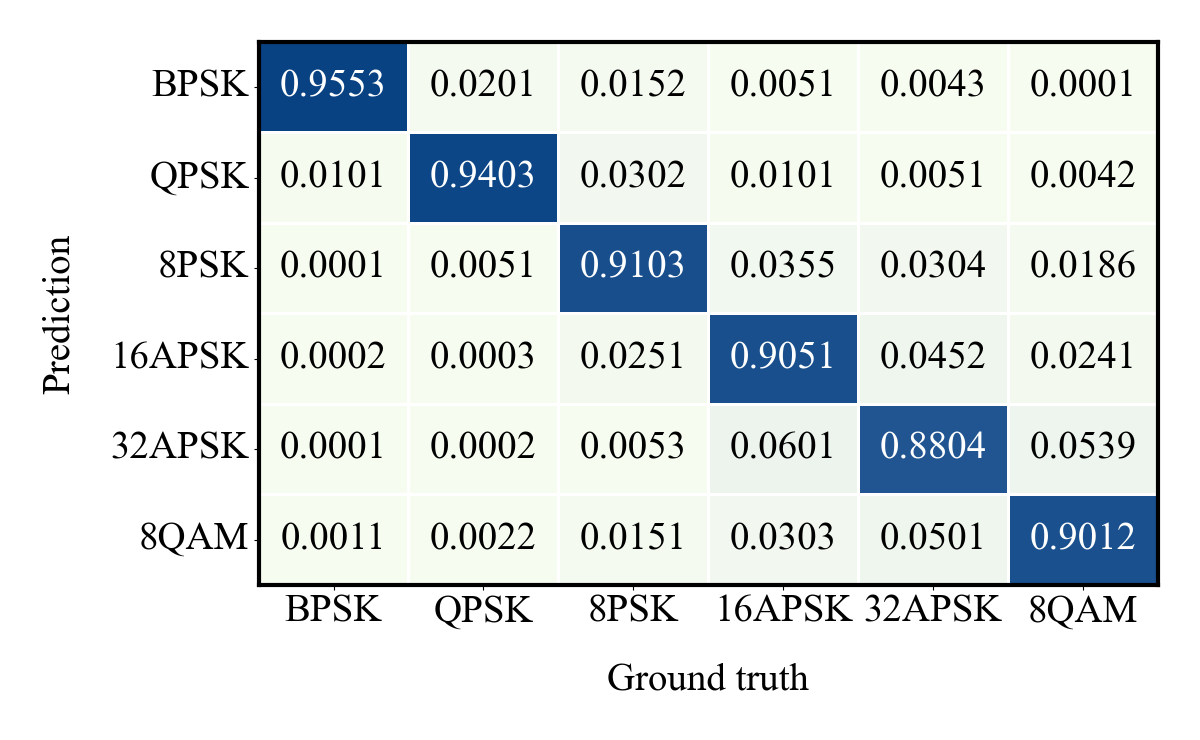}
\caption{Normalized confusion matrix for modulation classification.}
\label{fig15}
\end{figure}

The efficacy of the auxiliary Modulation Classification task $\mathcal{L}_{\mathrm{Mod}}$ is demonstrated by the DAT Encoder’s ability to reliably distinguish six representative NGSO modulation schemes. As shown by the normalized confusion matrix in Fig.~\ref{fig15}, the model achieves over 90\% accuracy for nearly all classes, indicating strong feature separability. This level of precision confirms that the DAT Encoder is learning intrinsic signal characteristics such as phase evolution patterns and constellation geometry, rather than depending on superficial energy cues. By promoting this fine-grained semantic understanding, the auxiliary task serves as a structural regularizer that prevents overfitting to overly simplistic binary features. Consequently, the aggregated feature representation $\mathbf{Z}_{\mathrm{Final}}$ becomes richer in physically meaningful and modulation-invariant information, which enhances the robustness and discriminative power of the primary interference detection task in complex and time-varying noise environments.

\begin{figure}[t]
\centering
\includegraphics[width=\columnwidth]{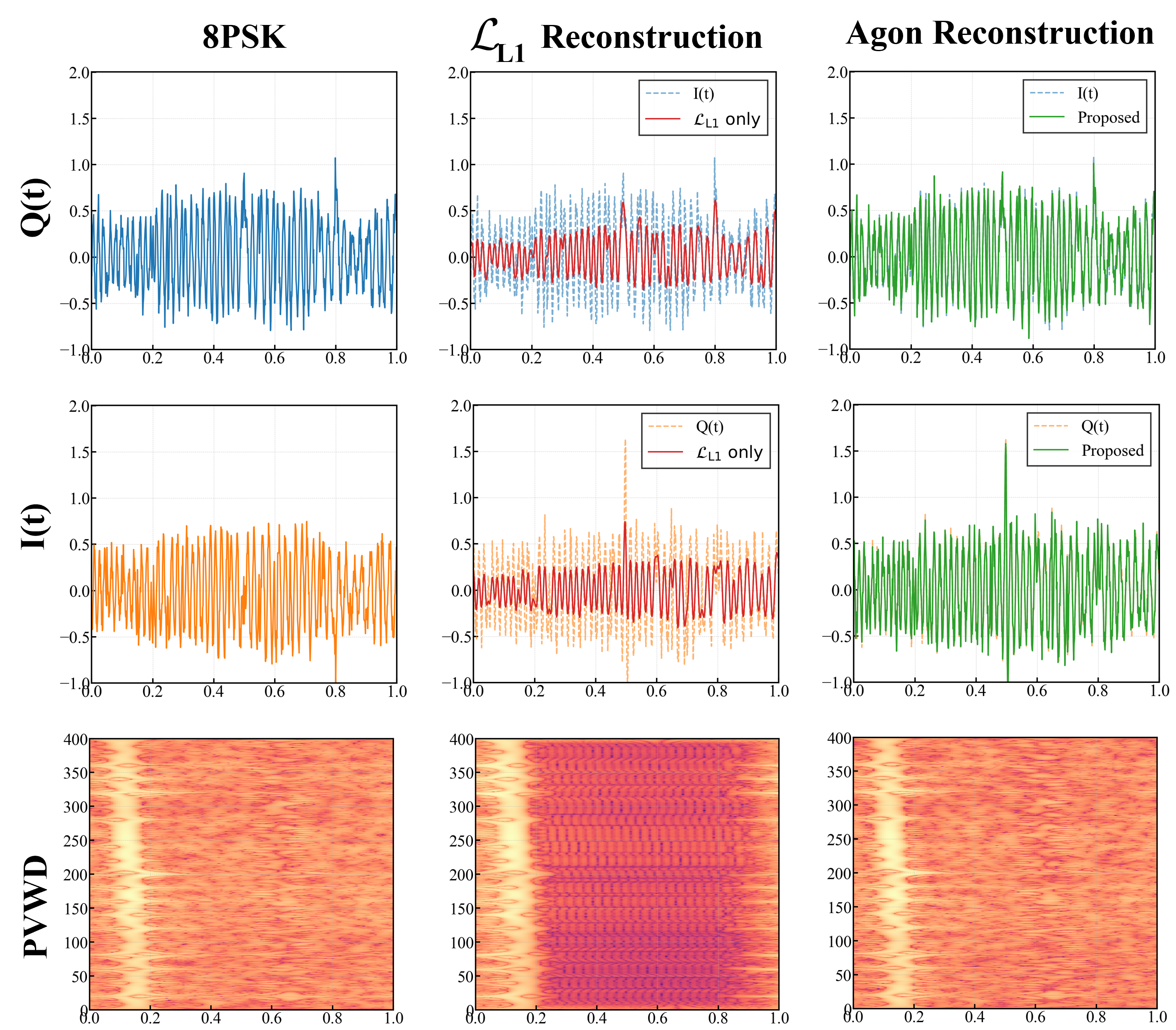}
\caption{Visual validation on signal reconstruction.}
\label{fig16}
\end{figure}

The structural integrity of the features is validated through assessing the impact of the structural regularization objective, $\mathcal{L}_{\mathrm{Rec}}$, which is confirmed by a direct visual comparison of signal reconstruction quality, as demonstrated in Fig.~\ref{fig16}. This figure contrasts the complex In-phase I(t) and Quadrature Q(t) transient features reconstructed by the proposed $\mathcal{L}_{\mathrm{Total}}$ loss against a baseline model trained only with a standard $\mathcal{L}_{\mathrm{L1}}$ loss. The visual superiority of the Agon reconstruction is apparent, particularly its accurate recovery of sharp, instantaneous features and high-frequency waveform details. This success confirms that the Wavelet Loss, a key component of $\mathcal{L}_{\mathrm{Rec}}$, enforces constraints in the DWT domain, effectively serving as a multi-scale regularization mechanism. By this method, the model is prevented from suffering catastrophic forgetting of essential low-level spectral details, enabling superior structural recovery compared to simple pixel-wise error minimization.

\subsection{Efficiency Assessment}
\label{sec67}
\begin{figure}[t]
  \centering
  \subfloat[Convergence on NGSO-GSO dataset.]{
    \includegraphics[width=0.49\linewidth]{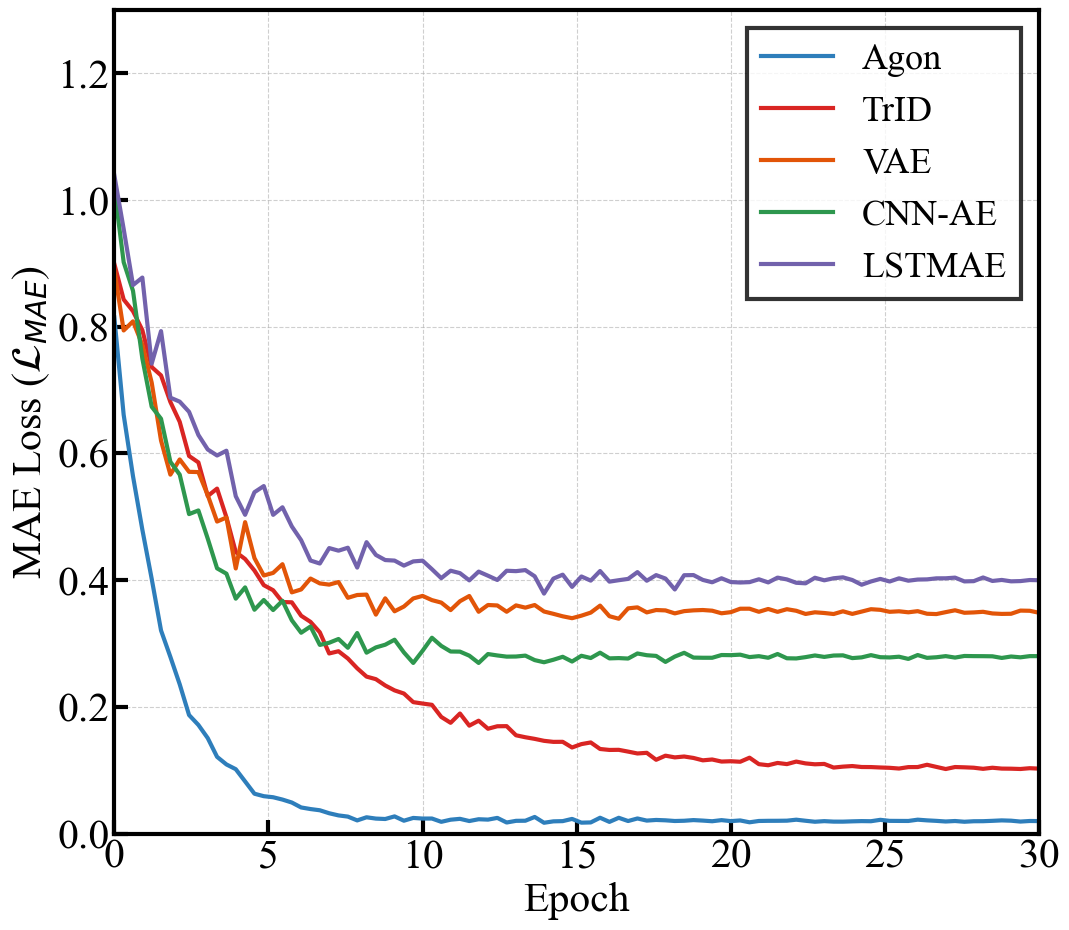}
    \label{fig17a}
  }
  \subfloat[Convergence on NGSO-NGSO dataset.]{
    \includegraphics[width=0.49\linewidth]{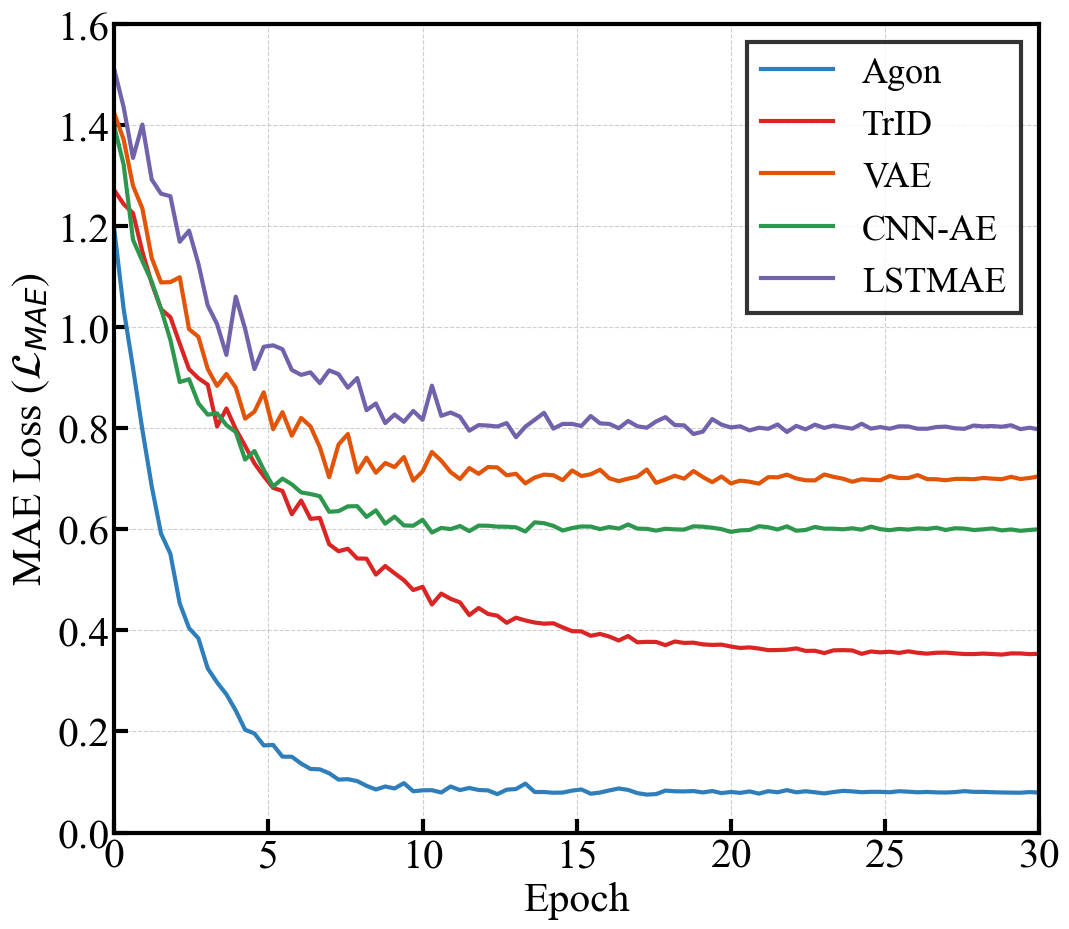}
    \label{fig17b}
  }
  \caption{Comparative analysis of model training stability.}
  \label{fig17}
  \vspace{-3ex}
\end{figure}

The training stability and feature learning capability of the Agon framework are demonstrated by the reconstruction loss convergence curves in Fig.~\ref{fig17}\subref{fig17a} and Fig.~\ref{fig17}\subref{fig17b}. Although all baselines achieve rapid early loss reduction, Agon consistently converges to the lowest final loss across both data environments. This pattern reflects the benefits of the MAE-based self-supervised initialization, which offers a strong foundation for representation learning, as well as the wavelet regularization loss, which promotes multi-scale structural coherence. Together, these components establish a more stable optimization trajectory and encourage the extraction of features that remain informative under diverse and noisy signal conditions, enabling the downstream classifier to achieve higher accuracy and stronger robustness.

\begin{figure}[t]
  \centering
  \subfloat[Number of parameters.]{
    \includegraphics[width=0.49\linewidth]{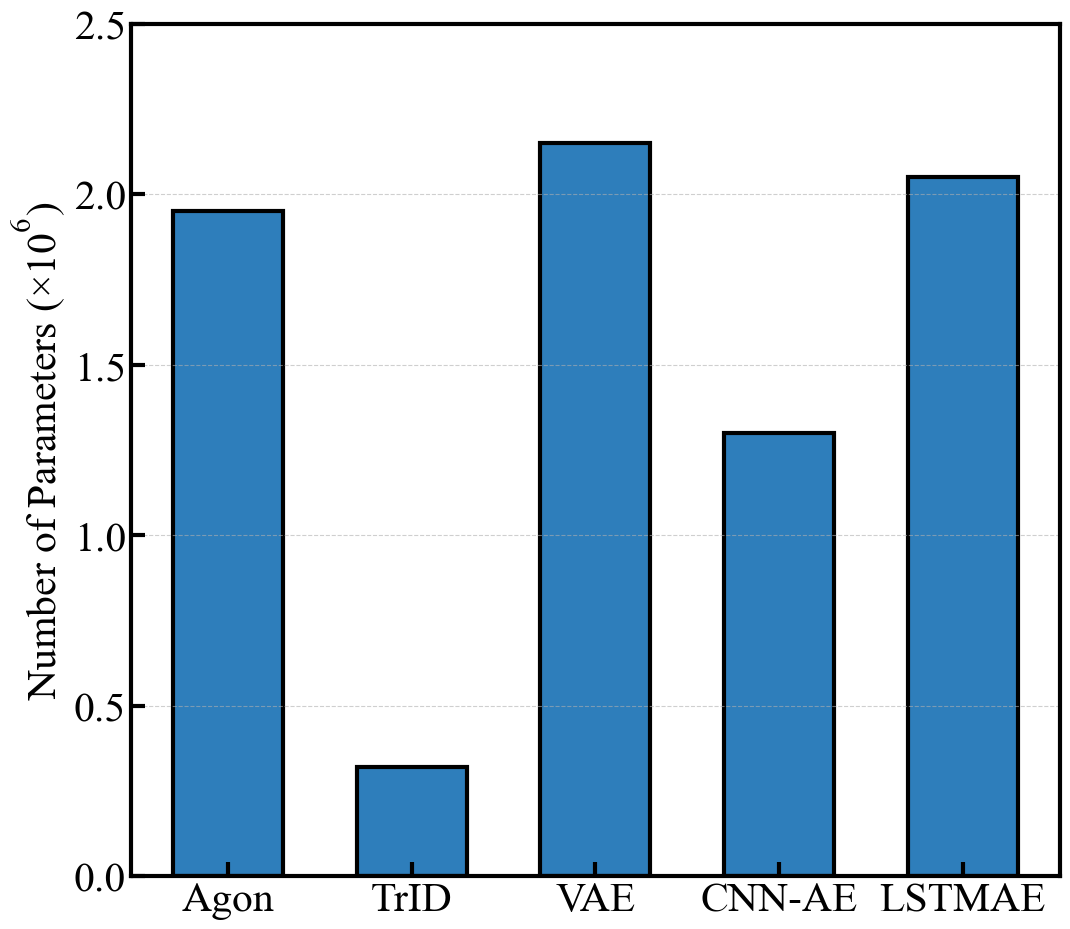}
    \label{fig18a}
  }
  \subfloat[Training and average inference time.]{
    \includegraphics[width=0.49\linewidth]{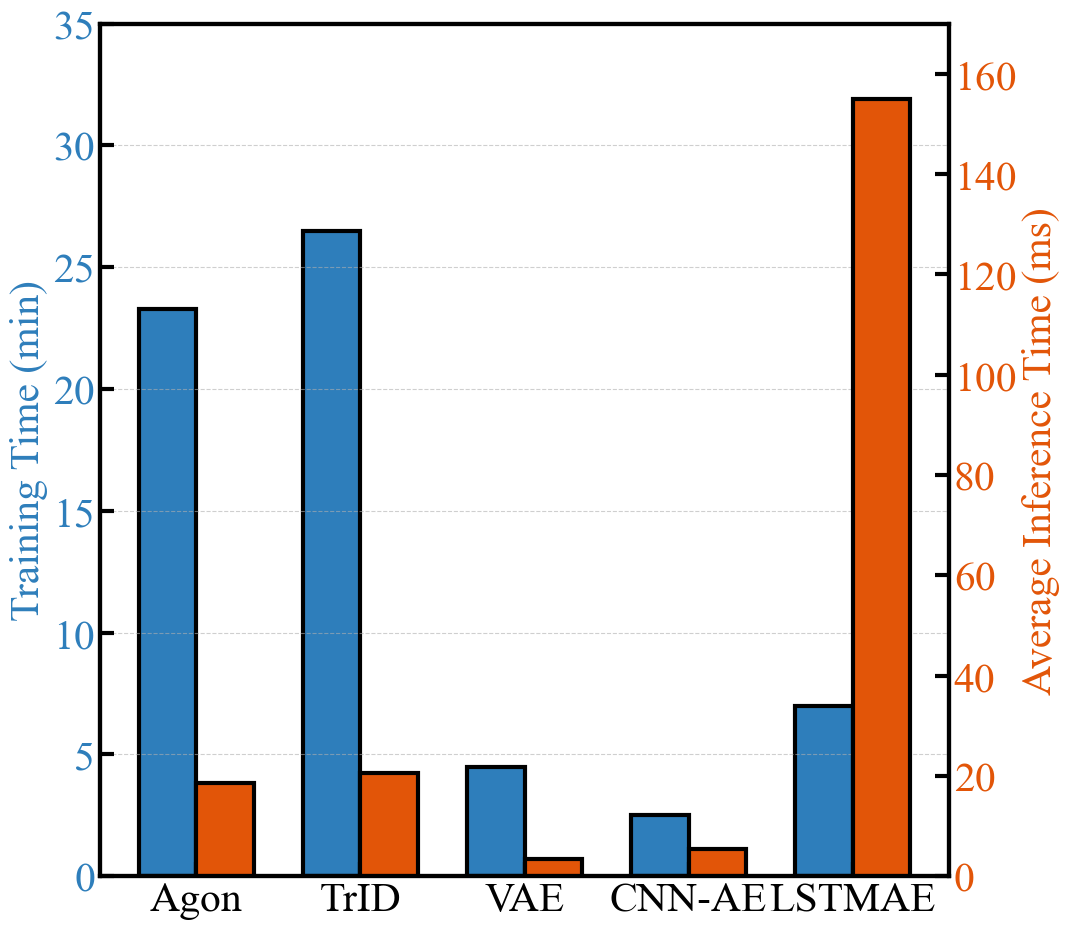}
    \label{fig18b}
  }
  \caption{Computational cost and model efficiency analysis.}
  \label{fig18}
  \vspace{-3ex}
\end{figure}

Unifying theoretical formulations with empirical measurements Fig.~\ref{fig18}\subref{fig18a} demonstrates the proposed architecture utilizes 1.96 million parameters avoiding heavy generative model bloat. Theoretically unpatched transformers like TrID exhibit a rigid $O(L \cdot ((P^2 N)^2 D + (P^2 N) D^2))$ dependency causing severe quadratic overhead. Convolutional baselines including CNN-AE and VAE require $O(L_c \cdot (P^2 N) \cdot C_c^2 \cdot K_c^2)$ operations scaling heavily with depth $L_c$ channels $C_c$ and kernel dimensions $K_c$ while recursive architectures like LSTMAE impose an irreducible $O(L_s \cdot d_h^2)$ complexity preventing hardware parallelization. Empirically Fig.~\ref{fig18}\subref{fig18b} confirms Agon achieves an 18 ms inference latency directly outperforming TrID while structural bottlenecks restrict LSTMAE to processing times exceeding 140 milliseconds rendering it fundamentally incompatible with dynamic satellite downlinks. These results rigorously prove the methodology achieves a Pareto-optimal balance under strict sub-second operational constraints.

Transitioning from high-performance graphics processing units to realistic resource-constrained environments we conduct comprehensive computational profiling utilizing the complex dual-channel representation specifically derived from the NGSO-NGSO dataset. We explicitly deploy the architectures directly on the Jetson Orin Nano edge AI platform \cite{nvidia_jetson_orin_nano} operating with FP16 precision and a strict unit batch size. As quantified in Table~\ref{tab4} the hardware metrics confirm that the proposed Agon framework sustains a highly efficient 18.6 ms average latency while maintaining a competitive 10.2 W power profile. Although recursive architectures like LSTMAE exhibit marginally lower memory footprints and power consumption their sequential computational nature generates a prohibitive 112.4 ms latency that fundamentally violates the sub-second operational constraints of dynamic NGSO links. In stark contrast the parallel attention mechanism within Agon achieves a sixfold speedup over LSTMAE by incurring a negligible increase in hardware resource overhead. These explicit physical deployment measurements substantiate that our methodology achieves the optimal Pareto-optimal balance between high-speed inference and energy efficiency on standard commercial edge computing devices.

\begin{table}[htbp]
\centering
\caption{Empirical Deployment Metrics on Jetson Orin Nano Edge AI Platform (Batch Size = 1)}
\label{tab4}
\setlength{\tabcolsep}{3pt} 
\renewcommand{\arraystretch}{1.2}
\begin{tabular}{@{}l c c c c@{}}
\toprule
Model & Precision & Latency (ms) & Memory (MB) & Power (W) \\
\midrule
\textbf{Agon (Proposed)} & \textbf{FP16} & \textbf{18.6} & 845 & 10.2 \\
TrID & FP16 & 48.2 & 1420 & 13.8 \\
CNN-AE & FP16 & 28.4 & 950 & 11.5 \\
VAE & FP16 & 54.3 & 1250 & 12.4 \\
LSTMAE & FP16 & 112.4 & \textbf{680} & \textbf{9.8} \\
\bottomrule
\end{tabular}
\end{table}

\begin{figure}[t]
  \centering
  \subfloat[Sample Efficiency on NGSO-GSO dataset.]{
    \includegraphics[width=0.49\linewidth]{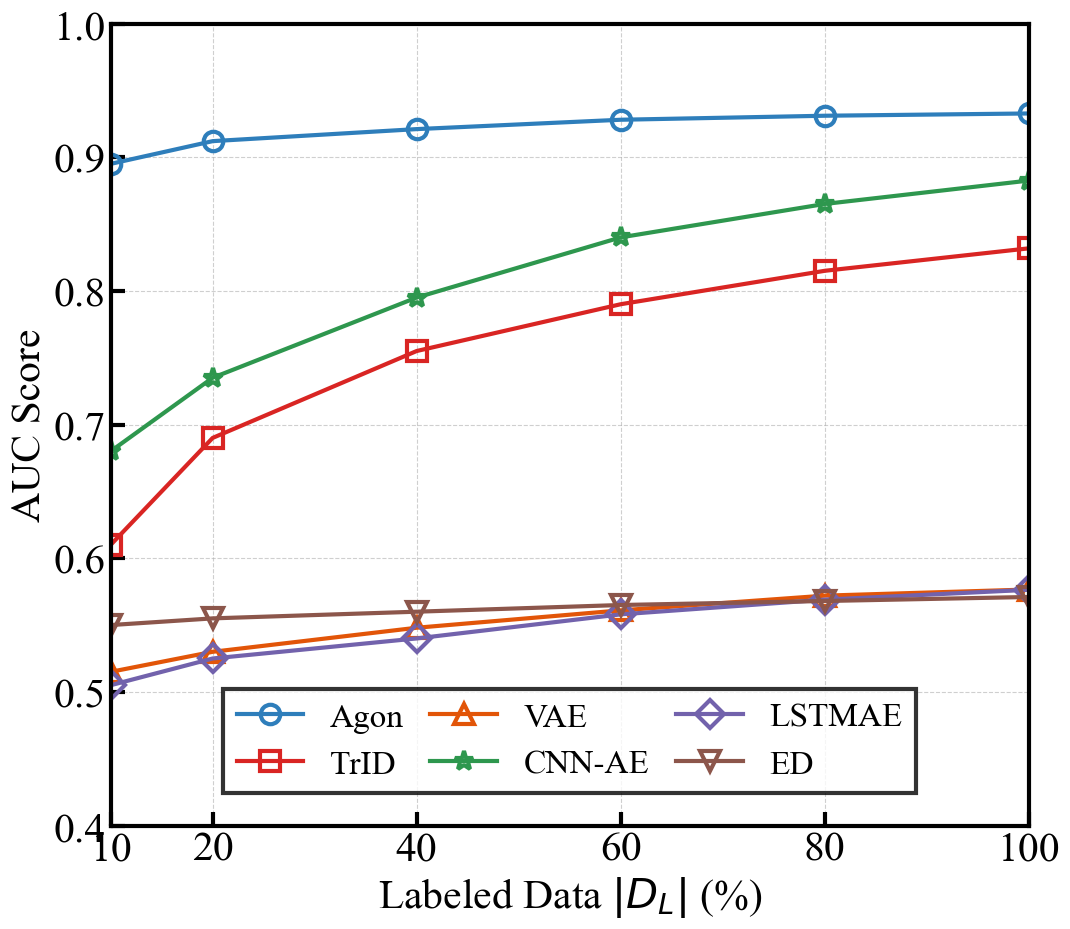}
    \label{fig19a}
  }
  \subfloat[Sample Efficiency on NGSO-NGSO dataset.]{
    \includegraphics[width=0.49\linewidth]{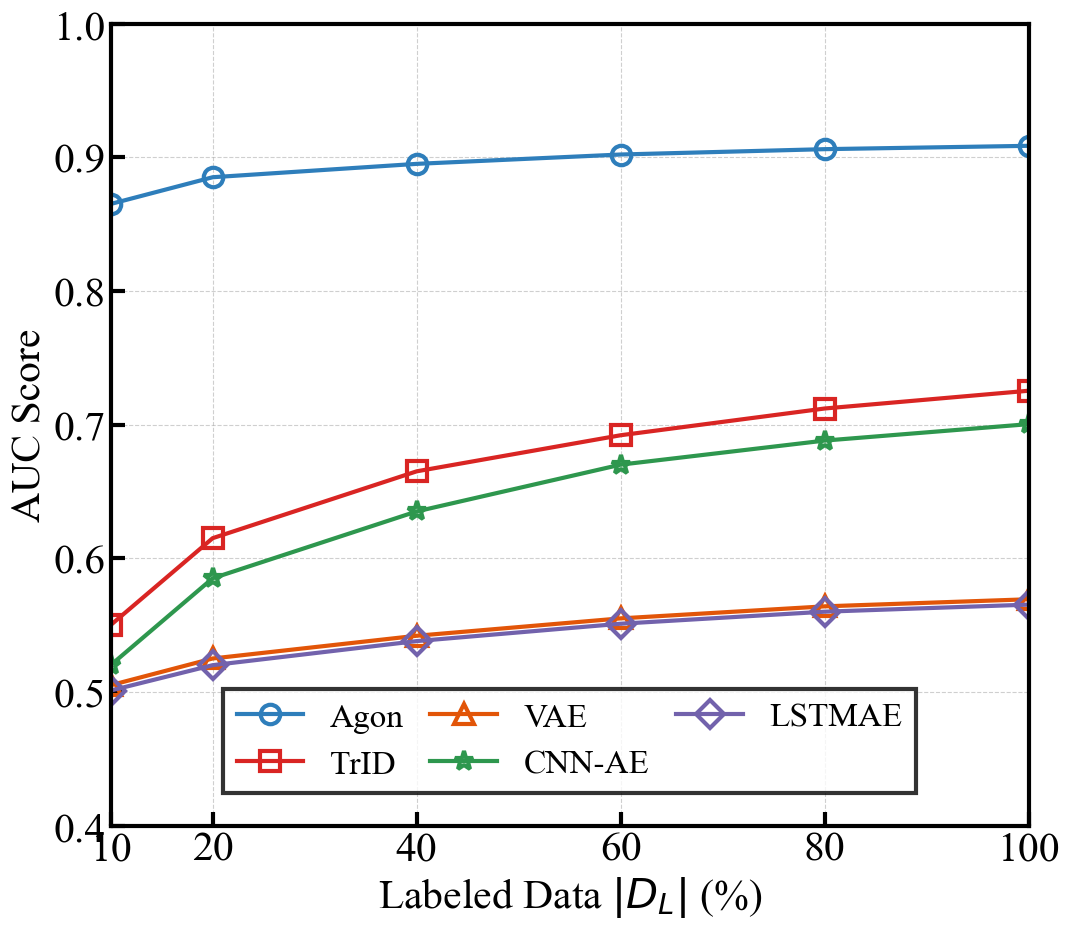}
    \label{fig19b}
  }
  \caption{Performance comparison of sample efficiency under varying proportions of the labeled dataset.}
  \label{fig19}
  \vspace{-3ex}
\end{figure}

Beyond computational efficiency, minimizing the communication overhead for model adaptation is equally critical for dynamic NGSO satellite networks. To evaluate this, Fig.~\ref{fig19} investigates the sample efficiency of the proposed framework across both the public NGSO-GSO dataset, shown in Fig.~\ref{fig19}\subref{fig19a}, and the complex NGSO-NGSO dataset, presented in Fig.~\ref{fig19}\subref{fig19b}, by sub-sampling the labeled dataset $|D_L|$ from 10\% to 100\%. It is important to emphasize that the full labeled dataset utilized in our fine-tuning stage, comprising 802 samples at 100\% of $|D_L|$, already represents a mere 6.48\% of the total available data pool. As demonstrated throughout Fig.~\ref{fig19}, the robust physical representations acquired during the self-supervised pre-training phase endow Agon with exceptional resilience to data scarcity. Even when the labeled data is reduced to 20\% of $|D_L|$, corresponding to approximately 160 samples, the framework maintains a highly competitive AUC of 0.8850 in the challenging NGSO-NGSO scenario. Furthermore, at an extreme 10\% labeling regime, Agon continues to achieve an AUC exceeding 0.86, significantly outperforming baseline models that lack pre-trained feature spaces. Consequently, adapting Agon to new orbital slots requires transmitting a negligible volume of signal samples, effectively mitigating the active communication burden.

\subsection{Ablation Study}
\label{sec68}

\begin{table}[b]
\centering
\caption{Ablation study of Agon performance components}
\label{tab5}
\renewcommand{\arraystretch}{1.2} 
\begin{tabular}{l c c c}
\toprule
Model Variant & Accuracy $\uparrow$ & F1 Score $\uparrow$ & AUC Score $\uparrow$ \\
\midrule
\textbf{Full Agon (Proposed)} & \textbf{0.9027} & \textbf{0.9055} & \textbf{0.9085} \\
\midrule
Agon w/o Wavelet Loss & 0.8718 & 0.8758 & 0.8762 \\
Agon w/o MAE Pretraining & 0.8542 & 0.8631 & 0.8685 \\ 
Agon w/o HOS & 0.8325 & 0.8552 & 0.8614 \\
\bottomrule
\end{tabular}
\end{table}
Comprehensive ablation study is performed to quantify the individual and combined contributions of the core components in the Agon framework. As shown in Table~\ref{tab5} the complete Agon model delivers the best performance reaching an AUC of 0.9085 and an accuracy of 0.9027 confirming the effectiveness of the integrated architecture. To validate the necessity of the self-supervised paradigm we introduced a variant trained from scratch without MAE pre-training. The results demonstrate that skipping the pre-training phase precipitates a significant performance degradation where the AUC drops to 0.8685 indicating that the model fails to internalize universal physical priors of clean signals. This gap proves that MAE pre-training is essential for capturing the underlying structural constants of the communication baseband which provides a superior initialization for subsequent interference detection. The largest performance drop occurs when the HOS Augmentation is removed with the AUC falling to 0.8614 highlighting the crucial role of higher-order statistical cues in improving feature robustness under complex noise conditions. Wavelet Regularization also provides a substantial contribution: eliminating the wavelet loss reduces the AUC to 0.8762 indicating that enforcing consistency in the wavelet domain is important for maintaining high-fidelity representations.

\begin{figure}[t]
  \centering
  \subfloat[Sensitivity analysis of task weights on NGSO-GSO dataset.]{
    \includegraphics[width=0.49\linewidth]{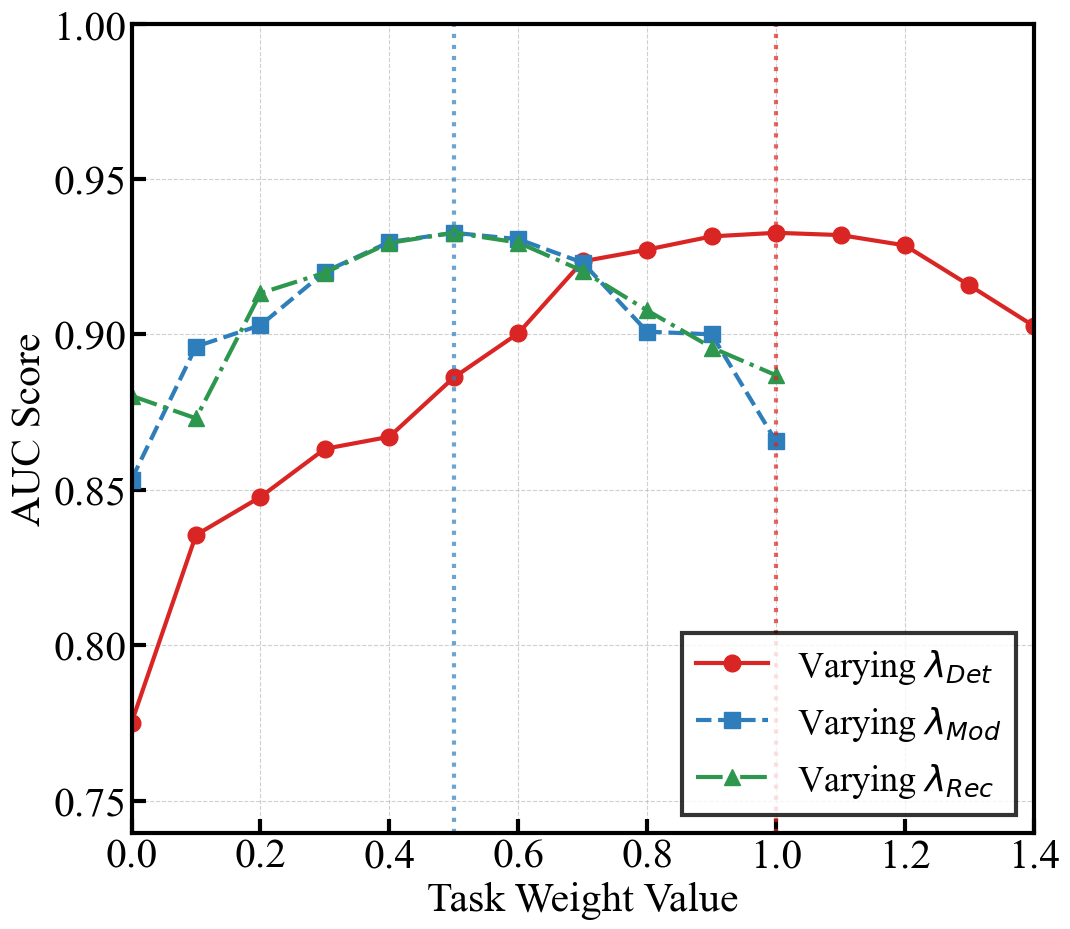}
    \label{fig20a}
  }
  \subfloat[Sensitivity analysis of task weights on NGSO-NGSO dataset.]{
    \includegraphics[width=0.49\linewidth]{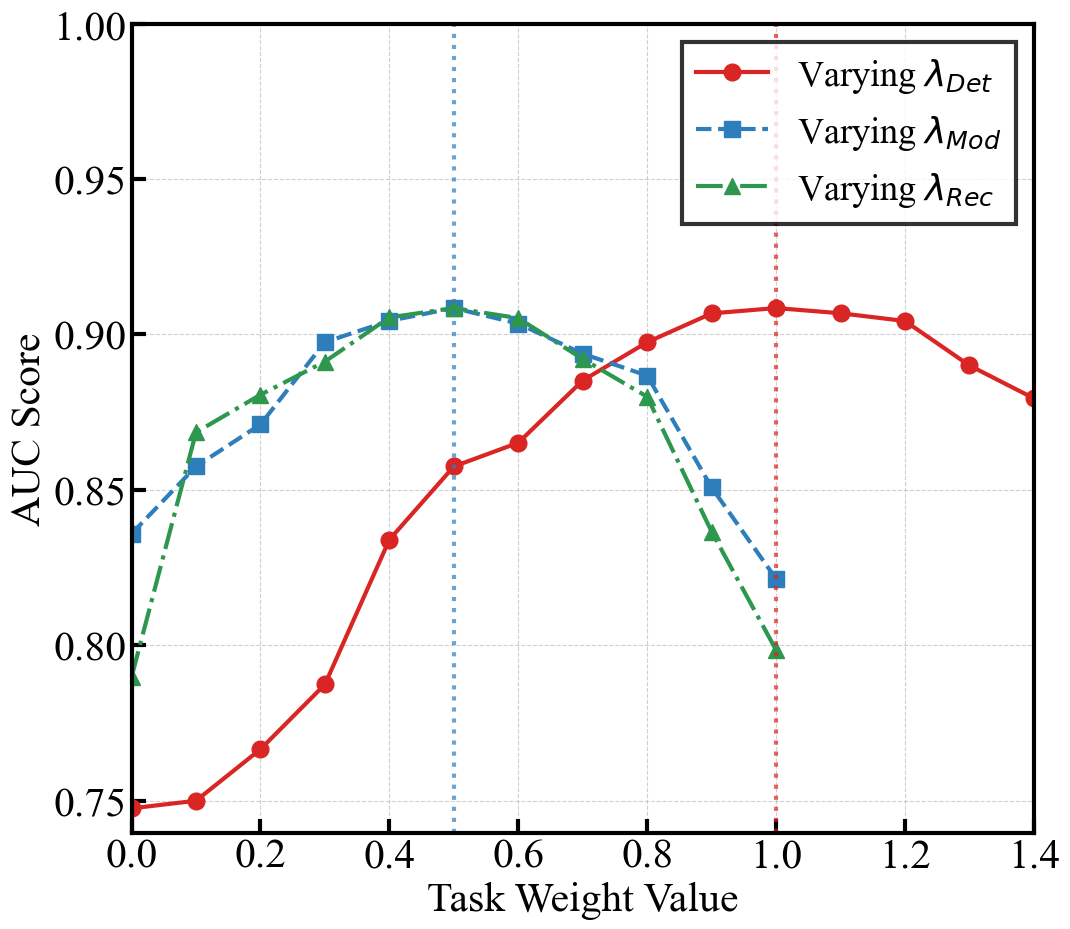}
    \label{fig20b}
  }
  \caption{Impact of individual task weights on detection performance.}
  \label{fig20}
  \vspace{-3ex}
\end{figure}

Furthermore, to empirically validate the theoretical multi-task compatibility and gradient alignment, we conducted a comprehensive sensitivity analysis. Rather than a heuristic parameter sweep, this control variable methodology systematically explores the empirical Pareto optimal front formed by the linear weighted-sum formulation. By independently varying each weight parameter in fine increments of 0.1 while freezing the complementary weights at their established baseline values of 1.0 or 0.5, the performance trajectories in Fig.~\ref{fig20}\subref{fig20a} and Fig.~\ref{fig20}\subref{fig20b} demonstrate that Agon consistently achieves peak AUC at the proposed coordinate configuration. This precise calibration provides empirical evidence that the primary binary detection task successfully governs the main gradient trajectory, while the auxiliary tasks fulfill the compatibility condition in (\ref{equ_gradient}) to provide optimal structural regularization. Any deviation from these theoretically justified coordinates breaches Pareto optimality, immediately inducing catastrophic gradient interference and severe detection degradation. For practical deployments in varying scenarios, these weight parameters should be calibrated via similar step-by-step parameter sweeping on a scenario-specific validation set.

\begin{figure}[t]
  \centering
  \subfloat[Robustness analysis on NGSO-GSO dataset.]{
    \includegraphics[width=0.46\linewidth]{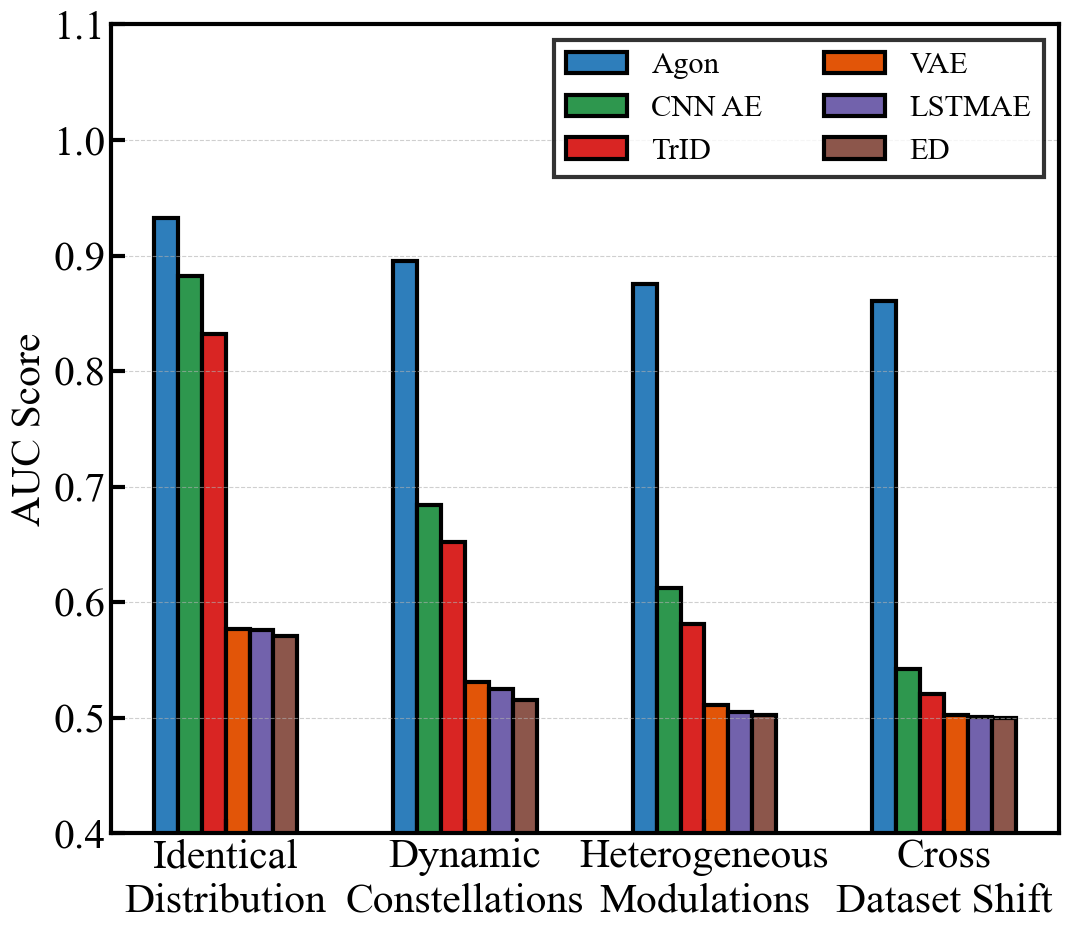}
    \label{fig21a}
  }
  \hfill
  \subfloat[Robustness analysis on NGSO-NGSO dataset.]{
    \includegraphics[width=0.46\linewidth]{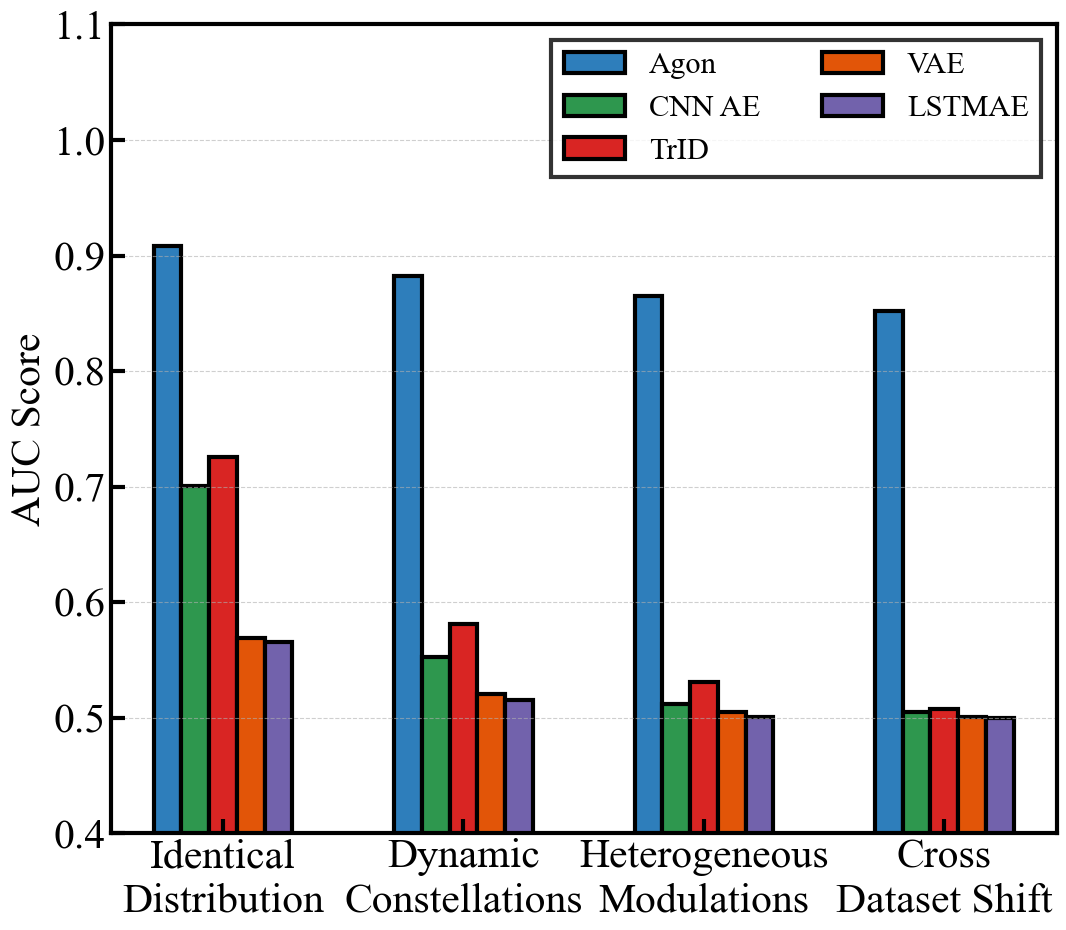}
    \label{fig21b}
  }
  \caption{Comprehensive performance evaluation comparing Identical Distribution deployments with three progressive levels of distribution shifts.}
  \label{fig21}
  \vspace{-3ex}
\end{figure}

Transitioning from internal parameter optimization to external environmental robustness, we extended the assessment from identical distribution scenarios to three progressive out of distribution shifts. As visualized in Fig.~\ref{fig21} the first scenario evaluates robustness against dynamic constellations by training exclusively on Starlink topologies and evaluating zero shot on unencountered OneWeb configurations. The second scenario assesses resilience against heterogeneous modulations by testing on novel high order signatures absent from the training manifold. Finally, a cross dataset shift evaluates zero shot transferability between our distinct datasets. Empirical results demonstrate that severe distribution shifts precipitate performance collapse across traditional baselines indicating overfitting. In stark contrast, Agon demonstrates generalization sustaining AUC scores above 0.85. This resilience originates from the semi supervised MAE formulation which compels the model to reconstruct universal invariant physical priors of clean communication signals rather than memorizing transient interference patterns. By capturing these underlying structural constants Agon effectively decouples its detection mechanism from fluctuating topological states and varying modulation characteristics thereby ensuring reliable deployment across dynamically evolving satellite networks.

\section{Conclusion}
\label{sec7}
This paper explores the application of a semi-supervised learning framework for satellite interference detection, aiming to manage interference in complex NGSO systems. We designed Agon, a two-stage hybrid learning solution, to compensate for the limitations of existing methods, such as reliance on unstable thresholds and high computational costs. Agon learns universal signal representations via MAE pre-training and then optimizes a direct classifier using a MTL fine-tuning strategy. This framework integrates a novel HOS-augmented attention mechanism for noise robustness and a wavelet regularization loss to preserve multi-scale spectral fidelity. Our approach was validated on both a public NGSO-GSO dataset and a high-fidelity NGSO-NGSO dataset. Experimental results show that Agon significantly improves accuracy with a 25.3\% AUC gain over SOTA on the complex NGSO-NGSO set and also enhances generalization and offers greater flexibility by removing threshold dependency and supporting auxiliary tasks.
Future research will prioritize a rigorous comparative analysis between large language models (LLMs) and specialized semi-supervised frameworks regarding their deployment efficiency and architectural value on actual satellite hardware. Evaluating whether the advanced cognitive reasoning of LLMs justifies substantial computational overhead in power-constrained orbital environments remains critical to determining the practical feasibility of next-generation autonomous satellite orchestration.


\bibliographystyle{IEEEtran}
\bibliography{ref}


\begin{IEEEbiography}[{\includegraphics[width=1in,height=1.25in,clip,keepaspectratio]{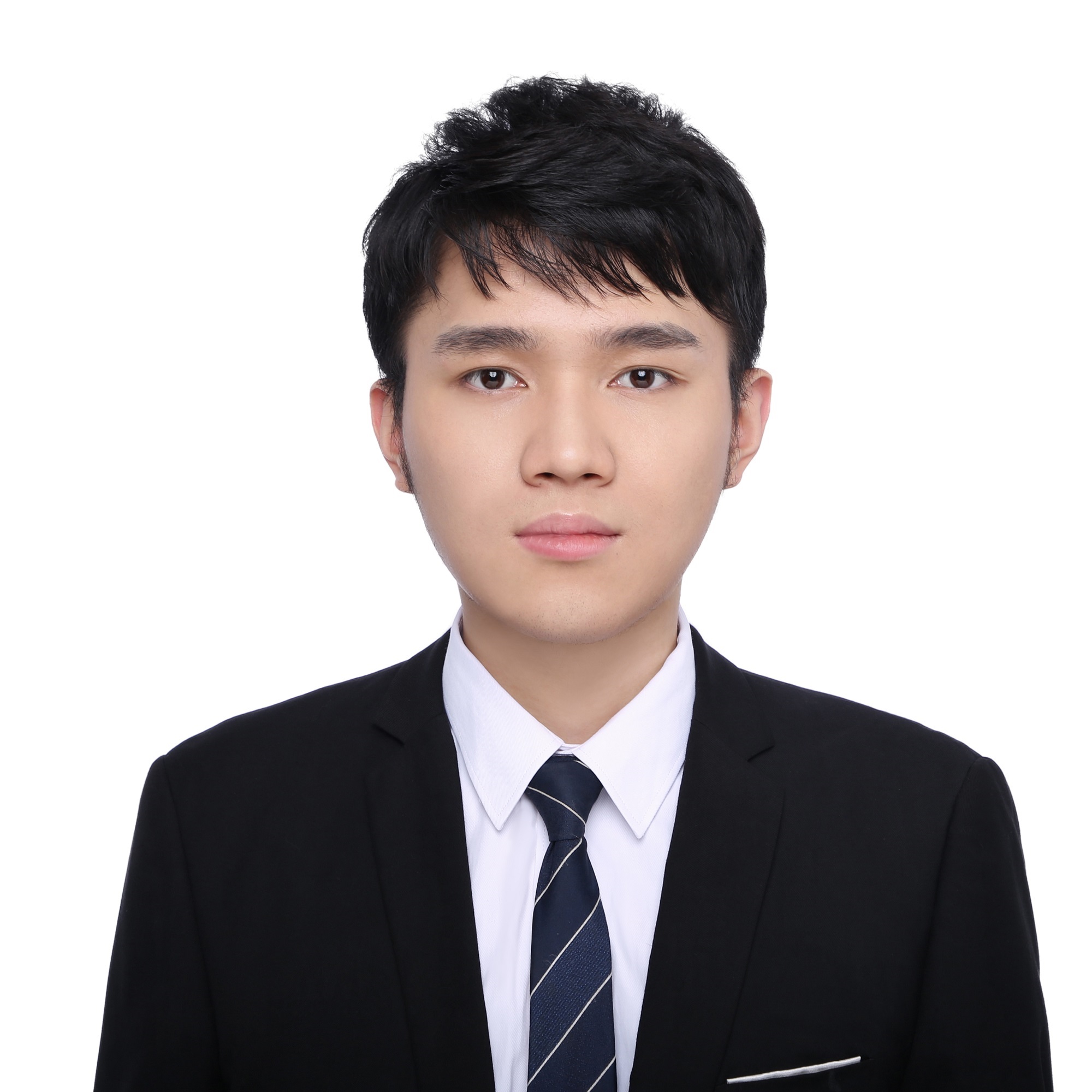}}]{Boyu Yang} is currently working toward the doctoral degree at the College of Computer Science and Artificial Intelligence, Fudan University, China. His research interests include satellite interference detection and machine learning.
\end{IEEEbiography}

\begin{IEEEbiography}[{\includegraphics[width=1in,height=1.25in,clip,keepaspectratio]{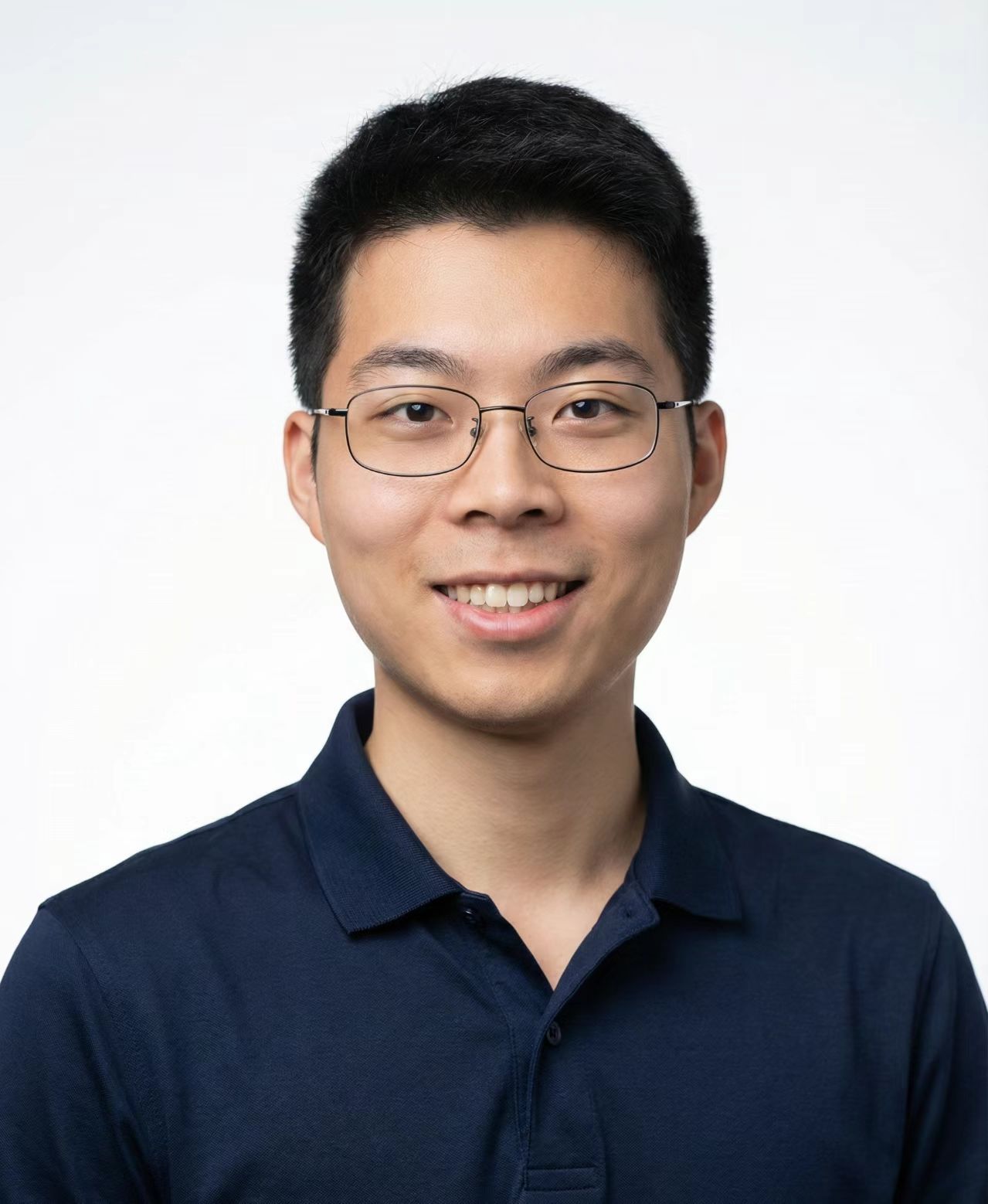}}]{Chunyu Yang} expects to receive his B.S. degree from Fudan University in 2026. His research interests primarily lie in Deep Learning and Embodied AI. 
\end{IEEEbiography}

\begin{IEEEbiography}[{\includegraphics[width=1in,height=1.25in,clip,keepaspectratio]{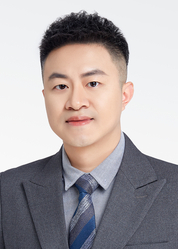}}]{Zhe Chen} received his Ph.D. degree in Computer Science from Fudan University, China, with a 2019 ACM SIGCOMM China Doctoral Dissertation Award. He is an assistant professor within the School of Computing and Intelligent Innovation at Fudan University, and the Co-Founder of AIWiSe Ltd. Inc. Before joining Fudan University, he worked as a research fellow in NTU for several years, and his research achievements, along with his efforts in launching products based on them, have thus earned him 2021 ACM SIGMOBILE China Rising Star Award recently.
\end{IEEEbiography}

\begin{IEEEbiography}[{\includegraphics[width=1in,height=1.25in,clip,keepaspectratio]{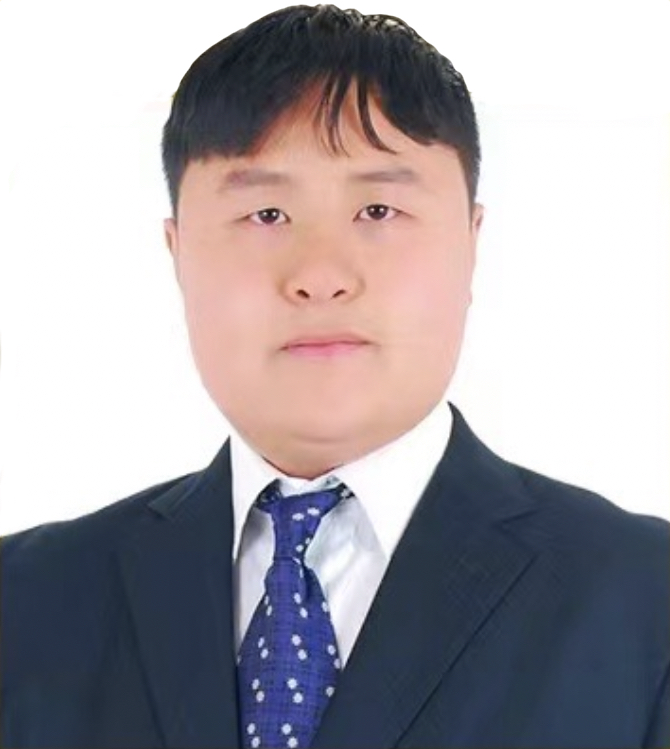}}]{Kun Qiu} [corresponding author] received his B.Sc. Degree from Fudan University in 2013, and received his Ph.D. Degree from Fudan University in 2019. He works for Intel as a software engineer from 2019 to 2023. Now he joined Fudan University in 2023 as an assistant professor. His research interests include computer networks and computer architecture. 
He is a member of ACM and a senior member of IEEE and CCF.
\end{IEEEbiography}

\begin{IEEEbiography}[{\includegraphics[width=1in,height=1.25in,clip,keepaspectratio]{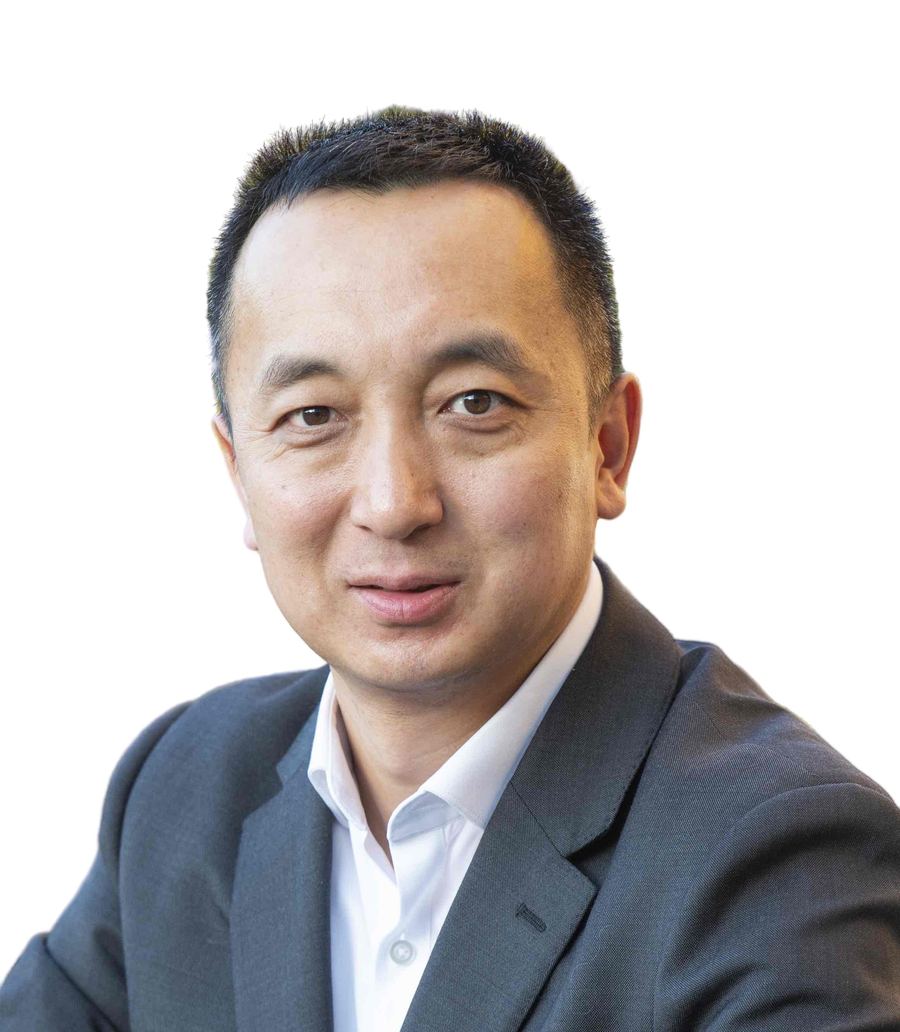}}]{Yue Gao} received his Ph.D. from the Queen Mary University of London (QMUL), U.K., in 2007. He is a Chair Professor at School of Computer Science, and Director of Space Internet Research Institute at Fudan University, China and a Visiting Professor at University of Surrey, UK. His research interests include smart antennas, sparse signal processing and cognitive networks for mobile and satellite systems. He is a fellow of IEEE.
\end{IEEEbiography}

\vspace{11pt}




\vfill

\end{document}